\documentclass[pra,superscriptaddress,twocolumn,showpacs,preprintnumbers,nofootinbib,amsmath,amssymb,longbibliography]{revtex4-2}
\usepackage{comment}
\usepackage{amsmath}
\usepackage{amssymb}
\usepackage{graphicx}
\usepackage{float}
\usepackage{bm}
\usepackage[english]{babel}
\usepackage{dsfont}
\usepackage{epstopdf}
\usepackage{textcmds}
\usepackage{soul}
\usepackage{color}
\usepackage{braket}
\usepackage[colorlinks=true,
            linkcolor=magenta,
            urlcolor=blue,
            citecolor=blue]{hyperref}

\usepackage[normalem]{ulem}
\usepackage{orcidlink}

\begin{document}

\title{Many-body quantum dimerization in 2D atomic arrays}
\author{Giuseppe Calaj\'o \orcidlink{0000-0002-5749-2224}}
\affiliation{Istituto Nazionale di Fisica Nucleare (INFN), Sezione di Padova, I-35131 Padova, Italy}

\author{Matija Te\v{c}er\orcidlink{0009-0009-5903-9499}}
\affiliation{Dipartimento di Fisica e Astronomia ``G. Galilei'', via Marzolo 8, I-35131 Padova, Italy}

\author{Simone Montangero\orcidlink{0000-0002-8882-2169}}  
\affiliation{Dipartimento di Fisica e Astronomia ``G. Galilei'', via Marzolo 8, I-35131 Padova, Italy}
\affiliation{Padua Quantum Technologies Research Center, Universit\'a degli Studi di Padova}
\affiliation{Istituto Nazionale di Fisica Nucleare (INFN), Sezione di Padova, I-35131 Padova, Italy}

\author{Pietro Silvi\orcidlink{0000-0001-5279-7064}}
\affiliation{Dipartimento di Fisica e Astronomia ``G. Galilei'', via Marzolo 8, I-35131 Padova, Italy}
\affiliation{Padua Quantum Technologies Research Center, Universit\'a degli Studi di Padova}
\affiliation{Istituto Nazionale di Fisica Nucleare (INFN), Sezione di Padova, I-35131 Padova, Italy}

\author{Marco Di Liberto\orcidlink{0000-0002-3574-7003}}
\affiliation{Dipartimento di Fisica e Astronomia ``G. Galilei'', via Marzolo 8, I-35131 Padova, Italy}
\affiliation{Padua Quantum Technologies Research Center, Universit\'a degli Studi di Padova}
\affiliation{Istituto Nazionale di Fisica Nucleare (INFN), Sezione di Padova, I-35131 Padova, Italy}

\date{\today}

\begin{abstract}
We consider a 2D atomic array coupled to different photonic environments, focusing on the half-filled excitation subspace, where strong photon interactions can give rise to complex many-body states.
In particular, we demonstrate that the least radiant state in this sector is well described by a coherent superposition of all possible quantum dimer coverings:  a resonating valence bond (RVB) liquid state. We discuss possible strategies to probe this exotic state, along with their limitations and challenges.
Finally, we show that such a quantum dimer covering can also emerge as the ground state of the coherent Hamiltonian describing a 2D atomic array coupled to a photonic band-gap material.

\end{abstract}

\maketitle

The study of photon-mediated atom-atom interactions in atomic arrays coupled to different photonic environments has been an extensive field of research in the past years. 
Two main distinct scenarios have been studied.
In the first one, atoms interact via a process of coherent emission and reabsorption leading to an effective coherent Hamiltonian, a topic extensively explored in the context of cavity QED~\cite{reiserer2015cavity}.
A similar effect can be achieved when atoms are coupled to a photonic band-gap material. 
In this case, the evanescent field in the photonic gap can be employed to engineer long-range interactions~\cite{PhysRevA.50.1764,kofman1994spontaneous,PhysRevA.87.033831,PhysRevA.55.1485,PhysRevA.93.033833}, which can be exploited for quantum simulation of long-range spin models~\cite{douglas2015quantum,gonzalez2015subwavelength,gonzalez2015subwavelength,gonzalez2017markovian,galve2018coherent,gonzalez2018anisotropic,PhysRevX.9.011021,PhysRevResearch.2.043307,gonzalez2024light,RevModPhys.90.031002,PhysRevLett.126.103603,PRXQuantum.4.030306,bello2019unconventional}.
The second one considers the coupling of emitters to radiative modes, thus being intrinsically dissipative. 
This approach began with Dicke's seminal work~\cite{PhysRev.93.99}, where multiple atoms located at the same position are coupled to the electromagnetic field, and predicted the occurrence of collective emission ruled by sub-radiant and super-radiant states.

More recently, significant interest has been directed toward exploring the fate of these collective effects when interactions among emitters acquire spatial dependence, as occurring in ordered atomic arrays coupled to an extended multimode photonic environment.
This  has been extensively studied in the context of 1D waveguide 
quantum electrodynamics (WQED)  ~\cite{sheremet2023waveguide,roy2017colloquium}, where multiple emitters couple to light confined within a one-dimensional channel, either at optical~\cite{lodahl2015interfacing,hood2016atom,corzo2019waveguide,tiranov2023collective}  or microwave frequencies~\cite{astafiev2010resonance,brehm2021waveguide,mirhosseini2019cavity,kannan2023demand}. 
In this case, light confinement gives rise to strong photon correlations, leading to the emergence of repulsive~\cite{albrecht2019subradiant,henriet2019critical,zhang2019theory,ostermann2019super,needham2019subradiance,zhong2020photon,kornovan2019extremely,Schrinski_polariton} and attractive~\cite{shen2007strongly,shen2007stronglyL,zheng2011cavity,mahmoodian2018strongly,prasad2020correlating,mahmoodian2020dynamics,le2022dynamical,tomm2023photon,zhang2020subradiant,poddubny2020quasiflat,bakkensen2021photonic,calajo2022emergence,PhysRevA.108.023707} sub-radiant photonic states.
While most studies have focused on either the few-excitation regime or the specular scenario of super-radiance~\cite{PhysRevLett.131.033605,PhysRevX.14.011020,goncalves2403driven}, where all emitters are initially excited, more complex many-body states are expected to emerge at finite filling fraction.
This question was recently addressed in  studies showing that, at half-filling of excitations, the most sub-radiant state is composed of a product of dimerized (singlet) states~\cite{PhysRevLett.127.173601,PhysRevA.110.053707,lonigro2021stationary}, a  result closely connected to the steady-state dimerization~\cite{stannigel2012driven,PhysRevA.91.042116}  recently experimentally observed in circuit QED~\cite{PRXQuantum.5.030346}.

\begin{figure}[t!]
 \includegraphics[width=0.99\linewidth]{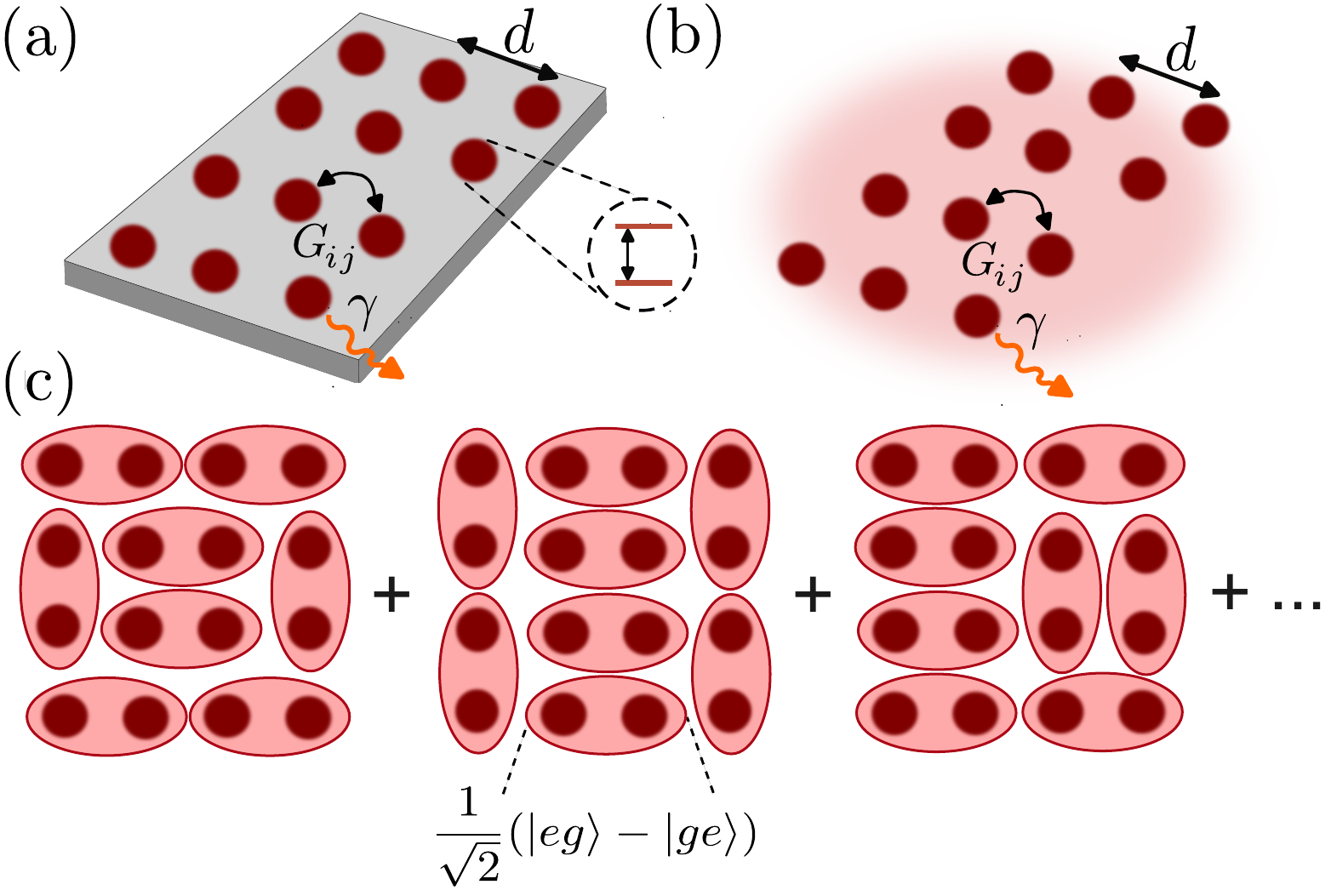}
    \caption{ (a)-(b) A square array of two-level atoms coupled to either (a) a 2D photonic waveguide or (b) the electromagnetic free-space environment.
(c) Schematic of the RVB state, representing an equal superposition of all nearest-neighbor quantum dimer (singlet state) coverings of the lattice.}\label{fig:dimer_covering}
\end{figure}

The recent advances in engineering
 scalable microwave resonator arrays coupled to superconducting qubits~\cite{zhang2023superconducting,scigliuzzo2022controlling,gong2021quantum,kollar2019hyperbolic}, in interfacing two-dimensional atomic arrays with 2D photonic waveguides~\cite{yu2019two} and in preparing ordered sub-wavelength atomic lattices in free space~\cite{rui2020subradiant, srakaew2023subwavelength, HUANG2023100470} have opened up exciting new possibilities for exploring correlated photon dynamics. 
In particular, the emergence of strong photon-photon interactions in two-dimensional arrays has been recently predicted in both platforms~\cite{marques2021bound, tevcer2024strongly, PhysRevResearch.6.043264}.  
This raises the question whether exotic many-body states of light could emerge at half-filling fraction in such higher-dimensional settings.

In this work, we consider a two-dimensional square lattice geometry and focus on the half-filled excitation subspace. 
We demonstrate that the interplay between photon repulsion and long-range, position-dependent dissipative interactions can cause the least radiant state in this sector to resemble a  resonating valence bond (RVB) liquid state~\cite{liang1988some}. 
This state, originally proposed by Anderson ~\cite{anderson1973resonating}, 
emerges in quantum dimer models~\cite{moessner2010quantum,PhysRevLett.61.2376,PhysRevB.35.8865} and it is known to give rise to a topological ordered~\cite{PhysRevB.41.9377} spin-liquid phases~\cite{savary2016quantum,Semeghini2021,pnas.2015785118,PhysRevResearch.7.L012006} in frustrated lattices~\cite{PhysRevLett.86.1881}. 
Here, we demonstrate and characterize the emergence of this state for atomic arrays coupled to a 2D photonic waveguide and for those coupled to a free-space electromagnetic environment. 
As an eigenstate of the non-Hermitian Hamiltonian governing the dissipative dynamics of these systems, this state has a finite lifetime. 
We thus propose a steady-state pumping scheme to probe this state in small lattices, discussing the limitations and challenges associated with extending this approach to larger system sizes. 
Finally, we also consider the case of an atomic array coupled to a photonic band-gap material described by a coherent long-range Hamiltonian and show that the RVB ansatz also efficiently captures its half-filling ground-state structure. 
This scenario may be considered as an alternative route to realize and probe this state and could serve as a valuable resource for quantum computation protocols~\cite{PhysRevLett.131.073602,PhysRevB.109.035128}.

\section{Model}

 We consider an ordered array with lattice constant $d$ made of $N$ two-level atoms, each with ground and excited states 
 $|g\rangle$ and $|e\rangle$, separated by a transition frequency $\omega_{eg}$.
 The atoms are coupled at position $\mathbf{x_i}$ to a generic photonic bath described by a collection of photonic modes indexed  by $n$ and having frequencies $\omega_n$.
The Hamiltonian that describes this coupled atom-light interacting system under the rotating-wave and dipole approximations reads ($\hbar=1$):
\begin{align}
    \label{eq: Full Hamiltionian}
    \begin{split}
        &\hat{H}=\hat{H}_0+\hat{H}_{\rm int}\,, \\
&\hat{H}_0=\sum_{i}\omega_{eg}\hat{\sigma}_{+}^i\hat{\sigma}_{-}^{i}+\sum_n\omega_n\hat a_n^{\dagger}\hat a_n \,, \\
 &\hat{H}_{\rm int}=\sum_{i}\sum_{n}\left[g_{n}(\mathbf{x_i})\hat a_{n}\hat\sigma_+^i+\text{H.c.}\right]\,,
    \end{split}
\end{align}
where  the index $i$ runs over all atoms in the array, and $\hat\sigma_+^i=(\hat\sigma_-^i)^{\dagger}=|e\rangle_i\langle g|$  and  
$\hat{a}_n^{\dagger}$ denote  the atomic and bosonic  creation   operators, respectively. 
The interaction between the atoms and the photonic bath is ruled by the atom-photon coupling constant
 \begin{equation}
    g_n(\mathbf{x}_i)=\sqrt{\frac{\omega_n} {2 \epsilon_0 } }\mathbf{d}^{i}_{eg}\cdot \mathbf{\Phi}_n(\mathbf{x}_i)\,,
\end{equation}
with $\epsilon_0$ being the vacuum permittivity, $\mathbf{d}^{i}_{eg}=\langle e|\hat{\mathbf{d}}^{i}|g\rangle$  the  dipole moment of the i-th atom and $\mathbf{\Phi}_n(\mathbf{x}_i)$  the spatial mode eigenfunction of the field normalized according:
\begin{equation}
    \int d\mathbf{x}\mathbf{\Phi}_n(\mathbf{x})\cdot\mathbf{\Phi}_{n'}(\mathbf{x})=\delta_{nn'}.
\end{equation}
  These functions  are solutions of the Helmholtz equation, determined by the boundary conditions and, consequently, by the geometry of the system. In the next sections, we examine how this influences the photon-mediated interactions between atoms in various geometries, within the framework of the Born-Markov approximation.

\subsection{Effective spin model}
To simplify the description of this complex interacting system, we assume that the standard conditions for applying the Born-Markov approximation are fulfilled. These conditions are satisfied when the density of states of the photonic bath is sufficiently smooth and when time-retardation effects—arising from the finite speed of light $c$—can be neglected. The latter can be formalized by requiring that the timescale of atom-photon dynamics, governed by the spontaneous emission rate $\gamma$ of a single atom into the photonic bath, is much longer than the time it takes for a photon to propagate across the system, i.e., $\gamma \ll c/(N d)$ .
In this regime, light-mediated interactions between atoms can be treated as effectively instantaneous. Equivalently, the strongly dispersive nature of the atoms ensures that the resulting light-matter dressed excitations are predominantly atomic in character. Under these assumptions, the photonic degrees of freedom can be adiabatically eliminated, yielding the following Lindblad master equation that governs the dynamics of the atomic system~\cite{Carmichael1999,Breuer2007}:
\begin{equation}\label{eq:ME_linbl}
  \dot{\hat\rho}(t)=-i[\hat H_S,\hat\rho(t)]+ \sum_{ij}\Gamma_{ij}\left(2 \hat \sigma_-^j\hat\rho \hat\sigma_+^i- \{\hat\sigma_+^i\hat\sigma_-^j,\hat\rho\}\right)\,,
  \end{equation}
where
\begin{equation}
 \hat H_S=  \sum_{i}\omega_{eg}\hat{\sigma}_{+}^i\hat{\sigma}_{-}^{i}+\sum_{i,j} {\rm Re}\{G_{ij}\}\hat{\sigma}_{+}^{i}\hat{\sigma}_{-}^j\,,
 \end{equation}
describes the coherent collective dynamics of the atomic system while the  matrix 
$\Gamma_{ij}=-{\rm Im}\{G_{ij}\}$ encodes the collective dissipation. Coherent and dissipative long-range photon-mediated interactions among the atoms are ruled by the  function $G_{ij}$, which  is directly related to the electromagnetic Green's function  of the photonic environment~\cite{asenjo2017exponential,asenjo2017atom} and is given by:
\begin{equation}
    \label{eq: interaction with integral}
    G_{ij}= \,\lim_{s \to 0^+}\sum_n \frac{g_n(\mathbf{x_i})^*g_n(\mathbf{x_j})}{is-(\omega_{a}-\omega_n)}.
\end{equation}
It is convenient to rewrite the Lindblad master equation written in Eq.~\eqref{eq:ME_linbl}  in the form
\begin{equation}
    \label{eq: Master equation}
    \frac{d\hat{\rho}}{dt}=-i\Big{[}\hat{H}_{\rm eff}\hat{\rho}-\hat{\rho} \hat{H}_{\rm eff}^{\dagger} \Big{]}+2\sum_{i,j} \Gamma_{ij} \hat{\sigma}_{-}^{i}\hat{\rho}\hat{\sigma}_{+}^j\,,
\end{equation}
where the anticommutator part of the master equation has been incorporated into an effective spin model non-Hermitian Hamiltonian, defined as:
\begin{equation}\label{eq:Heff}
    \hat{H}_{\rm eff}=\sum_{i,j}G_{ij}\hat{\sigma}_{+}^{i}\hat{\sigma}_{-}^j\,.
\end{equation}
This effective Hamiltonian  inherently encodes photon-photon interactions, as can be seen by recasting it as a hard-core boson model (see Ref.~\cite{sheremet2023waveguide} for the 1D waveguide case and Ref.~\cite{tevcer2024strongly} for the 2D arrays). 
In the following, we will explore various scenarios characterized by different forms of the interaction matrix $G_{ij}$.

\subsection{Interaction matrix}\label{Sec.interaction matrix}
In this work, we focus our discussion on 
the following scenarios. The first two are sketched in Fig.~\ref{fig:dimer_covering}.
\begin{itemize}
    \item In the first case, we consider a 2D atomic array  coupled to light confined within a 2D photonic waveguide~\cite{tevcer2024strongly,gonzalez2015subwavelength,gonzalez2017markovian,galve2018coherent,gonzalez2018anisotropic}.
We assume the waveguide displays a quadratic and isotropic continuous dispersion relation $\omega_\mathbf{k}=\omega_{c}+A|\mathbf{k}|^2$, where $\omega_\mathbf{k}$ are the field frequencies labeled by the wavevector $\mathbf{k}$,
$\omega_c$ is the dispersion cutoff and we consider $A>0$.
Under this assumption, the rate at which an atom emits a photon into the structure is given by
$\gamma=|g_{\mathbf{k}_0}(\mathbf{x_i})|^2/(4A)$
and the photon-mediated atom-atom interaction matrix reads~\cite{tevcer2024strongly,gonzalez2015subwavelength}:
\begin{equation}\label{eq:2DWQED}
  G^{2D}_{ij} = (\gamma/2)\left[\mathcal{Y}_0(k_{0}x_{ij})-i\mathcal{J}_0(k_{0}x_{ij})\right]\,
\end{equation}
where $k_{0}=2\pi/\lambda_0$ is the photon wavevector, whose corresponding frequency is resonant with the atomic transition of wavelenght $\lambda_0$.
The term $x_{ij}=|\mathbf{x}_i-\mathbf{x}_j|$ denotes the distance between atoms in the 2D array, and $\mathcal{J}_0$ and $\mathcal{Y}_0$ represent the zeroth-order Bessel functions of the first and second kinds, respectively.
\item In the second scenario, we consider a 2D atomic array  embedded in free space~\cite{PhysRevA.86.031602,
bettles2016enhanced,shahmoon2017cooperative,manzoni2018optimization,PhysRevLett.117.243601,PhysRevA.108.030101}.
In this case, atomic emission is governed by the free-space spontaneous decay rate,
$\gamma=d_{eg}^2k_0^3/(3\pi\epsilon_0)$.
Assuming that the atomic dipoles are polarized orthogonally to the array plane, the correlated free-space emission is given by~\cite{asenjo2017exponential,asenjo2017atom}: 
\begin{equation}\label{eq:3D}
G^{2D,\mathrm{free}}_{ij} = \frac{3\gamma\,e^{ik_{0}x_{ij}}}{4(k_{0}x_{ij})^3}[1 - ik_{0}x_{ij} - (k_{0}x_{ij})^2]
\end{equation}

\item Finally, for comparison with previous results, we  consider the case of multiple atoms coupled to a one-dimensional waveguide~\cite{sheremet2023waveguide}. In this case, assuming a linear dispersion relation for the photons of the form $\omega_k=c|k|$,  the decay rate of the atom into the waveguide is given by $\gamma=|g_{k_0}(x_i)|^2/(2c)$ and the effective interactions are described by the interaction matrix~\cite{sheremet2023waveguide}:
    \begin{equation}\label{eq:1DWQED}
      G^{1D}_{ij} = -i\frac{\gamma}{2}e^{ik_0x_{ij}}\, .
    \end{equation}
Note that for atomic distances $k_0|x_i-x_j|=2\pi n$ with $n \in \mathbb{N}$,
this last model reduces to the dissipative Dicke model with $G^{1D}_{ij} = -i\gamma/2$~\cite{albrecht2019subradiant}. This limit is crucial for understanding the occurrence of dimerization, as discussed in the following sections.
\end{itemize}

\subsection{Half-filling subspace}
Although the full open-system dynamics is governed by Eq.~\eqref{eq: Master equation}, in the following analysis we exclusively focus on the effective Hamiltonian in Eq.~\eqref{eq:Heff} at half-filling, \emph{i.e.} with $N/2$ excited atoms in the whole array. 
This approach offers a comprehensive description of the system within a fixed excitation sector, provided that no external pumping fields are present.

From the exact diagonalization of the half-filling sector, we obtain a set of eigenvalues of the form \begin{equation}\label{eq:eigenvalues}
E^{(N/2)}_s = \epsilon^{(N/2)}_s -i\gamma^{(N/2)}_s/2,
\end{equation} 
where the index $s$ orders the states by increasing decay rate. 
Here, $\epsilon^{(N/2)}_s$ and $\gamma^{(N/2)}_s$ denote the energy and collective decay rate of the $s$-th state, respectively.

Fixing the excitation sector also allows us to apply exact diagonalization methods within a reduced Hilbert space of dimension given by the binomial coefficient \( \chi = \binom{N}{N/2} \). This reduction enables us to access system sizes of up to \( N = 24 \) atoms.

\begin{figure*}[t!]
 \includegraphics[width=0.7\linewidth]{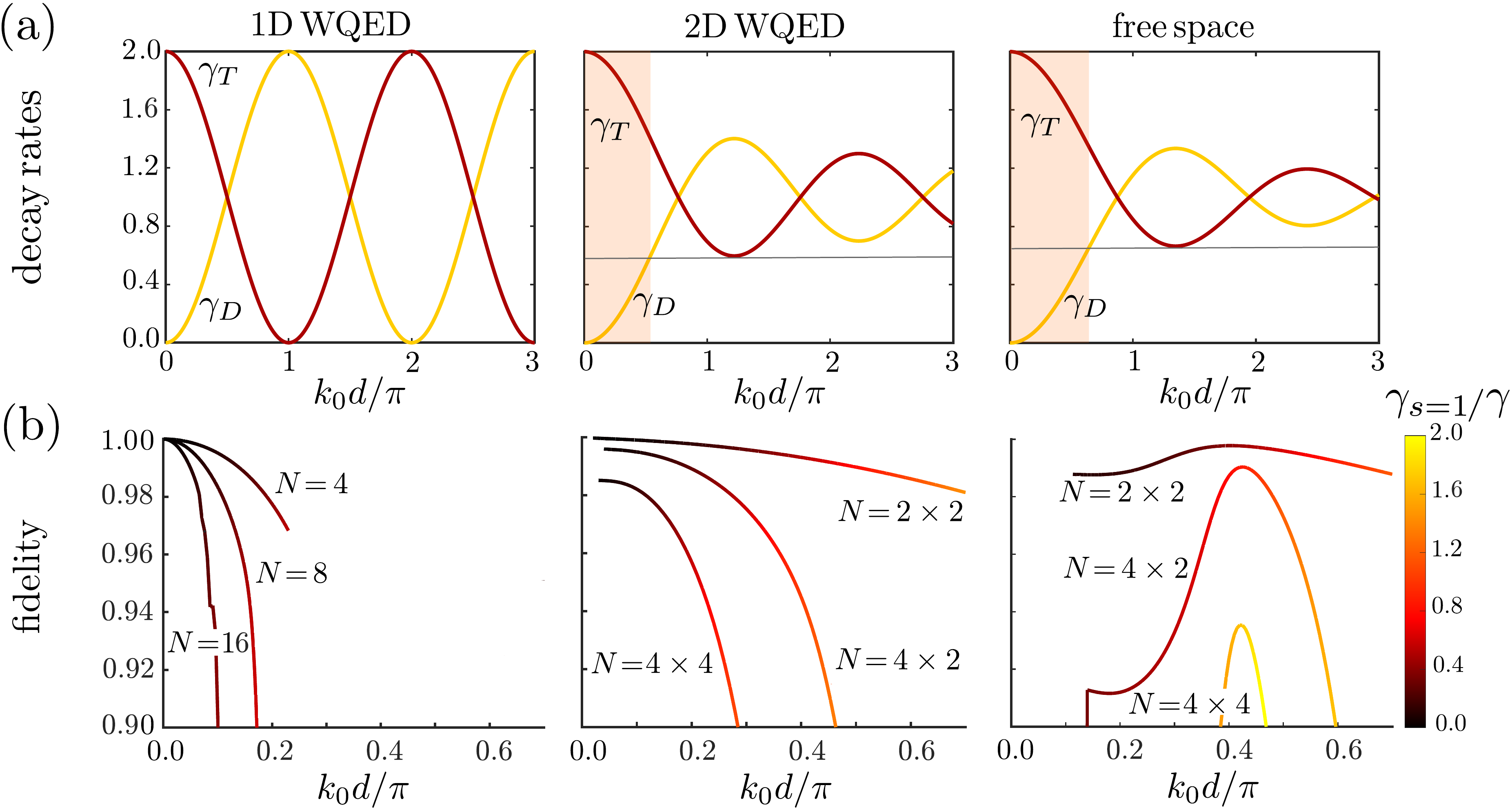}
    \caption{(a) Singlet and triplet decay rates in unit of $\gamma$ for two atoms as function of the inter-atomic distance for the three models of Eqs.~\eqref{eq:2DWQED}-\eqref{eq:1DWQED}. The shaded orange area indicate the inter-atomic distances where a NN dimer is expected to be less radiant than long range dimers and triplet states.
(b) Fidelity as function of the inter-atomic distance between the dimer covering ansatz  and the least radiant state within the $N/2$ subspace of the effective Hamiltonian \eqref{eq:Heff} for the three considered models. 
Note that we use Eq.~\eqref{eq:dimer1D} as the ansatz for the 1D waveguide case, and Eq.~\eqref{eq:RVB} for the two-dimensional setting.
We considered a $N=4,8,16$ chain for the 1D waveguide and a $N=2\times2,4\times2,4\times4$ square lattice for the other two cases. The color scale indicates the decay rate $\gamma_{s=1}$ of the  least radiant state. The results are obtained via exact diagonalization of the non-Hermitian Hamiltonian $\hat H_\text{eff}$. 
}\label{fig:Fvsdist}
\end{figure*}

\section{Many-body quantum dimerization}\label{Sec.dimer}

In this section, we discuss the occurrence of many-body dimerization across the different scenarios introduced in Sec.~\ref{Sec.interaction matrix}. We begin in Sec.~\ref{Sec.dim1D} with the 1D waveguide case, in order to build physical intuition for identifying the least radiant state of the system at half-filling as a product of dimer states. We then examine in Sec.~\ref{Sec.dim2D} how this mechanism can lead to the emergence of an RVB state as the least radiant state in a 2D array. Finally, in Sec.~\ref{Sec.dim_car}, we characterize this state in detail.

\subsection{Dimerization in 1D waveguide QED}\label{Sec.dim1D}

For an atomic array coupled to a 1D waveguide, it is well established that subradiant states in the two-excitation sector are typically composed of repulsive (or so-called \emph{fermionic}) excitations~\cite{asenjo2017exponential,albrecht2019subradiant,henriet2019critical,zhang2019theory,ostermann2019super,needham2019subradiance,zhong2020photon,kornovan2019extremely,Schrinski_polariton}.
For these states, the two particles are maximally far from each other and from the edge of the system as well, thus reducing the likelihood of photons scattering out of the system. 
This becomes particularly clear at the two-excitation level, by solving the two-body problem described by Hamiltonian~\eqref{eq:Heff} in the relative coordinate frame~\cite{sheremet2023waveguide,tevcer2024strongly}. In this framework, the hard-core photon-photon interactions are mapped onto an impurity term, which captures the atomic saturation, i.e., the inability of an atom to absorb more than one excitation. The repulsive behavior of the excitations thus emerges from the effective hard-wall condition imposed by this impurity potential.

At higher fillings, the mechanism is similar by shrinking the multiparticle distribution accordingly~\cite{PhysRevLett.127.173601}.
Instead, half-filling is the limiting case where excitations can spread the least and one might intuitively expect the least radiant state to emerge from each excitation being shared between, in principle, any pair of atoms. 
This intuition was formalized for 1D waveguide QED systems in Ref.~\cite{PhysRevLett.127.173601}, which demonstrated that the least radiant state at half-filling is represented by a product state of all nearest-neighbor (NN) dimers (singlet states):
\begin{equation}\label{eq:dimer1D}
 |\phi_D\rangle=\prod_{j\in \rm odd}\hat D^\dagger_{j,j+1}|0\rangle\,.
\end{equation}
where 
\begin{equation}
    \hat D^\dagger_{i,j}=\frac{1}{\sqrt{2}}(\hat\sigma^i_+-\hat\sigma^j_+)\,
\end{equation}
is the dimer creation operator.  
The formation of dimers only between nearest-neighbor atoms can be understood through the following argument. In the Dicke model limit of Eq.~\eqref{eq:1DWQED}, the dark subspace with zero decay rate consists of all possible dimers, including both short- and long-range configurations. 
As one moves smoothly away from this limit, the spatial dependence of $G_{ij}^{1D}$ induces a spatial dependence of the collective decay rates as well. As we will see below, short-range dimer coverings are favoured over long-range ones, as they minimize the decay rates of emitter pairs. Consequently, a state predominantly composed of nearest-neighbor dimers is filtered out from the degenerate singlet subspace.

To make these arguments more concrete and to assess whether they hold in higher dimensions, let us consider the simple case of two atoms, labeled by indices $i$ and $j$, coupled to a photonic bath.
In this case, the half-filling regime trivially corresponds to the single-excitation subspace, which is spanned by the singlet state, $|D_{ij}\rangle=\hat D^\dagger_{i,j}|0\rangle$, and the triplet state,  $|T_{ij}\rangle=\hat T^\dagger_{i,j}|0\rangle$, with associated collective decay rates:
\begin{equation}\label{rate_single_triplet}
  \gamma^{ij}_D=\gamma+2{\rm Im}\{G_{ij}\}\quad\quad\quad \gamma^{ij}_T=\gamma-2{\rm Im}\{G_{ij}\}\,,
\end{equation}
where \mbox{$\hat T^\dagger_{i,j}=(1/\sqrt{2})(\hat\sigma^i_+ +\hat\sigma^j_+)$} is the triplet creation operator.
In the dissipative Dicke model limit, where $G_{ij}=-i\gamma/2$, there is no spatial dependence in the atom-atom interaction and the singlet state is perfectly dark, $\gamma_D=0$, while the triplet state is super-radiant, $\gamma_T=2\gamma$.
When there is an explicit spatial dependence in the dissipative atomic interaction, $\text{Im} \{G_{ij}\}$, the dark states acquire a finite decay rate. The dependence of these two decay rates on the inter-atomic distance for atoms coupled to a 1D waveguide is shown in the first panel of Fig.~\ref{fig:Fvsdist}(a). 
The infinite-range interaction causes the two decay rates to fully swap between the dark and superradiant regimes at integer distances $k_0d=n\pi$, with $n\in \mathbb{N}$. 

Assuming a purely pairwise entanglement structure, as motivated above, we can thus pose a necessary condition for having a covering of nearest-neighbor dimers as the least radiant state in extended systems.
This condition requires that the decay rates of the dimer state involving any two NN atoms,  $\gamma^{\langle ij\rangle}_D$, should be smaller than the decay rate associated to all possible triplet pairs, $\gamma^{ij}_T$, and any other non NN dimer, $\gamma^{\rangle ij\langle}_D$:
\begin{equation}\label{eq:cond}
\gamma^{\langle ij\rangle}_D<  \{\gamma^{\rangle ij\langle}_D,\gamma^{ij}_T\}.
\end{equation}
If this condition is not met, the formation of triplets or long-range singlet pairings will result in a state with a lower decay rate.
 In the 1D waveguide scenario the condition given in Eq.~\eqref{eq:cond} needs to be fine tuned for sufficiently large arrays.
Indeed, since the spatial periodic dependence of the decay rate (see Fig.~\ref{fig:Fvsdist}(a)) is generally incommensurate with the lattice spacing, there can be an atom-atom distance at which the decay of long-range dimers or triplet states becomes arbitrarily smaller than that of nearest-neighbor  ones (see App.~\ref{AppA}).
At such distances, the least radiant state will start incorporating long-range dimers or triplet states, reducing its resemblance to the dimerized ansatz given in Eq.~\eqref{eq:dimer1D}.

This can be seen in the the first panel of Fig.~\ref{fig:Fvsdist}(b), where we compute the global fidelity, $F=|\langle \phi_D |\psi_{\rm s=1}\rangle|^2$, between the least radiant state of the $N/2$ excitation subspace, $|\psi_{\rm s=1}\rangle$, labeled by the index $s=1$ defined in Eq.~\eqref{eq:eigenvalues}, and the dimer product state given in Eq.~\eqref{eq:dimer1D}. The fidelity rapidly decreases with interatomic distance as the array size increases, due to the argument explained above.

\subsection{Dimerization in 2D arrays}\label{Sec.dim2D}

In two spatial dimensions, both for the 2D WQED and free-space cases, repulsive two-excitation sub-radiant states have been identified~\cite{tevcer2024strongly}. 
These findings suggest that a pairwise entanglement structure at half-filling fraction could also emerge in this context. 
The condition given in Eq.~\eqref{eq:cond} is satisfied within the distance range highlighted by the orange-shaded area in Fig.~\ref{fig:Fvsdist}(a). 
This region corresponds to where the NN dimer decay rate remains lower than the decay associated with long-range dimer pairs and the triplet state. This behavior arises due to the dampening of the collective decay rates, which scale as $1/\sqrt{k_0d}$ in a 2D waveguide and as $1/(k_0d)$ in free space.
This suggests  that nearest-neighbor dimers should be favored in forming the least radiant state.
However, in contrast with the 1D waveguide case, in this scenario there are different possible dimer coverings.
In the following discussion, we focus on the non-frustrated case of a square lattice consisting of $N=N_x\times N_y$ atoms.
Given that it is reasonable to assume no particular dimer covering is dissipatively favored over another,
we conjecture that, in this scenario, the least radiant state at half-filling is a coherent and equal superposition of all nearest-neighbor quantum dimers, as illustrated in Fig.~\ref{fig:dimer_covering}(c).  We thus propose the following ansatz:
\begin{equation}\label{eq:RVB}
 |\phi_{\rm RVB} \rangle=  \sum_\alpha\prod_{i,j\in \mathcal{D}_\alpha}\hat D^\dagger_{i,j}|0\rangle\,,
\end{equation}
where the sum runs over all possible NN dimer coverings $\mathcal{D}_\alpha=\{(i_1,i_2),(i_3,i_4)....\}_\alpha$  and the pair of indices $(i_a,i_{b})$ labels the position of the two atoms forming a given dimer. 
For instance, for a $N=2\times2$  lattice labeling the atoms in the first raw as $1,2$ and in the second as $3,4$ there are two possible dimer coverings given by $\mathcal{D}_1=\{(1,2),(3,4)\}_1$ and 
$\mathcal{D}_2=\{(1,3),(2,4)\}_2$. 
As the size of the array increases, the number of possible dimer coverings grows exponentially, but all the possible dimer coverings in the lattice can be found numerically (see App.~\ref{AppB}).
Note that this ansatz coincides with the  ``Resonating Valence bond'' (RVB) liquid phase wavefunction for a square lattice, as proposed by Anderson~\cite{anderson1973resonating,liang1988some} and also appearing in the Rokhsar-Kivelson (RK) quantum dimer model at the so-called RK point~\cite{moessner2010quantum,PhysRevLett.61.2376,PhysRevB.35.8865}.

To probe whether our conjecture holds for a 2D atomic array, we compute the global fidelity $F=|\langle \phi_{\rm RVB} |\psi_{\rm s=1}\rangle|^2$, between the least radiant state at half-filling, again labeled by the index $s=1$ defined in Eq.~\eqref{eq:eigenvalues}, and the RVB ansatz.
The global fidelity as a function of inter-atomic distance is shown in Fig.~\ref{fig:Fvsdist}(b) for the three cases described in   Eqs.~\eqref{eq:2DWQED}-\eqref{eq:1DWQED} and different system sizes. 
In all cases, the fidelity decreases as the system size increases. 
This drop in fidelity can be attributed to the fact that both dissipative and coherent interactions induce mixing among the various dimer configurations that span the original singlet subspace, thereby perturbing the RVB-like state.
Interestingly, the ansatz appears to be more robust for 2D arrays than the 1D scenario, with the maximum fidelity extending over a larger range of finite distances in the region where Eq.~\eqref{eq:cond} holds. 
Specifically, the fidelity maximum remains relatively flat at shorter distances in the 2D waveguide case.
In free space, instead, the maximum in fidelity is achieved at an intermediate distance $k_0d\sim0.4\pi$. 
This effect is most likely due to the fact that strong near-field coherent interactions occur at short distances, thus shifting the optimal regime to larger inter-atomic distances. 

In Fig.~\ref{fig:Fvsdist}(b), we also highlight the decay rate associated with this state. 
The results show that the least radiant state at half-filling fraction is actually quite radiant, particularly in the free-space setting. 
This presents an important limitation for the observation of this state, as we will discuss in Sec.~\ref{Sec.steady}.

\begin{figure}[!t]
\includegraphics[width=0.49\textwidth]{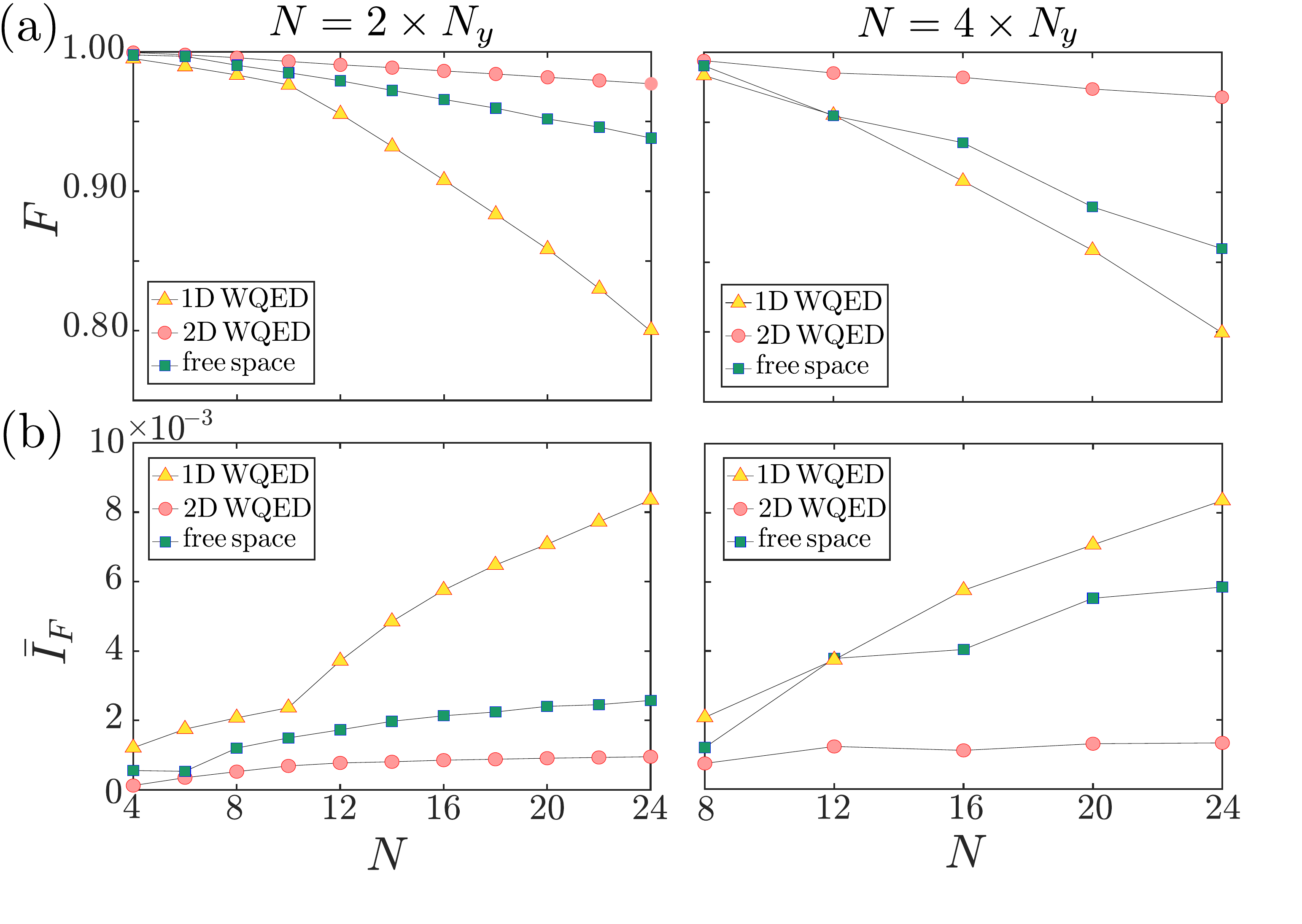}%
    \caption{(a) Global Fidelity, $F$, and (b) global Infidelity density, $\bar{I}_F$, between the dimer covering ansatz (Eq.~\eqref{eq:dimer1D} for the 1D WQED case and Eq.~\eqref{eq:RVB} for the 2D case) and the least radiant state for the models~\eqref{eq:2DWQED}-\eqref{eq:1DWQED}   as function of the system size for two different lattice geometries, as indicated. For the 1D and 2D waveguide case we fixed $k_0d=0.1\pi$ while for the free space case we fixed the inter-atomic distance to $k_0d=0.42\pi$.
    }\label{fig:F_scaling}
\end{figure}

\subsection{State characterization}\label{Sec.dim_car}

To better assess the performance of the RVB ansatz, we plot the scaling of the global fidelity with respect to the system size in Fig.~\ref{fig:F_scaling}(a). 
We focus on two different lattice geometries, namely ladders with $N_y=2$ and $N_y=4$, and compare the results with the 1D waveguide scenario for direct comparison.
In all cases, we observe a decrease in fidelity with increasing system size. However, while the decrease in the 1D case is quite steep, as expected from the previous discussion, the fidelity remains relatively robust for  two-dimensional arrays.
The decrease in fidelity is not surprising, as for a many-body state ansatz small deviations from the true 
state typically lead to a decrease of the global fidelity linearly with the system size.
This implies that the global infidelity density defined as $\bar I_F=(1-F)/N$ should saturate when increasing the system size.
In  Fig.~\ref{fig:F_scaling}(b), we show that this trend  holds for two-dimensional arrays, particularly in the 2D waveguide QED scenario.

The fact that the least radiant state is well approximated by the 2D RVB ansatz should be reflected in the system's entanglement structure and correlations. 
To verify this, and considering that state fidelity cannot be measured in many-body quantum systems, we compute several relevant observables benchmarked against the exact RVB ansatz. 

\begin{figure}[!t]
    \includegraphics[width=0.49\textwidth]{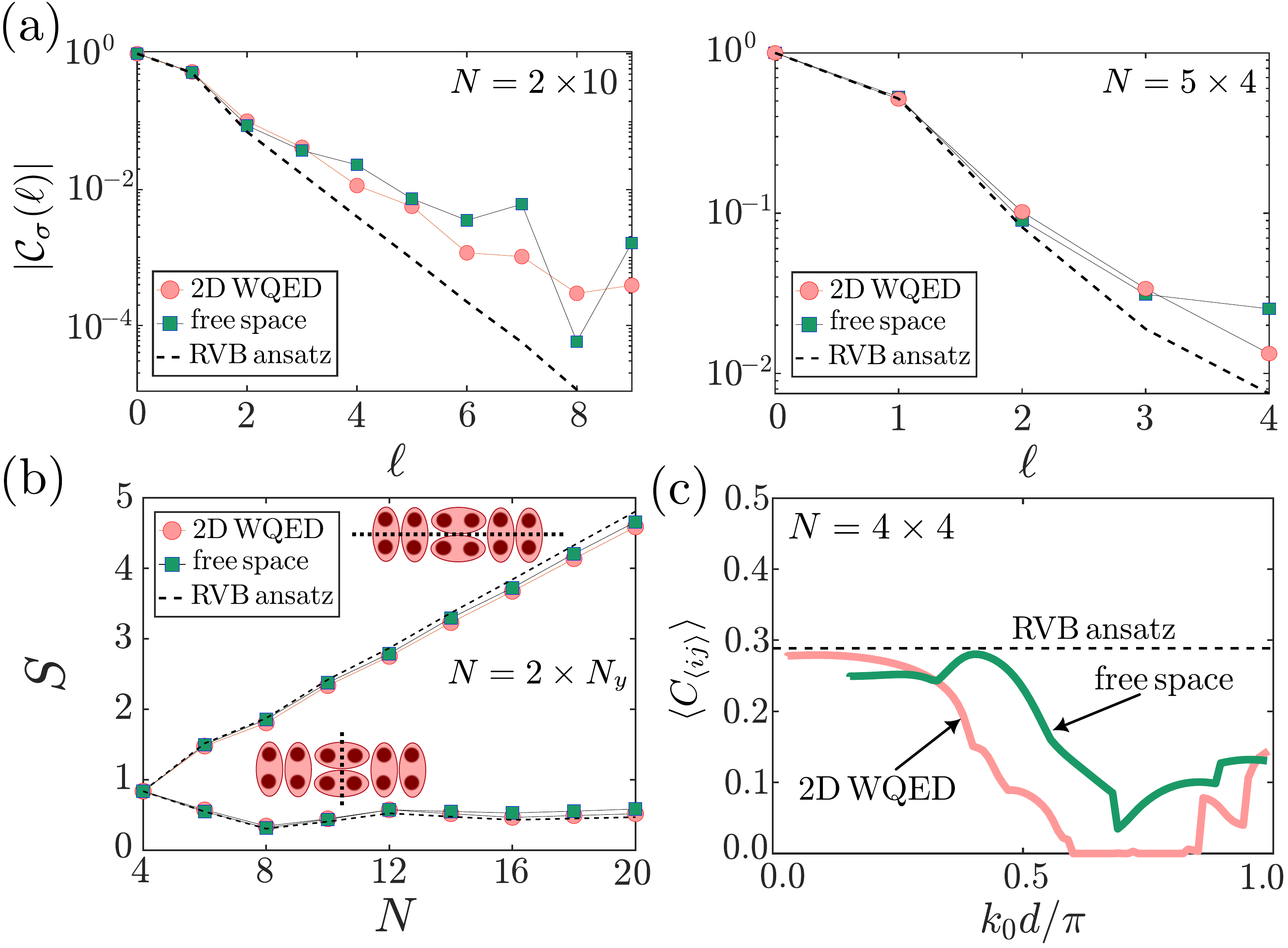}%
    \caption{(a) Absolute value of the spin-spin correlation function for two lattices of sizes $N=2\times 10$ and $N=5\times 4$. In the first geometry we fixed: $\mathbf{i}=(1,1)$ and $\boldsymbol{\ell}=(1,\ell)$ while in the second $\mathbf{i}=(2,1)$ and $\boldsymbol{\ell}=(2,\ell)$.
    (b) Bipartite entanglement entropy in a ladder as a function of system size  for two different partitions as indicated in the cartoon sketches.
    (c) Average nearest-neighbor concurrence as a function of inter-atomic distance for a $N=4\times 4$ lattice. 
    In all plots, we considered the  two models given in Eqs.~\eqref{eq:2DWQED} and~\eqref{eq:3D} fixing the inter-atomic distance to $k_0d=0.1\pi$ and $k_0d=0.42\pi$, respectively.
    The results obtained from the RVB ansatz are  reported with dashed line for direct comparison.
}\label{fig:liquid_phase}
\end{figure}

\begin{itemize}
    \item \textbf{Spin-spin correlations}. We start by evaluating the  spin-spin correlation function defined as 
\begin{equation}
   \mathcal C_\sigma(\boldsymbol{\ell})=\frac{1}{3}\langle \hat \sigma_x^{\mathbf{i}}\hat \sigma_x^{\mathbf{i}+\boldsymbol{\ell}}+\hat \sigma_y^{\mathbf{i}}\hat \sigma_y^{\mathbf{i}+\boldsymbol{\ell}}+\hat \sigma_z^{\mathbf{i}}\hat \sigma_z^{\mathbf{i}+\boldsymbol{\ell}}\rangle \,,
\end{equation}
where the expectation value is taken with respect to the target state, the vector  $\mathbf{i}=(i_x,i_y)$ defines the position of the first atom and the vector $\boldsymbol{\ell}=(\ell_1,\ell_2)$ represents the relative distance of the second atom from the first. 
The absolute value of this quantity is shown in Fig.~\ref{fig:liquid_phase}(a) for two different 2D lattice arrangements as indicated in Fig.~\ref{fig:liquid_phase}. 
We compute this quantity by fixing the position of one of the spin, and by varying   the second as indicated in the caption of the figure.
For all the considered cases we observe short range correlations, with an approximate exponential decay. 
This signals the absence of long-range order as expected for a liquid phase.

\item \textbf{Bipartite entanglement entropy.} We next consider the  bipartite entanglement entropy, defined as $S=-{\rm Tr}[\varrho_A\log \varrho_A]$, where $A$ and $B$ represents the two partitions of the lattice and $\varrho_A={\rm Tr}_B[\varrho_{AB}]$ is the density operator reduced to the A subsystem. We focus our analysis on  a $N=2\times N_y$ ladder, though similar considerations apply to other lattice geometries and partitioning schemes. 
For this ladder, we consider two bi-partitions.
The first bi-partition divides the ladder along its long edge by cutting every rung, effectively cutting every rung and separating the two legs (see the schematic illustration in Fig.~\ref{fig:liquid_phase}(b)). If the system is in an RVB-like state, the entanglement entropy is expected to grow linearly with system size, as the number of dimers cut by this partition increases proportionally.
The second bi-partition divides the ladder along its short edge. In this case, the partition cuts at most two dimers, leading to an entanglement entropy that remains constant as the system size increases.
In Fig.~\ref{fig:liquid_phase}(b), we present results for both the 2D waveguide and free-space scenarios. These results well align with the expected behavior and  follow the predictions of the RVB ansatz.

\item \textbf{Concurrence.} Finally, to locally probe the  entanglement  structure of the state under consideration  we  use   the Wootters concurrence, a monotone entanglement measurement \cite{PhysRevLett.80.2245}. 
The concurrence for the density operator of a two qubits system is defined as:
\begin{equation}
C(\varrho)=\rm max(0,\lambda_1-\lambda_2-\lambda_3-\lambda_4)
\end{equation}
where $\lambda_i$ are the eigenvalues in decreasing order of the operator $R=\sqrt{\sqrt{\varrho}\tilde\varrho\sqrt{\varrho}}$ with  \mbox{$\tilde\varrho=(\sigma_y\otimes\sigma_y)\varrho^*(\sigma_y\otimes\sigma_y)$} and $\varrho$ is the corresponding density matrix. 
This measurement  ranges from $0$ for non-entangled states to $1$ for maximally entangled configurations. 
Since an RVB liquid ansatz consists of a covering of nearest-neighbor dimers, we expect the two-qubit concurrence for this state to be finite only between nearest-neighbor sites. 
To quantify this, we define the average nearest-neighbour concurrence
\begin{equation}
    \langle C_{\langle ij\rangle}\rangle=\frac{1}{N_{\langle ij\rangle}}\sum_{\langle ij\rangle}C(\varrho_{\langle ij\rangle})\,,
\end{equation}
where $N_{\langle ij\rangle}$ is the number  of nearest neighbour atoms and $C(\varrho_{\langle ij\rangle})$ is the two-qubit concurrence computed for the   density operator reduced to a given nearest neighbour pair $\langle ij\rangle$. 
The dependence of this quantity from the inter-atomic distance is shown in Fig.~\ref{fig:liquid_phase}(c). 
The figure shows that there exists certain ranges of interatomic distances, consistent with those highlighted in Fig.~\ref{fig:Fvsdist}, where the average nearest-neighbor concurrence remains close to the value predicted by the RVB ansatz (dashed line). This behavior confirms the formation of a dimer-like entanglement structure within the system for specific interatomic distance ranges.
\end{itemize}

We thus conclude that the RVB ansatz serves as an excellent approximation for the least radiant state of  half-filled 2D atomic arrays. As we will discuss in the next section, the main challenge lies in identifying a strategy to access and prepare this state.

\section{Steady state preparation}\label{Sec.steady}

\begin{figure}[!t]
\includegraphics[width=0.49\textwidth]{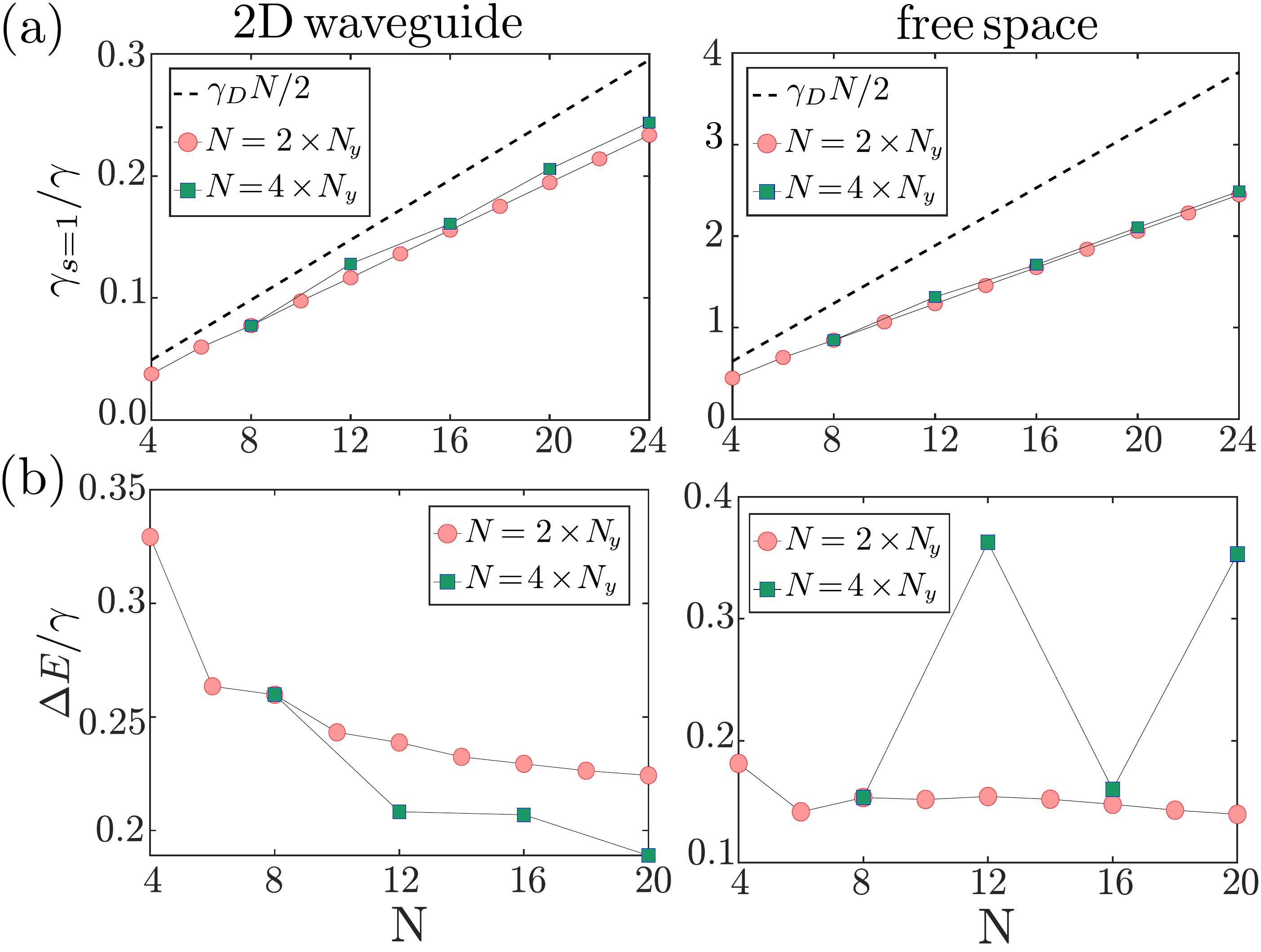}%
    \caption{(a) Scaling with system size of the decay rate and (b) of the spectral gap, $\Delta E$, of the least radiant state, $\gamma_{s=1}$,   for model~\eqref{eq:2DWQED} (first panel) and model~\eqref{eq:3D} (second panel). In the first case, the interatomic distance is fixed at $k_0d=0.1\pi$, while in the second case, it is set to  $k_0d=0.42\pi$. In all cases,  we compare two different lattice geometries and in (a) we included the decay rate scaling associated with $N/2$ dimers.
    }\label{fig:decay}
\end{figure}

In this section, we explore possible strategies to probe the RVB state discussed in the previous section. In Sec.~\ref{Sec.dim_lifetime}, we discuss the finite lifetime and spectral isolation of the target state, which hinder its  preparation via relaxation dynamics.  In Sec.~\ref{Sec.steady}, we instead propose a steady-state scheme that enables probing of the RVB state for small system sizes.

\subsection{Lifetime and spectral isolation of the dimerized state}\label{Sec.dim_lifetime}

As previously discussed, the least radiant state at half-filling is an eigenstate of the non-Hermitian Hamiltonian given in Eq.~\eqref{eq:Heff} and thus it has a finite decay rate. 
The scaling of this decay rate with the system size for the two cases in Eqs.~\eqref{eq:2DWQED}-\eqref{eq:3D} is shown in Fig.~\ref{fig:decay}(a) for two different geometrical configurations. 
In all cases, we observe that the decay rate is upper-bounded by the single dimer decay rate times the number of excited emitters, $\gamma_D N/2$, where $\gamma_D$ is the decay rate of a single dimer, and remains relatively close to this bound.
This implies that as the system size increases, even though this state remains the least radiant within the half-filling subspace, its decay rate can surpass that of an independent single emitter.
This  presents a significant challenge for preparing this state via a relaxation process. 
An initial half-filled configuration will rapidly radiate into excitation subspaces with lower filling, and the desired state will only be prepared with high probability under the condition that $N/2$ excitations remain throughout the system’s evolution, an event that becomes exponentially unlikely in time.

Another potential limitation for the preparation of the target RVB state arises from its spectral isolation. To quantify this, Fig.~\ref{fig:decay}(b) shows the energy difference between the least radiant state in the half-filling sector and the closest state in energy within the same sector, plotted as a function of system size. In the 2D waveguide scenario, we observe a gradual decrease of this spectral gap with increasing system size, whereas in the free-space case, the gap remains comparatively stable. Based on these observations, we expect that the primary limitation in preparing the target state is related to its finite lifetime. This is confirmed in the following section, where we propose a possible driving scheme to address this challenge.

\subsection{Steady-state protocol}\label{Sec.dim_lifetime}
Another possible approach to prepare the dimerized state is to employ a steady state protocol. In Ref.~\cite{PhysRevA.91.042116}, two minimal conditions were identified for realizing a dimerized steady state: (i) the state must be completely dark, i.e., it must have a zero imaginary part in its eigenvalue, or equivalently, be annihilated by all jump operators, and (ii) it must be an eigenstate of the full coherent Hamiltonian, which includes both the photon-mediated interactions and the driving term. In the one-dimensional waveguide case, these conditions can be met under specific driving patterns, leading to a fully dark steady state composed of nearest-neighbor dimers (as given in Eq.~\eqref{eq:dimer1D}). However, in higher dimensions, these two conditions generally cannot be fulfilled simultaneously. In particular, in our case, as discussed above, the target state is not dark to begin with.

\begin{figure}[!t]
    \includegraphics[width=0.5\textwidth]{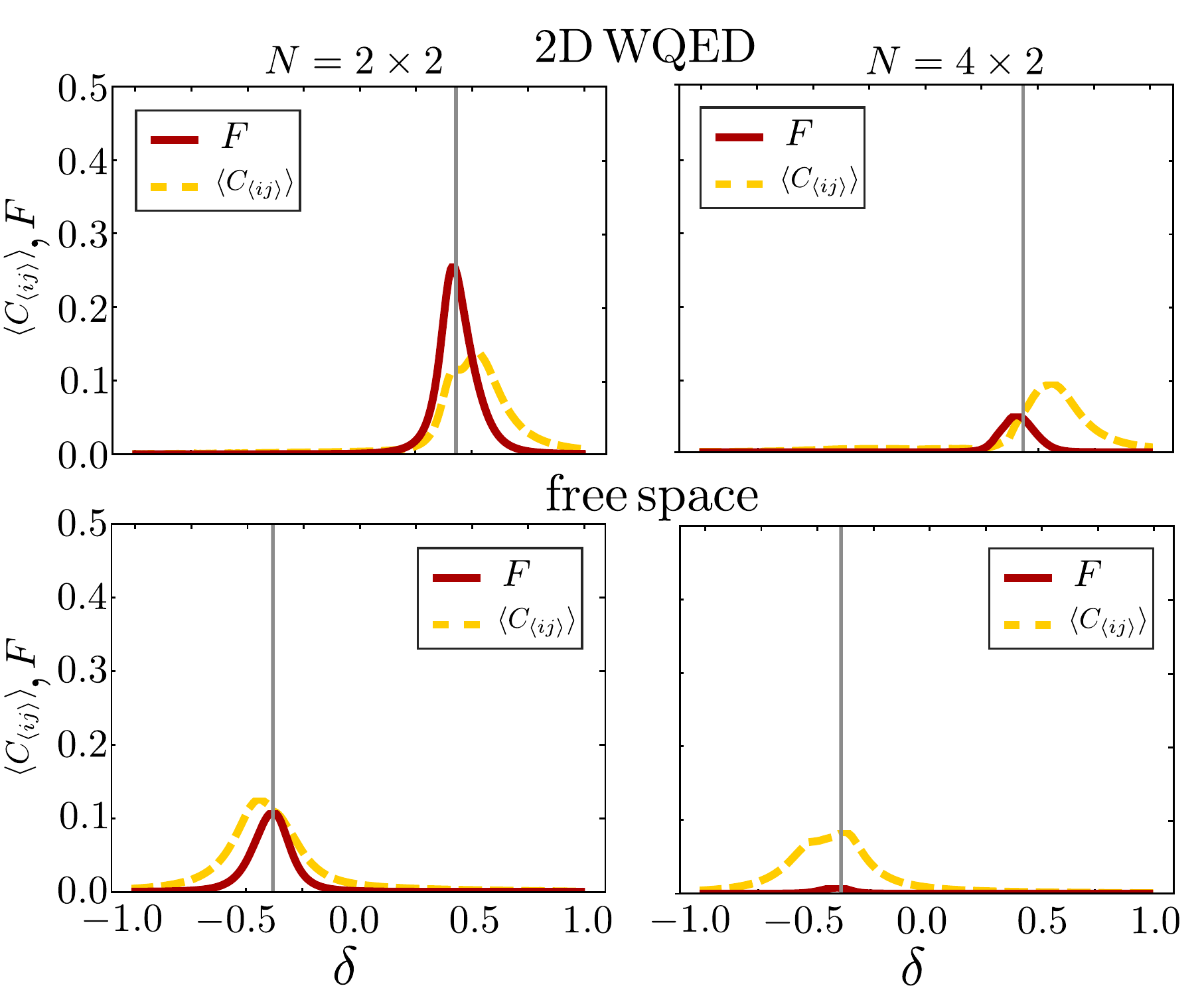}%
    \caption{Steady-state response under coherent driving with a Rabi frequency of  $\Omega=0.05$. 
    The state Fidelity (solid red line) and the average nearest-neighbor concurrence (dashed yellow line)  are plotted against laser detuning for the 2D waveguide case (first row) and the free-space array (second row). The vertical gray line marks the energy of the target  quantum dimer covering state as obtained via exact diagonalization. For the 2D waveguide we set $k_0d=0.1\pi$ while for the free space array $k_0d=0.42\pi$.
    }\label{fig:steady_state}
\end{figure}

We thus employ a different  steady-state protocol that targets the dimerized  state
by matching the energy and symmetry of the many-body wavefunction with the frequency and symmetry pattern of a driving field~\cite{plankensteiner2015selective,PhysRevLett.106.020501,rusconi2021exploiting,van2025resonant}.
More specifically, we assume that each atom in the array can be individually driven by a coherent field. 
This is incorporated by adding the following driving Hamiltonian (in the rotating wave approximation) to the master equation given in Eq.~\eqref{eq: Master equation}:
\begin{equation}
    \hat H_d=\sum_i\Omega_i\hat\sigma_x^i-\sum_i\delta_i\hat\sigma_+^i\hat\sigma_-^i\,,
\end{equation}
where $\Omega_i$ is the driving strength on the $i$-th atom and $\delta_i$ is the detuning between the laser and the $i$-th atomic frequency. 
The pattern of the driving strength is chosen to have  staggered (opposite) phases over the lattice to match the symmetry of the target many-body wave-function.
This choice is motivated by the fact that,  driving two atoms with opposite phase enhances the overlap with the singlet state, as shown in Ref.~\cite{PhysRevLett.106.020501} for a 1D waveguide. Since the RVB target state is a superposition of singlet configurations, one might expect that applying a staggered driving phase across the array on a larger ensembles of emitters would yield a finite overlap with the desired state. This expectation is indeed confirmed  by our simulations a posteriori, as shown in the following.
For a finite-sized array, dipole-dipole interactions give rise to an energetically separated spectrum. We thus set the atom-laser detuning to be uniform across the lattice, $\delta_i=\delta$, and resonant with the target state, while keeping the driving Rabi frequency weak with respect  to the energy gap with off-resonant states to minimize their unwanted excitation.
We then compute the average nearest-neighbor concurrence and the state fidelity, $F=\langle \phi_{\rm RVB}|\rho_{st}|\phi_{\rm RVB}\rangle$, for the steady state $\rho_{st}$ predicted by the master equation~\eqref{eq: Master equation}. This analysis is performed for both the 2D waveguide and free-space scenarios while varying the atom-laser detuning. The results are plotted in Fig.~\ref{fig:steady_state} for different lattice configurations.
In both settings, fidelity and concurrence peak around the target state
energy, corresponding to the least radiant state at half-filling obtained from the exact diagonalization of Eq.~\eqref{eq:Heff}, with the 2D waveguide case performing better.
Notably, while the fidelity peak rapidly decreases with increasing system size, indicating that the prepared state is likely mixed, the concurrence retains a distinct pattern, signaling the persistence of a dimerized entangled structure.

These results demonstrate that certain features of the quantum dimer covering state can be probed in the steady state for small system sizes via a simple pumping scheme. 
As the system size increases, we expect that exciting the target state will become progressively more challenging.

Given the scaling of the spectral isolation of the target state shown in Fig. \ref{fig:decay}(b), this difficulty appears to be mainly attributed to the fact that the state acquires a finite decay rate, causing it to decay rapidly once excited.
This also explains why state preparation in the free-space scenario performs significantly worse: the target state's lifetime in that case is approximately an order of magnitude shorter than in the 2D waveguide setup.

Finally, we note that the relative robustness of the concurrence with increasing system size suggests that dimerization likely persists in the steady state, not as a coherent superposition of all dimer configurations, but rather as a mixed state composed of multiple dimer arrangements. In particular, we could expect such dimerization to  better persists in the bulk of larger lattices. 
Indeed, in the 2D waveguide scenario, dissipation is expected to occur primarily through the edges of the lattice. Thus, even if the RVB state is not retained across the entire lattice, it could still dominate the steady state in the bulk region.

\section{Quantum dimerization in photonic band-gap materials}

So far, we have considered scenarios where the emitters are dissipatively coupled to a photonic environment. As discussed, in this case, the least radiant state at half-filling is well described by an RVB ansatz. However, identifying an efficient scheme to selectively prepare this state remains an open challenge, as its lifetime decreases with increasing system size.
To get around this issue, in this section we consider a different scenario in which a two-dimensional array of atoms is coupled to a 2D band-gap photonic structure~\cite{gonzalez2015subwavelength,yu2019two}. 
In this case, if the atomic frequencies lie within the photonic band gap, the atoms can effectively interact coherently via the induced evanescent field~\cite{PhysRevA.93.033833,douglas2015quantum}. This setting, being intrinsically coherent up to  implementation imperfections, has the advantage of not requiring a driven steady-state scheme.   

The interaction among the atoms is then given by the Hermitian Hamiltonian~\cite{gonzalez2015subwavelength}:
  \begin{equation}\label{eq:PCW}
    \hat{H}_{\rm eff}=J\sum_{i,j}\mathcal{K}_0(\xi x_{ij})\hat{\sigma}_{+}^{i}\hat{\sigma}_{-}^j\,,
\end{equation}  
 where again $x_{ij}=|\mathbf{x}_i-\mathbf{x}_j|$   and  $\mathcal{K}_0$ is the  modified Bessel function of the second kind. Here $J$ and $\xi$ respectively define the strength and range of the photon-mediated atom-atom interaction, which generally depend on the atom-band detuning, atom-field coupling, and band curvature~\cite{PhysRevA.93.033833,douglas2015quantum}. In the following, we assume $J$ to be positive to ensure an antiferromagnetic-like Hamiltonian.

\begin{figure}[!t]
    \includegraphics[width=0.5\textwidth]{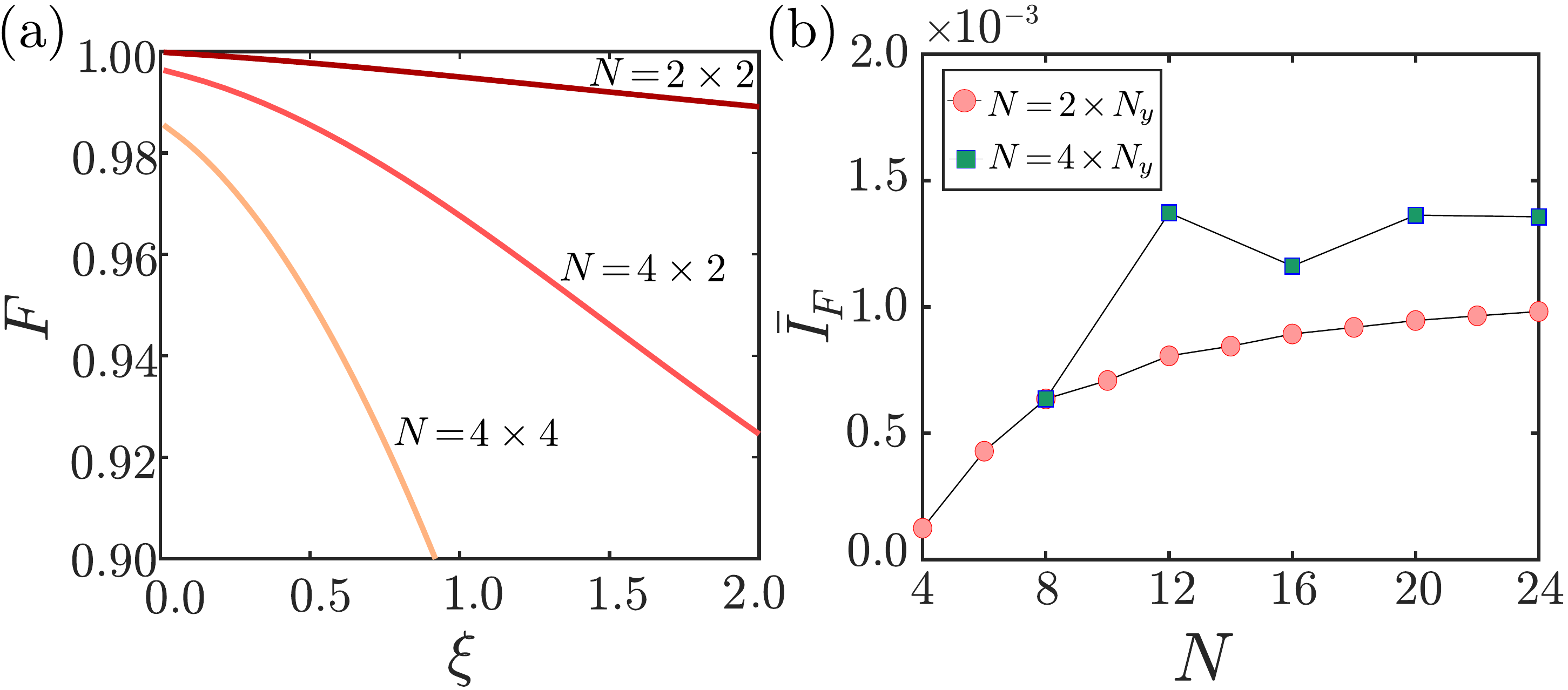}%
    \caption{(a) Fidelity between the dimer covering ansatz and  the half-filling ground state of Hamiltonian \eqref{eq:PCW} as a function of the localization  parameter $\xi$ for  different lattice sizes, as indicated. (b) Infidelity density as a function of system size for two different ladder geometries and $\xi=0.1$. 
   }\label{fig:coherent}
\end{figure}
We concentrate our focus to the half-filling  fraction   subspace made of  $N/2$ excitations. Unlike the open-system case discussed in the previous sections, this subspace can be addressed using an external laser, which introduces a chemical-potential-like term in the Hamiltonian. This term can be tuned via the atom-laser detuning, enabling ground state preparation through adiabatic methods or quantum optimal control techniques~\cite{PhysRevA.84.012312}, similar to those used in Rydberg tweezer arrays~\cite{science.aax9743,levine2018probing}.

In this scenario, we expect the half-filling  ground state of Hamiltonian~\eqref{eq:PCW}  to be well approximated by an RVB ansatz, if the interactions are sufficiently long-range.
This intuition follows a similar argument to that presented in Sec.~\ref{Sec.dimer}. 
Specifically, the emergence of a nearest-neighbor  dimer covering can be understood as a smooth transition from the cavity QED limit, where the ground state lies in the singlet subspace spanned by both short- and long-range dimers. 
In this regime, long-range interactions provide a higher energy penalty to the formation of long-range dimers, making a short-range dimer covering energetically more favorable. This reasoning is also consistent with recent predictions of a spin-liquid phase in Rydberg tweezers arrays with long-range $XY$ dipole interactions~\cite{bintz2024dirac}. 

To test this intuition, in Fig.~\ref{fig:coherent}(a) we plotted the fidelity between the ground state and the RVB ansatz, $F=|\langle \phi_{\rm RVB} |\psi_{\rm GS}|^2\rangle$,
 as a function of the localization parameter $\xi$ for different lattice sizes.
For small values of the localization parameter, $\xi\lesssim 1$, the interactions become long-range, and we observe excellent agreement between the two states. 
This is further supported by the scaling of the infidelity density with system size for $\xi=0.1$, plotted in Fig.~\ref{fig:coherent}(b)  which appears to approach saturation, confirming the robustness of the ansatz. Interestingly, the infidelity seems to saturate more effectively for a larger ladder ($N=4\times N_y$), suggesting that the RVB ansatz could be more robust in an extended, fully 2D arrangement.

\section{Conclusion}
In this work, we demonstrate that the interplay between photon repulsion and long-range, position-dependent dissipative interactions can favor the formation of a superposition of nearest-neighbor quantum dimers (singlet states) in 2D atomic arrays as the least radiant state at half-filling fraction. 
This state is equivalent to a resonating valence bond liquid, which is known to give rise to a spin-liquid phase in frustrated lattices~\cite{PhysRevLett.86.1881}. 
We show that such a state can emerge both in 2D atomic arrays coupled to a two-dimensional waveguide and in those coupled to a free-space electromagnetic environment and we characterized its emergent correlations. 

Our results lead to the fascinating conclusion that the  simple  standard model for photon-mediated atom-atom interactions in Markov approximation
encode such a complex entangled state as an RVB. 
However, a major open challenge remains: how to efficiently prepare and address this state. 
Here, we demonstrated that the key features of the quantum dimer covering can be probed through steady-state pumping in small lattices. 
However, as the system size increases, these correlations appear to degrade. It would be interesting to explore whether such correlations can persist in the bulk of larger systems or if more efficient schemes, such as Raman-based approaches~\cite{PhysRevLett.106.090502,Reiter_2012}, non-Markovian reservoir engineering~\cite{Lebreuilly2017, ma2019dissipatively} or Floquet protocols~\cite{periwal2021programmable}, could be identified to better address this state.

We also emphasize that dimerization emerges as the ground state at half-filling of the coherent Hamiltonian describing an array of atoms coupled to a photonic band-gap material, provided the interactions among the emitters are long-range.
The fact that the ground state of this Hamiltonian exhibits such a rich entanglement structure could present an interesting potential resource for variational eigensolvers, as recently proposed for atoms coupled to 1D photonic crystals.~\cite{PhysRevLett.131.073602}.

\begin{figure}[!t]
    \includegraphics[width=0.49\textwidth]{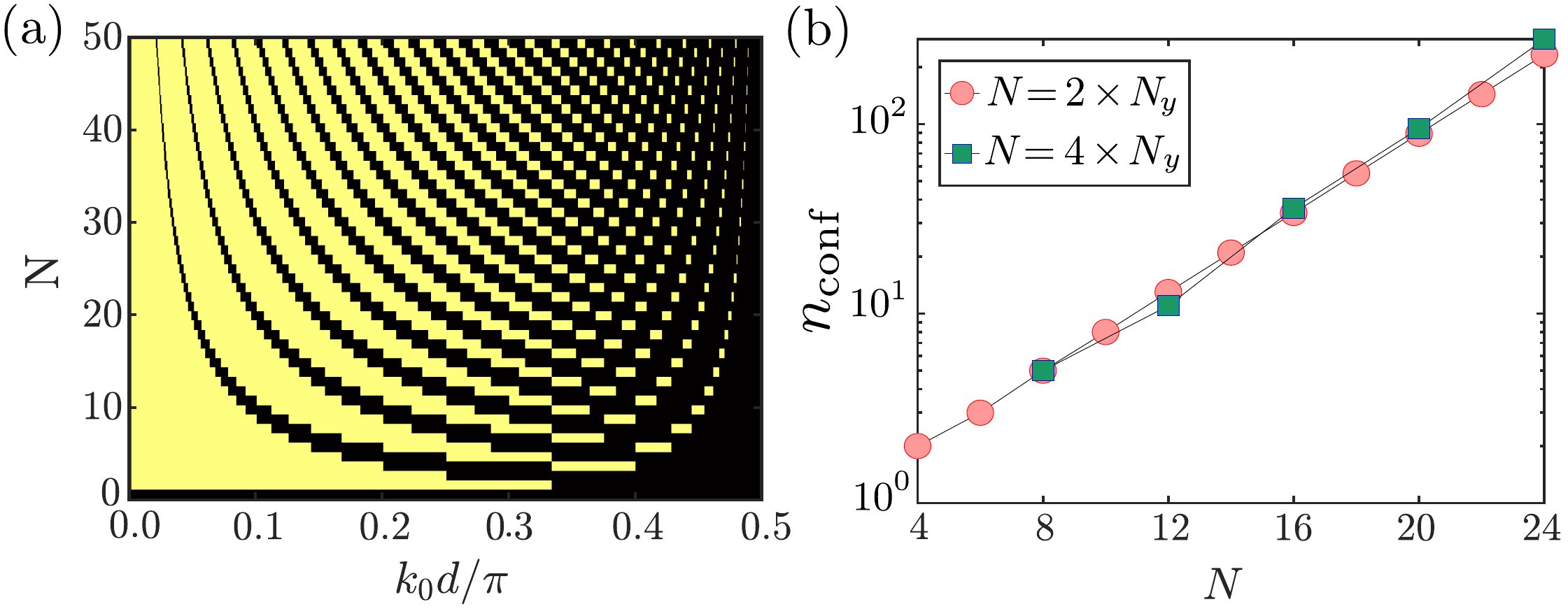}
    \caption{(a) Validity of condition~\eqref{eq:cond} for a 1D waveguide as a function of the number of emitters and inter-atomic distance. The region where the condition is satisfied is highlighted in bright color.
    (b) Number of all possible dimer covering configurations $n_{\rm conf}$ as a function of the system size for two different lattice geometries. 
   }\label{fig:nconf}
\end{figure}

\section{Acknowledgments}
The authors thank Darrick Chang and Charlie-Ray
Mann for inspiring discussions. 
This work was supported by the EU QuantERA2021 project T-NiSQ, the  Quantum Technology Flagship project PASQuanS2, the NextGenerationEU project CN00000013, the Italian Research Center on HPC, Big Data and Quantum Computing, the  Quantum Computing and Simulation Center of Padova University, the INFN project Iniziativa Specifica IS-Quantum
and by the Italian Ministry of University and Research via PRIN2022-PNRR project TANQU, the NQSTI Bandi a Cascata PNRR project OPTIMISTIQ and the Rita Levi-Montalcini program.
The authors also acknowledge computational resources by Cloud Veneto and the \texttt{QuSpin} library for exact diagonalization \cite{10.21468/SciPostPhys.2.1.003,10.21468/SciPostPhys.7.2.020}.

\appendix

\section{Dimerization in 1D waveguide}\label{AppA}
As discussed in the main text, the minimal condition for the dimerization of the least radiant state at half filling, given in Eq.~\eqref{eq:cond}, can be violated in a 1D waveguide at any inter-atomic distance if the array is sufficiently large. This can be seen in Fig.~\ref{fig:nconf}(a), where we plot the region of validity of Eq.~\eqref{eq:cond}, highlighted in bright color, as a function of interatomic distance and the number of emitters. As shown, for a given distance, the incommensurability of the spatial dependence of the two-atom collective decay rates leads to an infinite number of array sizes where the condition is not satisfied. This is reflected in the scaling of state fidelity with the inter-atomic distance,  Fig.~\ref{fig:Fvsdist}, and system size, Fig.~\ref{fig:F_scaling}.
To ensure that the least radiant state remains a covering of nearest-neighbor dimers for any chain size, the atomic spacing must be increasingly close to $k_0d=2\pi n$ with $n\in \mathbb{N}$, otherwise, the formation of triplets or longer-range dimers may become favorable.


\section{Quantum dimer covering configurations in 2D}\label{AppB}
To determine all possible dimer coverings on a given square lattice, we begin with a specific covering and explore all possible configurations using the basic move: flipping a pair of vertical nearest-neighbor dimers with a pair of horizontal ones, and vice versa. This process can be easily implemented numerically. In Fig.~\ref{fig:nconf}(b), we show the scaling of the number of coverings with the system size for the two lattice geometries considered in the main text. We observe that in both cases the number of configurations grows exponentially with system size signalling the complexity of the RVB ansatz.

\bibliography{refs}

\begin{thebibliography}{109}%
\makeatletter
\providecommand \@ifxundefined [1]{%
 \@ifx{#1\undefined}
}%
\providecommand \@ifnum [1]{%
 \ifnum #1\expandafter \@firstoftwo
 \else \expandafter \@secondoftwo
 \fi
}%
\providecommand \@ifx [1]{%
 \ifx #1\expandafter \@firstoftwo
 \else \expandafter \@secondoftwo
 \fi
}%
\providecommand \natexlab [1]{#1}%
\providecommand \enquote  [1]{``#1''}%
\providecommand \bibnamefont  [1]{#1}%
\providecommand \bibfnamefont [1]{#1}%
\providecommand \citenamefont [1]{#1}%
\providecommand \href@noop [0]{\@secondoftwo}%
\providecommand \href [0]{\begingroup \@sanitize@url \@href}%
\providecommand \@href[1]{\@@startlink{#1}\@@href}%
\providecommand \@@href[1]{\endgroup#1\@@endlink}%
\providecommand \@sanitize@url [0]{\catcode `\\12\catcode `\$12\catcode
  `\&12\catcode `\#12\catcode `\^12\catcode `\_12\catcode `\%12\relax}%
\providecommand \@@startlink[1]{}%
\providecommand \@@endlink[0]{}%
\providecommand \url  [0]{\begingroup\@sanitize@url \@url }%
\providecommand \@url [1]{\endgroup\@href {#1}{\urlprefix }}%
\providecommand \urlprefix  [0]{URL }%
\providecommand \Eprint [0]{\href }%
\providecommand \doibase [0]{https://doi.org/}%
\providecommand \selectlanguage [0]{\@gobble}%
\providecommand \bibinfo  [0]{\@secondoftwo}%
\providecommand \bibfield  [0]{\@secondoftwo}%
\providecommand \translation [1]{[#1]}%
\providecommand \BibitemOpen [0]{}%
\providecommand \bibitemStop [0]{}%
\providecommand \bibitemNoStop [0]{.\EOS\space}%
\providecommand \EOS [0]{\spacefactor3000\relax}%
\providecommand \BibitemShut  [1]{\csname bibitem#1\endcsname}%
\let\auto@bib@innerbib\@empty
\bibitem [{\citenamefont {Reiserer}\ and\ \citenamefont
  {Rempe}(2015)}]{reiserer2015cavity}%
  \BibitemOpen
  \bibfield  {author} {\bibinfo {author} {\bibfnamefont {A.}~\bibnamefont
  {Reiserer}}\ and\ \bibinfo {author} {\bibfnamefont {G.}~\bibnamefont
  {Rempe}},\ }\bibfield  {title} {\bibinfo {title} {Cavity-based quantum
  networks with single atoms and optical photons},\ }\href
  {https://doi.org/10.1103/RevModPhys.87.1379} {\bibfield  {journal} {\bibinfo
  {journal} {Rev. Mod. Phys.}\ }\textbf {\bibinfo {volume} {87}},\ \bibinfo
  {pages} {1379} (\bibinfo {year} {2015})}\BibitemShut {NoStop}%
\bibitem [{\citenamefont {John}\ and\ \citenamefont
  {Quang}(1994)}]{PhysRevA.50.1764}%
  \BibitemOpen
  \bibfield  {author} {\bibinfo {author} {\bibfnamefont {S.}~\bibnamefont
  {John}}\ and\ \bibinfo {author} {\bibfnamefont {T.}~\bibnamefont {Quang}},\
  }\bibfield  {title} {\bibinfo {title} {Spontaneous emission near the edge of
  a photonic band gap},\ }\href {https://doi.org/10.1103/PhysRevA.50.1764}
  {\bibfield  {journal} {\bibinfo  {journal} {Phys. Rev. A}\ }\textbf {\bibinfo
  {volume} {50}},\ \bibinfo {pages} {1764} (\bibinfo {year}
  {1994})}\BibitemShut {NoStop}%
\bibitem [{\citenamefont {Kofman}\ \emph {et~al.}(1994)\citenamefont {Kofman},
  \citenamefont {Kurizki},\ and\ \citenamefont
  {Sherman}}]{kofman1994spontaneous}%
  \BibitemOpen
  \bibfield  {author} {\bibinfo {author} {\bibfnamefont {A.}~\bibnamefont
  {Kofman}}, \bibinfo {author} {\bibfnamefont {G.}~\bibnamefont {Kurizki}},\
  and\ \bibinfo {author} {\bibfnamefont {B.}~\bibnamefont {Sherman}},\
  }\bibfield  {title} {\bibinfo {title} {Spontaneous and induced atomic decay
  in photonic band structures},\ }\href
  {https://www.tandfonline.com/doi/abs/10.1080/09500349414550381?casa_token=gfc1pvvFMfgAAAAA:LfK2jK2f1bleH_v63m5mgkA8wz6vybFpUG6tf2zJwIrnfj_FLLULghuVcUmHommszoMQSd6S26rVttY}
  {\bibfield  {journal} {\bibinfo  {journal} {Journal of Modern Optics}\
  }\textbf {\bibinfo {volume} {41}},\ \bibinfo {pages} {353} (\bibinfo {year}
  {1994})}\BibitemShut {NoStop}%
\bibitem [{\citenamefont {Shahmoon}\ and\ \citenamefont
  {Kurizki}(2013)}]{PhysRevA.87.033831}%
  \BibitemOpen
  \bibfield  {author} {\bibinfo {author} {\bibfnamefont {E.}~\bibnamefont
  {Shahmoon}}\ and\ \bibinfo {author} {\bibfnamefont {G.}~\bibnamefont
  {Kurizki}},\ }\bibfield  {title} {\bibinfo {title} {Nonradiative interaction
  and entanglement between distant atoms},\ }\href
  {https://doi.org/10.1103/PhysRevA.87.033831} {\bibfield  {journal} {\bibinfo
  {journal} {Phys. Rev. A}\ }\textbf {\bibinfo {volume} {87}},\ \bibinfo
  {pages} {033831} (\bibinfo {year} {2013})}\BibitemShut {NoStop}%
\bibitem [{\citenamefont {Bay}\ \emph {et~al.}(1997)\citenamefont {Bay},
  \citenamefont {Lambropoulos},\ and\ \citenamefont
  {M\o{}lmer}}]{PhysRevA.55.1485}%
  \BibitemOpen
  \bibfield  {author} {\bibinfo {author} {\bibfnamefont {S.}~\bibnamefont
  {Bay}}, \bibinfo {author} {\bibfnamefont {P.}~\bibnamefont {Lambropoulos}},\
  and\ \bibinfo {author} {\bibfnamefont {K.}~\bibnamefont {M\o{}lmer}},\
  }\bibfield  {title} {\bibinfo {title} {Atom-atom interaction in strongly
  modified reservoirs},\ }\href {https://doi.org/10.1103/PhysRevA.55.1485}
  {\bibfield  {journal} {\bibinfo  {journal} {Phys. Rev. A}\ }\textbf {\bibinfo
  {volume} {55}},\ \bibinfo {pages} {1485} (\bibinfo {year}
  {1997})}\BibitemShut {NoStop}%
\bibitem [{\citenamefont {Calaj\'o}\ \emph {et~al.}(2016)\citenamefont
  {Calaj\'o}, \citenamefont {Ciccarello}, \citenamefont {Chang},\ and\
  \citenamefont {Rabl}}]{PhysRevA.93.033833}%
  \BibitemOpen
  \bibfield  {author} {\bibinfo {author} {\bibfnamefont {G.}~\bibnamefont
  {Calaj\'o}}, \bibinfo {author} {\bibfnamefont {F.}~\bibnamefont
  {Ciccarello}}, \bibinfo {author} {\bibfnamefont {D.}~\bibnamefont {Chang}},\
  and\ \bibinfo {author} {\bibfnamefont {P.}~\bibnamefont {Rabl}},\ }\bibfield
  {title} {\bibinfo {title} {Atom-field dressed states in slow-light waveguide
  qed},\ }\href {https://doi.org/10.1103/PhysRevA.93.033833} {\bibfield
  {journal} {\bibinfo  {journal} {Phys. Rev. A}\ }\textbf {\bibinfo {volume}
  {93}},\ \bibinfo {pages} {033833} (\bibinfo {year} {2016})}\BibitemShut
  {NoStop}%
\bibitem [{\citenamefont {Douglas}\ \emph {et~al.}(2015)\citenamefont
  {Douglas}, \citenamefont {Habibian}, \citenamefont {Hung}, \citenamefont
  {Gorshkov}, \citenamefont {Kimble},\ and\ \citenamefont
  {Chang}}]{douglas2015quantum}%
  \BibitemOpen
  \bibfield  {author} {\bibinfo {author} {\bibfnamefont {J.~S.}\ \bibnamefont
  {Douglas}}, \bibinfo {author} {\bibfnamefont {H.}~\bibnamefont {Habibian}},
  \bibinfo {author} {\bibfnamefont {C.-L.}\ \bibnamefont {Hung}}, \bibinfo
  {author} {\bibfnamefont {A.~V.}\ \bibnamefont {Gorshkov}}, \bibinfo {author}
  {\bibfnamefont {H.~J.}\ \bibnamefont {Kimble}},\ and\ \bibinfo {author}
  {\bibfnamefont {D.~E.}\ \bibnamefont {Chang}},\ }\bibfield  {title} {\bibinfo
  {title} {Quantum many-body models with cold atoms coupled to photonic
  crystals},\ }\href {https://www.nature.com/articles/nphoton.2015.57.pdf}
  {\bibfield  {journal} {\bibinfo  {journal} {Nature Photonics}\ }\textbf
  {\bibinfo {volume} {9}},\ \bibinfo {pages} {326} (\bibinfo {year}
  {2015})}\BibitemShut {NoStop}%
\bibitem [{\citenamefont {González-Tudela}\ \emph {et~al.}(2015)\citenamefont
  {González-Tudela}, \citenamefont {Hung}, \citenamefont {Chang},
  \citenamefont {Cirac},\ and\ \citenamefont
  {Kimble}}]{gonzalez2015subwavelength}%
  \BibitemOpen
  \bibfield  {author} {\bibinfo {author} {\bibfnamefont {A.}~\bibnamefont
  {González-Tudela}}, \bibinfo {author} {\bibfnamefont {C.-L.}\ \bibnamefont
  {Hung}}, \bibinfo {author} {\bibfnamefont {D.~E.}\ \bibnamefont {Chang}},
  \bibinfo {author} {\bibfnamefont {J.~I.}\ \bibnamefont {Cirac}},\ and\
  \bibinfo {author} {\bibfnamefont {H.~J.}\ \bibnamefont {Kimble}},\ }\bibfield
   {title} {\bibinfo {title} {Subwavelength vacuum lattices and atom–atom
  interactions in two-dimensional photonic crystals},\ }\href
  {https://doi.org/10.1038/nphoton.2015.54} {\bibfield  {journal} {\bibinfo
  {journal} {Nature Photonics}\ }\textbf {\bibinfo {volume} {9}},\ \bibinfo
  {pages} {320–325} (\bibinfo {year} {2015})}\BibitemShut {NoStop}%
\bibitem [{\citenamefont {Gonz\'alez-Tudela}\ and\ \citenamefont
  {Cirac}(2017)}]{gonzalez2017markovian}%
  \BibitemOpen
  \bibfield  {author} {\bibinfo {author} {\bibfnamefont {A.}~\bibnamefont
  {Gonz\'alez-Tudela}}\ and\ \bibinfo {author} {\bibfnamefont {J.~I.}\
  \bibnamefont {Cirac}},\ }\bibfield  {title} {\bibinfo {title} {Markovian and
  non-markovian dynamics of quantum emitters coupled to two-dimensional
  structured reservoirs},\ }\href {https://doi.org/10.1103/PhysRevA.96.043811}
  {\bibfield  {journal} {\bibinfo  {journal} {Phys. Rev. A}\ }\textbf {\bibinfo
  {volume} {96}},\ \bibinfo {pages} {043811} (\bibinfo {year}
  {2017})}\BibitemShut {NoStop}%
\bibitem [{\citenamefont {Galve}\ and\ \citenamefont
  {Zambrini}(2018)}]{galve2018coherent}%
  \BibitemOpen
  \bibfield  {author} {\bibinfo {author} {\bibfnamefont {F.}~\bibnamefont
  {Galve}}\ and\ \bibinfo {author} {\bibfnamefont {R.}~\bibnamefont
  {Zambrini}},\ }\bibfield  {title} {\bibinfo {title} {Coherent and radiative
  couplings through two-dimensional structured environments},\ }\href
  {https://doi.org/10.1103/PhysRevA.97.033846} {\bibfield  {journal} {\bibinfo
  {journal} {Phys. Rev. A}\ }\textbf {\bibinfo {volume} {97}},\ \bibinfo
  {pages} {033846} (\bibinfo {year} {2018})}\BibitemShut {NoStop}%
\bibitem [{\citenamefont {González-Tudela}\ and\ \citenamefont
  {Galve}(2018)}]{gonzalez2018anisotropic}%
  \BibitemOpen
  \bibfield  {author} {\bibinfo {author} {\bibfnamefont {A.}~\bibnamefont
  {González-Tudela}}\ and\ \bibinfo {author} {\bibfnamefont {F.}~\bibnamefont
  {Galve}},\ }\bibfield  {title} {\bibinfo {title} {Anisotropic quantum emitter
  interactions in two-dimensional photonic-crystal baths},\ }\href
  {https://doi.org/10.1021/acsphotonics.8b01455} {\bibfield  {journal}
  {\bibinfo  {journal} {ACS Photonics}\ }\textbf {\bibinfo {volume} {6}},\
  \bibinfo {pages} {221–229} (\bibinfo {year} {2018})}\BibitemShut {NoStop}%
\bibitem [{\citenamefont {Sundaresan}\ \emph {et~al.}(2019)\citenamefont
  {Sundaresan}, \citenamefont {Lundgren}, \citenamefont {Zhu}, \citenamefont
  {Gorshkov},\ and\ \citenamefont {Houck}}]{PhysRevX.9.011021}%
  \BibitemOpen
  \bibfield  {author} {\bibinfo {author} {\bibfnamefont {N.~M.}\ \bibnamefont
  {Sundaresan}}, \bibinfo {author} {\bibfnamefont {R.}~\bibnamefont
  {Lundgren}}, \bibinfo {author} {\bibfnamefont {G.}~\bibnamefont {Zhu}},
  \bibinfo {author} {\bibfnamefont {A.~V.}\ \bibnamefont {Gorshkov}},\ and\
  \bibinfo {author} {\bibfnamefont {A.~A.}\ \bibnamefont {Houck}},\ }\bibfield
  {title} {\bibinfo {title} {Interacting qubit-photon bound states with
  superconducting circuits},\ }\href
  {https://doi.org/10.1103/PhysRevX.9.011021} {\bibfield  {journal} {\bibinfo
  {journal} {Phys. Rev. X}\ }\textbf {\bibinfo {volume} {9}},\ \bibinfo {pages}
  {011021} (\bibinfo {year} {2019})}\BibitemShut {NoStop}%
\bibitem [{\citenamefont {Stewart}\ \emph {et~al.}(2020)\citenamefont
  {Stewart}, \citenamefont {Kwon}, \citenamefont {Lanuza},\ and\ \citenamefont
  {Schneble}}]{PhysRevResearch.2.043307}%
  \BibitemOpen
  \bibfield  {author} {\bibinfo {author} {\bibfnamefont {M.}~\bibnamefont
  {Stewart}}, \bibinfo {author} {\bibfnamefont {J.}~\bibnamefont {Kwon}},
  \bibinfo {author} {\bibfnamefont {A.}~\bibnamefont {Lanuza}},\ and\ \bibinfo
  {author} {\bibfnamefont {D.}~\bibnamefont {Schneble}},\ }\bibfield  {title}
  {\bibinfo {title} {Dynamics of matter-wave quantum emitters in a structured
  vacuum},\ }\href {https://doi.org/10.1103/PhysRevResearch.2.043307}
  {\bibfield  {journal} {\bibinfo  {journal} {Phys. Rev. Res.}\ }\textbf
  {\bibinfo {volume} {2}},\ \bibinfo {pages} {043307} (\bibinfo {year}
  {2020})}\BibitemShut {NoStop}%
\bibitem [{\citenamefont {Gonz{\'a}lez-Tudela}\ \emph
  {et~al.}(2024)\citenamefont {Gonz{\'a}lez-Tudela}, \citenamefont {Reiserer},
  \citenamefont {Garc{\'\i}a-Ripoll},\ and\ \citenamefont
  {Garc{\'\i}a-Vidal}}]{gonzalez2024light}%
  \BibitemOpen
  \bibfield  {author} {\bibinfo {author} {\bibfnamefont {A.}~\bibnamefont
  {Gonz{\'a}lez-Tudela}}, \bibinfo {author} {\bibfnamefont {A.}~\bibnamefont
  {Reiserer}}, \bibinfo {author} {\bibfnamefont {J.~J.}\ \bibnamefont
  {Garc{\'\i}a-Ripoll}},\ and\ \bibinfo {author} {\bibfnamefont {F.~J.}\
  \bibnamefont {Garc{\'\i}a-Vidal}},\ }\bibfield  {title} {\bibinfo {title}
  {Light--matter interactions in quantum nanophotonic devices},\ }\href
  {https://doi.org/10.1038/s42254-023-00681-1} {\bibfield  {journal} {\bibinfo
  {journal} {Nature Reviews Physics}\ }\textbf {\bibinfo {volume} {6}},\
  \bibinfo {pages} {166} (\bibinfo {year} {2024})}\BibitemShut {NoStop}%
\bibitem [{\citenamefont {Chang}\ \emph {et~al.}(2018)\citenamefont {Chang},
  \citenamefont {Douglas}, \citenamefont {Gonz\'alez-Tudela}, \citenamefont
  {Hung},\ and\ \citenamefont {Kimble}}]{RevModPhys.90.031002}%
  \BibitemOpen
  \bibfield  {author} {\bibinfo {author} {\bibfnamefont {D.~E.}\ \bibnamefont
  {Chang}}, \bibinfo {author} {\bibfnamefont {J.~S.}\ \bibnamefont {Douglas}},
  \bibinfo {author} {\bibfnamefont {A.}~\bibnamefont {Gonz\'alez-Tudela}},
  \bibinfo {author} {\bibfnamefont {C.-L.}\ \bibnamefont {Hung}},\ and\
  \bibinfo {author} {\bibfnamefont {H.~J.}\ \bibnamefont {Kimble}},\ }\bibfield
   {title} {\bibinfo {title} {Colloquium: Quantum matter built from nanoscopic
  lattices of atoms and photons},\ }\href
  {https://doi.org/10.1103/RevModPhys.90.031002} {\bibfield  {journal}
  {\bibinfo  {journal} {Rev. Mod. Phys.}\ }\textbf {\bibinfo {volume} {90}},\
  \bibinfo {pages} {031002} (\bibinfo {year} {2018})}\BibitemShut {NoStop}%
\bibitem [{\citenamefont {De~Bernardis}\ \emph {et~al.}(2021)\citenamefont
  {De~Bernardis}, \citenamefont {Cian}, \citenamefont {Carusotto},
  \citenamefont {Hafezi},\ and\ \citenamefont {Rabl}}]{PhysRevLett.126.103603}%
  \BibitemOpen
  \bibfield  {author} {\bibinfo {author} {\bibfnamefont {D.}~\bibnamefont
  {De~Bernardis}}, \bibinfo {author} {\bibfnamefont {Z.-P.}\ \bibnamefont
  {Cian}}, \bibinfo {author} {\bibfnamefont {I.}~\bibnamefont {Carusotto}},
  \bibinfo {author} {\bibfnamefont {M.}~\bibnamefont {Hafezi}},\ and\ \bibinfo
  {author} {\bibfnamefont {P.}~\bibnamefont {Rabl}},\ }\bibfield  {title}
  {\bibinfo {title} {Light-matter interactions in synthetic magnetic fields:
  Landau-photon polaritons},\ }\href
  {https://doi.org/10.1103/PhysRevLett.126.103603} {\bibfield  {journal}
  {\bibinfo  {journal} {Phys. Rev. Lett.}\ }\textbf {\bibinfo {volume} {126}},\
  \bibinfo {pages} {103603} (\bibinfo {year} {2021})}\BibitemShut {NoStop}%
\bibitem [{\citenamefont {De~Bernardis}\ \emph {et~al.}(2023)\citenamefont
  {De~Bernardis}, \citenamefont {Piccioli}, \citenamefont {Rabl},\ and\
  \citenamefont {Carusotto}}]{PRXQuantum.4.030306}%
  \BibitemOpen
  \bibfield  {author} {\bibinfo {author} {\bibfnamefont {D.}~\bibnamefont
  {De~Bernardis}}, \bibinfo {author} {\bibfnamefont {F.~S.}\ \bibnamefont
  {Piccioli}}, \bibinfo {author} {\bibfnamefont {P.}~\bibnamefont {Rabl}},\
  and\ \bibinfo {author} {\bibfnamefont {I.}~\bibnamefont {Carusotto}},\
  }\bibfield  {title} {\bibinfo {title} {Chiral quantum optics in the bulk of
  photonic quantum hall systems},\ }\href
  {https://doi.org/10.1103/PRXQuantum.4.030306} {\bibfield  {journal} {\bibinfo
   {journal} {PRX Quantum}\ }\textbf {\bibinfo {volume} {4}},\ \bibinfo {pages}
  {030306} (\bibinfo {year} {2023})}\BibitemShut {NoStop}%
\bibitem [{\citenamefont {Bello}\ \emph {et~al.}(2019)\citenamefont {Bello},
  \citenamefont {Platero}, \citenamefont {Cirac},\ and\ \citenamefont
  {Gonz{\'a}lez-Tudela}}]{bello2019unconventional}%
  \BibitemOpen
  \bibfield  {author} {\bibinfo {author} {\bibfnamefont {M.}~\bibnamefont
  {Bello}}, \bibinfo {author} {\bibfnamefont {G.}~\bibnamefont {Platero}},
  \bibinfo {author} {\bibfnamefont {J.~I.}\ \bibnamefont {Cirac}},\ and\
  \bibinfo {author} {\bibfnamefont {A.}~\bibnamefont {Gonz{\'a}lez-Tudela}},\
  }\bibfield  {title} {\bibinfo {title} {Unconventional quantum optics in
  topological waveguide qed},\ }\href@noop {} {\bibfield  {journal} {\bibinfo
  {journal} {Science advances}\ }\textbf {\bibinfo {volume} {5}},\ \bibinfo
  {pages} {eaaw0297} (\bibinfo {year} {2019})}\BibitemShut {NoStop}%
\bibitem [{\citenamefont {Dicke}(1954)}]{PhysRev.93.99}%
  \BibitemOpen
  \bibfield  {author} {\bibinfo {author} {\bibfnamefont {R.~H.}\ \bibnamefont
  {Dicke}},\ }\bibfield  {title} {\bibinfo {title} {Coherence in spontaneous
  radiation processes},\ }\href {https://doi.org/10.1103/PhysRev.93.99}
  {\bibfield  {journal} {\bibinfo  {journal} {Phys. Rev.}\ }\textbf {\bibinfo
  {volume} {93}},\ \bibinfo {pages} {99} (\bibinfo {year} {1954})}\BibitemShut
  {NoStop}%
\bibitem [{\citenamefont {Sheremet}\ \emph {et~al.}(2023)\citenamefont
  {Sheremet}, \citenamefont {Petrov}, \citenamefont {Iorsh}, \citenamefont
  {Poshakinskiy},\ and\ \citenamefont {Poddubny}}]{sheremet2023waveguide}%
  \BibitemOpen
  \bibfield  {author} {\bibinfo {author} {\bibfnamefont {A.~S.}\ \bibnamefont
  {Sheremet}}, \bibinfo {author} {\bibfnamefont {M.~I.}\ \bibnamefont
  {Petrov}}, \bibinfo {author} {\bibfnamefont {I.~V.}\ \bibnamefont {Iorsh}},
  \bibinfo {author} {\bibfnamefont {A.~V.}\ \bibnamefont {Poshakinskiy}},\ and\
  \bibinfo {author} {\bibfnamefont {A.~N.}\ \bibnamefont {Poddubny}},\
  }\bibfield  {title} {\bibinfo {title} {Waveguide quantum electrodynamics:
  Collective radiance and photon-photon correlations},\ }\href
  {https://doi.org/10.1103/RevModPhys.95.015002} {\bibfield  {journal}
  {\bibinfo  {journal} {Rev. Mod. Phys.}\ }\textbf {\bibinfo {volume} {95}},\
  \bibinfo {pages} {015002} (\bibinfo {year} {2023})}\BibitemShut {NoStop}%
\bibitem [{\citenamefont {Roy}\ \emph {et~al.}(2017)\citenamefont {Roy},
  \citenamefont {Wilson},\ and\ \citenamefont
  {Firstenberg}}]{roy2017colloquium}%
  \BibitemOpen
  \bibfield  {author} {\bibinfo {author} {\bibfnamefont {D.}~\bibnamefont
  {Roy}}, \bibinfo {author} {\bibfnamefont {C.~M.}\ \bibnamefont {Wilson}},\
  and\ \bibinfo {author} {\bibfnamefont {O.}~\bibnamefont {Firstenberg}},\
  }\bibfield  {title} {\bibinfo {title} {Colloquium: Strongly interacting
  photons in one-dimensional continuum},\ }\href
  {https://doi.org/10.1103/RevModPhys.89.021001} {\bibfield  {journal}
  {\bibinfo  {journal} {Rev. Mod. Phys.}\ }\textbf {\bibinfo {volume} {89}},\
  \bibinfo {pages} {021001} (\bibinfo {year} {2017})}\BibitemShut {NoStop}%
\bibitem [{\citenamefont {Lodahl}\ \emph {et~al.}(2015)\citenamefont {Lodahl},
  \citenamefont {Mahmoodian},\ and\ \citenamefont
  {Stobbe}}]{lodahl2015interfacing}%
  \BibitemOpen
  \bibfield  {author} {\bibinfo {author} {\bibfnamefont {P.}~\bibnamefont
  {Lodahl}}, \bibinfo {author} {\bibfnamefont {S.}~\bibnamefont {Mahmoodian}},\
  and\ \bibinfo {author} {\bibfnamefont {S.}~\bibnamefont {Stobbe}},\
  }\bibfield  {title} {\bibinfo {title} {Interfacing single photons and single
  quantum dots with photonic nanostructures},\ }\href
  {https://doi.org/10.1103/RevModPhys.87.347} {\bibfield  {journal} {\bibinfo
  {journal} {Rev. Mod. Phys.}\ }\textbf {\bibinfo {volume} {87}},\ \bibinfo
  {pages} {347} (\bibinfo {year} {2015})}\BibitemShut {NoStop}%
\bibitem [{\citenamefont {Hood}\ \emph {et~al.}(2016)\citenamefont {Hood},
  \citenamefont {Goban}, \citenamefont {Asenjo-Garcia}, \citenamefont {Lu},
  \citenamefont {Yu}, \citenamefont {Chang},\ and\ \citenamefont
  {Kimble}}]{hood2016atom}%
  \BibitemOpen
  \bibfield  {author} {\bibinfo {author} {\bibfnamefont {J.~D.}\ \bibnamefont
  {Hood}}, \bibinfo {author} {\bibfnamefont {A.}~\bibnamefont {Goban}},
  \bibinfo {author} {\bibfnamefont {A.}~\bibnamefont {Asenjo-Garcia}}, \bibinfo
  {author} {\bibfnamefont {M.}~\bibnamefont {Lu}}, \bibinfo {author}
  {\bibfnamefont {S.-P.}\ \bibnamefont {Yu}}, \bibinfo {author} {\bibfnamefont
  {D.~E.}\ \bibnamefont {Chang}},\ and\ \bibinfo {author} {\bibfnamefont
  {H.~J.}\ \bibnamefont {Kimble}},\ }\bibfield  {title} {\bibinfo {title}
  {Atom–atom interactions around the band edge of a photonic crystal
  waveguide},\ }\href {https://doi.org/10.1073/pnas.1603788113} {\bibfield
  {journal} {\bibinfo  {journal} {Proceedings of the National Academy of
  Sciences}\ }\textbf {\bibinfo {volume} {113}},\ \bibinfo {pages}
  {10507–10512} (\bibinfo {year} {2016})}\BibitemShut {NoStop}%
\bibitem [{\citenamefont {Corzo}\ \emph {et~al.}(2019)\citenamefont {Corzo},
  \citenamefont {Raskop}, \citenamefont {Chandra}, \citenamefont {Sheremet},
  \citenamefont {Gouraud},\ and\ \citenamefont {Laurat}}]{corzo2019waveguide}%
  \BibitemOpen
  \bibfield  {author} {\bibinfo {author} {\bibfnamefont {N.~V.}\ \bibnamefont
  {Corzo}}, \bibinfo {author} {\bibfnamefont {J.}~\bibnamefont {Raskop}},
  \bibinfo {author} {\bibfnamefont {A.}~\bibnamefont {Chandra}}, \bibinfo
  {author} {\bibfnamefont {A.~S.}\ \bibnamefont {Sheremet}}, \bibinfo {author}
  {\bibfnamefont {B.}~\bibnamefont {Gouraud}},\ and\ \bibinfo {author}
  {\bibfnamefont {J.}~\bibnamefont {Laurat}},\ }\bibfield  {title} {\bibinfo
  {title} {Waveguide-coupled single collective excitation of atomic arrays},\
  }\href {https://doi.org/10.1038/s41586-019-0902-3} {\bibfield  {journal}
  {\bibinfo  {journal} {Nature}\ }\textbf {\bibinfo {volume} {566}},\ \bibinfo
  {pages} {359–362} (\bibinfo {year} {2019})}\BibitemShut {NoStop}%
\bibitem [{\citenamefont {Tiranov}\ \emph {et~al.}(2023)\citenamefont
  {Tiranov}, \citenamefont {Angelopoulou}, \citenamefont {van Diepen},
  \citenamefont {Schrinski}, \citenamefont {Sandberg}, \citenamefont {Wang},
  \citenamefont {Midolo}, \citenamefont {Scholz}, \citenamefont {Wieck},
  \citenamefont {Ludwig}, \citenamefont {Sørensen},\ and\ \citenamefont
  {Lodahl}}]{tiranov2023collective}%
  \BibitemOpen
  \bibfield  {author} {\bibinfo {author} {\bibfnamefont {A.}~\bibnamefont
  {Tiranov}}, \bibinfo {author} {\bibfnamefont {V.}~\bibnamefont
  {Angelopoulou}}, \bibinfo {author} {\bibfnamefont {C.~J.}\ \bibnamefont {van
  Diepen}}, \bibinfo {author} {\bibfnamefont {B.}~\bibnamefont {Schrinski}},
  \bibinfo {author} {\bibfnamefont {O.~A.~D.}\ \bibnamefont {Sandberg}},
  \bibinfo {author} {\bibfnamefont {Y.}~\bibnamefont {Wang}}, \bibinfo {author}
  {\bibfnamefont {L.}~\bibnamefont {Midolo}}, \bibinfo {author} {\bibfnamefont
  {S.}~\bibnamefont {Scholz}}, \bibinfo {author} {\bibfnamefont {A.~D.}\
  \bibnamefont {Wieck}}, \bibinfo {author} {\bibfnamefont {A.}~\bibnamefont
  {Ludwig}}, \bibinfo {author} {\bibfnamefont {A.~S.}\ \bibnamefont
  {Sørensen}},\ and\ \bibinfo {author} {\bibfnamefont {P.}~\bibnamefont
  {Lodahl}},\ }\bibfield  {title} {\bibinfo {title} {Collective super- and
  subradiant dynamics between distant optical quantum emitters},\ }\href
  {https://doi.org/10.1126/science.ade9324} {\bibfield  {journal} {\bibinfo
  {journal} {Science}\ }\textbf {\bibinfo {volume} {379}},\ \bibinfo {pages}
  {389–393} (\bibinfo {year} {2023})}\BibitemShut {NoStop}%
\bibitem [{\citenamefont {Astafiev}\ \emph {et~al.}(2010)\citenamefont
  {Astafiev}, \citenamefont {Zagoskin}, \citenamefont {Abdumalikov},
  \citenamefont {Pashkin}, \citenamefont {Yamamoto}, \citenamefont {Inomata},
  \citenamefont {Nakamura},\ and\ \citenamefont
  {Tsai}}]{astafiev2010resonance}%
  \BibitemOpen
  \bibfield  {author} {\bibinfo {author} {\bibfnamefont {O.}~\bibnamefont
  {Astafiev}}, \bibinfo {author} {\bibfnamefont {A.~M.}\ \bibnamefont
  {Zagoskin}}, \bibinfo {author} {\bibfnamefont {A.~A.}\ \bibnamefont
  {Abdumalikov}}, \bibinfo {author} {\bibfnamefont {Y.~A.}\ \bibnamefont
  {Pashkin}}, \bibinfo {author} {\bibfnamefont {T.}~\bibnamefont {Yamamoto}},
  \bibinfo {author} {\bibfnamefont {K.}~\bibnamefont {Inomata}}, \bibinfo
  {author} {\bibfnamefont {Y.}~\bibnamefont {Nakamura}},\ and\ \bibinfo
  {author} {\bibfnamefont {J.~S.}\ \bibnamefont {Tsai}},\ }\bibfield  {title}
  {\bibinfo {title} {Resonance fluorescence of a single artificial atom},\
  }\href {https://doi.org/10.1126/science.1181918} {\bibfield  {journal}
  {\bibinfo  {journal} {Science}\ }\textbf {\bibinfo {volume} {327}},\ \bibinfo
  {pages} {840–843} (\bibinfo {year} {2010})}\BibitemShut {NoStop}%
\bibitem [{\citenamefont {Brehm}\ \emph {et~al.}(2021)\citenamefont {Brehm},
  \citenamefont {Poddubny}, \citenamefont {Stehli}, \citenamefont {Wolz},
  \citenamefont {Rotzinger},\ and\ \citenamefont
  {Ustinov}}]{brehm2021waveguide}%
  \BibitemOpen
  \bibfield  {author} {\bibinfo {author} {\bibfnamefont {J.~D.}\ \bibnamefont
  {Brehm}}, \bibinfo {author} {\bibfnamefont {A.~N.}\ \bibnamefont {Poddubny}},
  \bibinfo {author} {\bibfnamefont {A.}~\bibnamefont {Stehli}}, \bibinfo
  {author} {\bibfnamefont {T.}~\bibnamefont {Wolz}}, \bibinfo {author}
  {\bibfnamefont {H.}~\bibnamefont {Rotzinger}},\ and\ \bibinfo {author}
  {\bibfnamefont {A.~V.}\ \bibnamefont {Ustinov}},\ }\bibfield  {title}
  {\bibinfo {title} {Waveguide bandgap engineering with an array of
  superconducting qubits},\ }\href
  {http://dx.doi.org/10.1038/s41535-021-00310-z} {\bibfield  {journal}
  {\bibinfo  {journal} {npj Quantum Materials}\ }\textbf {\bibinfo {volume}
  {6}} (\bibinfo {year} {2021})}\BibitemShut {NoStop}%
\bibitem [{\citenamefont {Mirhosseini}\ \emph {et~al.}(2019)\citenamefont
  {Mirhosseini}, \citenamefont {Kim}, \citenamefont {Zhang}, \citenamefont
  {Sipahigil}, \citenamefont {Dieterle}, \citenamefont {Keller}, \citenamefont
  {Asenjo-Garcia}, \citenamefont {Chang},\ and\ \citenamefont
  {Painter}}]{mirhosseini2019cavity}%
  \BibitemOpen
  \bibfield  {author} {\bibinfo {author} {\bibfnamefont {M.}~\bibnamefont
  {Mirhosseini}}, \bibinfo {author} {\bibfnamefont {E.}~\bibnamefont {Kim}},
  \bibinfo {author} {\bibfnamefont {X.}~\bibnamefont {Zhang}}, \bibinfo
  {author} {\bibfnamefont {A.}~\bibnamefont {Sipahigil}}, \bibinfo {author}
  {\bibfnamefont {P.~B.}\ \bibnamefont {Dieterle}}, \bibinfo {author}
  {\bibfnamefont {A.~J.}\ \bibnamefont {Keller}}, \bibinfo {author}
  {\bibfnamefont {A.}~\bibnamefont {Asenjo-Garcia}}, \bibinfo {author}
  {\bibfnamefont {D.~E.}\ \bibnamefont {Chang}},\ and\ \bibinfo {author}
  {\bibfnamefont {O.}~\bibnamefont {Painter}},\ }\bibfield  {title} {\bibinfo
  {title} {Cavity quantum electrodynamics with atom-like mirrors},\ }\href
  {https://doi.org/10.1038/s41586-019-1196-1} {\bibfield  {journal} {\bibinfo
  {journal} {Nature}\ }\textbf {\bibinfo {volume} {569}},\ \bibinfo {pages}
  {692–697} (\bibinfo {year} {2019})}\BibitemShut {NoStop}%
\bibitem [{\citenamefont {Kannan}\ \emph {et~al.}(2023)\citenamefont {Kannan},
  \citenamefont {Almanakly}, \citenamefont {Sung}, \citenamefont {Di~Paolo},
  \citenamefont {Rower}, \citenamefont {Braumüller}, \citenamefont {Melville},
  \citenamefont {Niedzielski}, \citenamefont {Karamlou}, \citenamefont
  {Serniak}, \citenamefont {Vepsäläinen}, \citenamefont {Schwartz},
  \citenamefont {Yoder}, \citenamefont {Winik}, \citenamefont {Wang},
  \citenamefont {Orlando}, \citenamefont {Gustavsson}, \citenamefont {Grover},\
  and\ \citenamefont {Oliver}}]{kannan2023demand}%
  \BibitemOpen
  \bibfield  {author} {\bibinfo {author} {\bibfnamefont {B.}~\bibnamefont
  {Kannan}}, \bibinfo {author} {\bibfnamefont {A.}~\bibnamefont {Almanakly}},
  \bibinfo {author} {\bibfnamefont {Y.}~\bibnamefont {Sung}}, \bibinfo {author}
  {\bibfnamefont {A.}~\bibnamefont {Di~Paolo}}, \bibinfo {author}
  {\bibfnamefont {D.~A.}\ \bibnamefont {Rower}}, \bibinfo {author}
  {\bibfnamefont {J.}~\bibnamefont {Braumüller}}, \bibinfo {author}
  {\bibfnamefont {A.}~\bibnamefont {Melville}}, \bibinfo {author}
  {\bibfnamefont {B.~M.}\ \bibnamefont {Niedzielski}}, \bibinfo {author}
  {\bibfnamefont {A.}~\bibnamefont {Karamlou}}, \bibinfo {author}
  {\bibfnamefont {K.}~\bibnamefont {Serniak}}, \bibinfo {author} {\bibfnamefont
  {A.}~\bibnamefont {Vepsäläinen}}, \bibinfo {author} {\bibfnamefont {M.~E.}\
  \bibnamefont {Schwartz}}, \bibinfo {author} {\bibfnamefont {J.~L.}\
  \bibnamefont {Yoder}}, \bibinfo {author} {\bibfnamefont {R.}~\bibnamefont
  {Winik}}, \bibinfo {author} {\bibfnamefont {J.~I.-J.}\ \bibnamefont {Wang}},
  \bibinfo {author} {\bibfnamefont {T.~P.}\ \bibnamefont {Orlando}}, \bibinfo
  {author} {\bibfnamefont {S.}~\bibnamefont {Gustavsson}}, \bibinfo {author}
  {\bibfnamefont {J.~A.}\ \bibnamefont {Grover}},\ and\ \bibinfo {author}
  {\bibfnamefont {W.~D.}\ \bibnamefont {Oliver}},\ }\bibfield  {title}
  {\bibinfo {title} {On-demand directional microwave photon emission using
  waveguide quantum electrodynamics},\ }\href
  {https://doi.org/10.1038/s41567-022-01869-5} {\bibfield  {journal} {\bibinfo
  {journal} {Nature Physics}\ }\textbf {\bibinfo {volume} {19}},\ \bibinfo
  {pages} {394–400} (\bibinfo {year} {2023})}\BibitemShut {NoStop}%
\bibitem [{\citenamefont {Albrecht}\ \emph {et~al.}(2019)\citenamefont
  {Albrecht}, \citenamefont {Henriet}, \citenamefont {Asenjo-Garcia},
  \citenamefont {Dieterle}, \citenamefont {Painter},\ and\ \citenamefont
  {Chang}}]{albrecht2019subradiant}%
  \BibitemOpen
  \bibfield  {author} {\bibinfo {author} {\bibfnamefont {A.}~\bibnamefont
  {Albrecht}}, \bibinfo {author} {\bibfnamefont {L.}~\bibnamefont {Henriet}},
  \bibinfo {author} {\bibfnamefont {A.}~\bibnamefont {Asenjo-Garcia}}, \bibinfo
  {author} {\bibfnamefont {P.~B.}\ \bibnamefont {Dieterle}}, \bibinfo {author}
  {\bibfnamefont {O.}~\bibnamefont {Painter}},\ and\ \bibinfo {author}
  {\bibfnamefont {D.~E.}\ \bibnamefont {Chang}},\ }\bibfield  {title} {\bibinfo
  {title} {Subradiant states of quantum bits coupled to a one-dimensional
  waveguide},\ }\href {https://doi.org/10.1088/1367-2630/ab0134} {\bibfield
  {journal} {\bibinfo  {journal} {New Journal of Physics}\ }\textbf {\bibinfo
  {volume} {21}},\ \bibinfo {pages} {025003} (\bibinfo {year}
  {2019})}\BibitemShut {NoStop}%
\bibitem [{\citenamefont {Henriet}\ \emph {et~al.}(2019)\citenamefont
  {Henriet}, \citenamefont {Douglas}, \citenamefont {Chang},\ and\
  \citenamefont {Albrecht}}]{henriet2019critical}%
  \BibitemOpen
  \bibfield  {author} {\bibinfo {author} {\bibfnamefont {L.}~\bibnamefont
  {Henriet}}, \bibinfo {author} {\bibfnamefont {J.~S.}\ \bibnamefont
  {Douglas}}, \bibinfo {author} {\bibfnamefont {D.~E.}\ \bibnamefont {Chang}},\
  and\ \bibinfo {author} {\bibfnamefont {A.}~\bibnamefont {Albrecht}},\
  }\bibfield  {title} {\bibinfo {title} {Critical open-system dynamics in a
  one-dimensional optical-lattice clock},\ }\href
  {https://doi.org/10.1103/PhysRevA.99.023802} {\bibfield  {journal} {\bibinfo
  {journal} {Phys. Rev. A}\ }\textbf {\bibinfo {volume} {99}},\ \bibinfo
  {pages} {023802} (\bibinfo {year} {2019})}\BibitemShut {NoStop}%
\bibitem [{\citenamefont {Zhang}\ and\ \citenamefont
  {M\o{}lmer}(2019)}]{zhang2019theory}%
  \BibitemOpen
  \bibfield  {author} {\bibinfo {author} {\bibfnamefont {Y.-X.}\ \bibnamefont
  {Zhang}}\ and\ \bibinfo {author} {\bibfnamefont {K.}~\bibnamefont
  {M\o{}lmer}},\ }\bibfield  {title} {\bibinfo {title} {Theory of subradiant
  states of a one-dimensional two-level atom chain},\ }\href
  {https://doi.org/10.1103/PhysRevLett.122.203605} {\bibfield  {journal}
  {\bibinfo  {journal} {Phys. Rev. Lett.}\ }\textbf {\bibinfo {volume} {122}},\
  \bibinfo {pages} {203605} (\bibinfo {year} {2019})}\BibitemShut {NoStop}%
\bibitem [{\citenamefont {Ostermann}\ \emph {et~al.}(2019)\citenamefont
  {Ostermann}, \citenamefont {Meignant}, \citenamefont {Genes},\ and\
  \citenamefont {Ritsch}}]{ostermann2019super}%
  \BibitemOpen
  \bibfield  {author} {\bibinfo {author} {\bibfnamefont {L.}~\bibnamefont
  {Ostermann}}, \bibinfo {author} {\bibfnamefont {C.}~\bibnamefont {Meignant}},
  \bibinfo {author} {\bibfnamefont {C.}~\bibnamefont {Genes}},\ and\ \bibinfo
  {author} {\bibfnamefont {H.}~\bibnamefont {Ritsch}},\ }\bibfield  {title}
  {\bibinfo {title} {Super- and subradiance of clock atoms in multimode optical
  waveguides},\ }\href {https://doi.org/10.1088/1367-2630/ab05fb} {\bibfield
  {journal} {\bibinfo  {journal} {New Journal of Physics}\ }\textbf {\bibinfo
  {volume} {21}},\ \bibinfo {pages} {025004} (\bibinfo {year}
  {2019})}\BibitemShut {NoStop}%
\bibitem [{\citenamefont {Needham}\ \emph {et~al.}(2019)\citenamefont
  {Needham}, \citenamefont {Lesanovsky},\ and\ \citenamefont
  {Olmos}}]{needham2019subradiance}%
  \BibitemOpen
  \bibfield  {author} {\bibinfo {author} {\bibfnamefont {J.~A.}\ \bibnamefont
  {Needham}}, \bibinfo {author} {\bibfnamefont {I.}~\bibnamefont
  {Lesanovsky}},\ and\ \bibinfo {author} {\bibfnamefont {B.}~\bibnamefont
  {Olmos}},\ }\bibfield  {title} {\bibinfo {title} {Subradiance-protected
  excitation transport},\ }\href {https://doi.org/10.1088/1367-2630/ab31e8}
  {\bibfield  {journal} {\bibinfo  {journal} {New Journal of Physics}\ }\textbf
  {\bibinfo {volume} {21}},\ \bibinfo {pages} {073061} (\bibinfo {year}
  {2019})}\BibitemShut {NoStop}%
\bibitem [{\citenamefont {Zhong}\ \emph {et~al.}(2020)\citenamefont {Zhong},
  \citenamefont {Olekhno}, \citenamefont {Ke}, \citenamefont {Poshakinskiy},
  \citenamefont {Lee}, \citenamefont {Kivshar},\ and\ \citenamefont
  {Poddubny}}]{zhong2020photon}%
  \BibitemOpen
  \bibfield  {author} {\bibinfo {author} {\bibfnamefont {J.}~\bibnamefont
  {Zhong}}, \bibinfo {author} {\bibfnamefont {N.~A.}\ \bibnamefont {Olekhno}},
  \bibinfo {author} {\bibfnamefont {Y.}~\bibnamefont {Ke}}, \bibinfo {author}
  {\bibfnamefont {A.~V.}\ \bibnamefont {Poshakinskiy}}, \bibinfo {author}
  {\bibfnamefont {C.}~\bibnamefont {Lee}}, \bibinfo {author} {\bibfnamefont
  {Y.~S.}\ \bibnamefont {Kivshar}},\ and\ \bibinfo {author} {\bibfnamefont
  {A.~N.}\ \bibnamefont {Poddubny}},\ }\bibfield  {title} {\bibinfo {title}
  {Photon-mediated localization in two-level qubit arrays},\ }\href
  {https://doi.org/10.1103/PhysRevLett.124.093604} {\bibfield  {journal}
  {\bibinfo  {journal} {Phys. Rev. Lett.}\ }\textbf {\bibinfo {volume} {124}},\
  \bibinfo {pages} {093604} (\bibinfo {year} {2020})}\BibitemShut {NoStop}%
\bibitem [{\citenamefont {Kornovan}\ \emph {et~al.}(2019)\citenamefont
  {Kornovan}, \citenamefont {Corzo}, \citenamefont {Laurat},\ and\
  \citenamefont {Sheremet}}]{kornovan2019extremely}%
  \BibitemOpen
  \bibfield  {author} {\bibinfo {author} {\bibfnamefont {D.~F.}\ \bibnamefont
  {Kornovan}}, \bibinfo {author} {\bibfnamefont {N.~V.}\ \bibnamefont {Corzo}},
  \bibinfo {author} {\bibfnamefont {J.}~\bibnamefont {Laurat}},\ and\ \bibinfo
  {author} {\bibfnamefont {A.~S.}\ \bibnamefont {Sheremet}},\ }\bibfield
  {title} {\bibinfo {title} {Extremely subradiant states in a periodic
  one-dimensional atomic array},\ }\href
  {https://doi.org/10.1103/PhysRevA.100.063832} {\bibfield  {journal} {\bibinfo
   {journal} {Phys. Rev. A}\ }\textbf {\bibinfo {volume} {100}},\ \bibinfo
  {pages} {063832} (\bibinfo {year} {2019})}\BibitemShut {NoStop}%
\bibitem [{\citenamefont {Schrinski}\ and\ \citenamefont
  {Sørensen}(2022)}]{Schrinski_polariton}%
  \BibitemOpen
  \bibfield  {author} {\bibinfo {author} {\bibfnamefont {B.}~\bibnamefont
  {Schrinski}}\ and\ \bibinfo {author} {\bibfnamefont {A.~S.}\ \bibnamefont
  {Sørensen}},\ }\bibfield  {title} {\bibinfo {title} {Polariton dynamics in
  one-dimensional arrays of atoms coupled to waveguides},\ }\href
  {https://doi.org/10.1088/1367-2630/acaa4f} {\bibfield  {journal} {\bibinfo
  {journal} {New Journal of Physics}\ }\textbf {\bibinfo {volume} {24}},\
  \bibinfo {pages} {123023} (\bibinfo {year} {2022})}\BibitemShut {NoStop}%
\bibitem [{\citenamefont {Shen}\ and\ \citenamefont
  {Fan}(2007{\natexlab{a}})}]{shen2007strongly}%
  \BibitemOpen
  \bibfield  {author} {\bibinfo {author} {\bibfnamefont {J.-T.}\ \bibnamefont
  {Shen}}\ and\ \bibinfo {author} {\bibfnamefont {S.}~\bibnamefont {Fan}},\
  }\bibfield  {title} {\bibinfo {title} {Strongly correlated multiparticle
  transport in one dimension through a quantum impurity},\ }\href
  {https://doi.org/10.1103/PhysRevA.76.062709} {\bibfield  {journal} {\bibinfo
  {journal} {Phys. Rev. A}\ }\textbf {\bibinfo {volume} {76}},\ \bibinfo
  {pages} {062709} (\bibinfo {year} {2007}{\natexlab{a}})}\BibitemShut
  {NoStop}%
\bibitem [{\citenamefont {Shen}\ and\ \citenamefont
  {Fan}(2007{\natexlab{b}})}]{shen2007stronglyL}%
  \BibitemOpen
  \bibfield  {author} {\bibinfo {author} {\bibfnamefont {J.-T.}\ \bibnamefont
  {Shen}}\ and\ \bibinfo {author} {\bibfnamefont {S.}~\bibnamefont {Fan}},\
  }\bibfield  {title} {\bibinfo {title} {Strongly correlated two-photon
  transport in a one-dimensional waveguide coupled to a two-level system},\
  }\href {https://doi.org/10.1103/PhysRevLett.98.153003} {\bibfield  {journal}
  {\bibinfo  {journal} {Phys. Rev. Lett.}\ }\textbf {\bibinfo {volume} {98}},\
  \bibinfo {pages} {153003} (\bibinfo {year} {2007}{\natexlab{b}})}\BibitemShut
  {NoStop}%
\bibitem [{\citenamefont {Zheng}\ \emph {et~al.}(2011)\citenamefont {Zheng},
  \citenamefont {Gauthier},\ and\ \citenamefont {Baranger}}]{zheng2011cavity}%
  \BibitemOpen
  \bibfield  {author} {\bibinfo {author} {\bibfnamefont {H.}~\bibnamefont
  {Zheng}}, \bibinfo {author} {\bibfnamefont {D.~J.}\ \bibnamefont
  {Gauthier}},\ and\ \bibinfo {author} {\bibfnamefont {H.~U.}\ \bibnamefont
  {Baranger}},\ }\bibfield  {title} {\bibinfo {title} {Cavity-free photon
  blockade induced by many-body bound states},\ }\href
  {http://dx.doi.org/10.1103/PhysRevLett.107.223601} {\bibfield  {journal}
  {\bibinfo  {journal} {Physical Review Letters}\ }\textbf {\bibinfo {volume}
  {107}} (\bibinfo {year} {2011})}\BibitemShut {NoStop}%
\bibitem [{\citenamefont {Mahmoodian}\ \emph {et~al.}(2018)\citenamefont
  {Mahmoodian}, \citenamefont {\ifmmode~\check{C}\else \v{C}\fi{}epulkovskis},
  \citenamefont {Das}, \citenamefont {Lodahl}, \citenamefont {Hammerer},\ and\
  \citenamefont {S\o{}rensen}}]{mahmoodian2018strongly}%
  \BibitemOpen
  \bibfield  {author} {\bibinfo {author} {\bibfnamefont {S.}~\bibnamefont
  {Mahmoodian}}, \bibinfo {author} {\bibfnamefont {M.}~\bibnamefont
  {\ifmmode~\check{C}\else \v{C}\fi{}epulkovskis}}, \bibinfo {author}
  {\bibfnamefont {S.}~\bibnamefont {Das}}, \bibinfo {author} {\bibfnamefont
  {P.}~\bibnamefont {Lodahl}}, \bibinfo {author} {\bibfnamefont
  {K.}~\bibnamefont {Hammerer}},\ and\ \bibinfo {author} {\bibfnamefont
  {A.~S.}\ \bibnamefont {S\o{}rensen}},\ }\bibfield  {title} {\bibinfo {title}
  {Strongly correlated photon transport in waveguide quantum electrodynamics
  with weakly coupled emitters},\ }\href
  {https://doi.org/10.1103/PhysRevLett.121.143601} {\bibfield  {journal}
  {\bibinfo  {journal} {Phys. Rev. Lett.}\ }\textbf {\bibinfo {volume} {121}},\
  \bibinfo {pages} {143601} (\bibinfo {year} {2018})}\BibitemShut {NoStop}%
\bibitem [{\citenamefont {Prasad}\ \emph {et~al.}(2020)\citenamefont {Prasad},
  \citenamefont {Hinney}, \citenamefont {Mahmoodian}, \citenamefont {Hammerer},
  \citenamefont {Rind}, \citenamefont {Schneeweiss}, \citenamefont {Sørensen},
  \citenamefont {Volz},\ and\ \citenamefont
  {Rauschenbeutel}}]{prasad2020correlating}%
  \BibitemOpen
  \bibfield  {author} {\bibinfo {author} {\bibfnamefont {A.~S.}\ \bibnamefont
  {Prasad}}, \bibinfo {author} {\bibfnamefont {J.}~\bibnamefont {Hinney}},
  \bibinfo {author} {\bibfnamefont {S.}~\bibnamefont {Mahmoodian}}, \bibinfo
  {author} {\bibfnamefont {K.}~\bibnamefont {Hammerer}}, \bibinfo {author}
  {\bibfnamefont {S.}~\bibnamefont {Rind}}, \bibinfo {author} {\bibfnamefont
  {P.}~\bibnamefont {Schneeweiss}}, \bibinfo {author} {\bibfnamefont {A.~S.}\
  \bibnamefont {Sørensen}}, \bibinfo {author} {\bibfnamefont {J.}~\bibnamefont
  {Volz}},\ and\ \bibinfo {author} {\bibfnamefont {A.}~\bibnamefont
  {Rauschenbeutel}},\ }\bibfield  {title} {\bibinfo {title} {Correlating
  photons using the collective nonlinear response of atoms weakly coupled to an
  optical mode},\ }\href {https://doi.org/10.1038/s41566-020-0692-z} {\bibfield
   {journal} {\bibinfo  {journal} {Nature Photonics}\ }\textbf {\bibinfo
  {volume} {14}},\ \bibinfo {pages} {719–722} (\bibinfo {year}
  {2020})}\BibitemShut {NoStop}%
\bibitem [{\citenamefont {Mahmoodian}\ \emph {et~al.}(2020)\citenamefont
  {Mahmoodian}, \citenamefont {Calaj\'o}, \citenamefont {Chang}, \citenamefont
  {Hammerer},\ and\ \citenamefont {S\o{}rensen}}]{mahmoodian2020dynamics}%
  \BibitemOpen
  \bibfield  {author} {\bibinfo {author} {\bibfnamefont {S.}~\bibnamefont
  {Mahmoodian}}, \bibinfo {author} {\bibfnamefont {G.}~\bibnamefont
  {Calaj\'o}}, \bibinfo {author} {\bibfnamefont {D.~E.}\ \bibnamefont {Chang}},
  \bibinfo {author} {\bibfnamefont {K.}~\bibnamefont {Hammerer}},\ and\
  \bibinfo {author} {\bibfnamefont {A.~S.}\ \bibnamefont {S\o{}rensen}},\
  }\bibfield  {title} {\bibinfo {title} {Dynamics of many-body photon bound
  states in chiral waveguide qed},\ }\href
  {https://doi.org/10.1103/PhysRevX.10.031011} {\bibfield  {journal} {\bibinfo
  {journal} {Phys. Rev. X}\ }\textbf {\bibinfo {volume} {10}},\ \bibinfo
  {pages} {031011} (\bibinfo {year} {2020})}\BibitemShut {NoStop}%
\bibitem [{\citenamefont {Le~Jeannic}\ \emph {et~al.}(2022)\citenamefont
  {Le~Jeannic}, \citenamefont {Tiranov}, \citenamefont {Carolan}, \citenamefont
  {Ramos}, \citenamefont {Wang}, \citenamefont {Appel}, \citenamefont {Scholz},
  \citenamefont {Wieck}, \citenamefont {Ludwig}, \citenamefont {Rotenberg},
  \citenamefont {Midolo}, \citenamefont {García-Ripoll}, \citenamefont
  {Sørensen},\ and\ \citenamefont {Lodahl}}]{le2022dynamical}%
  \BibitemOpen
  \bibfield  {author} {\bibinfo {author} {\bibfnamefont {H.}~\bibnamefont
  {Le~Jeannic}}, \bibinfo {author} {\bibfnamefont {A.}~\bibnamefont {Tiranov}},
  \bibinfo {author} {\bibfnamefont {J.}~\bibnamefont {Carolan}}, \bibinfo
  {author} {\bibfnamefont {T.}~\bibnamefont {Ramos}}, \bibinfo {author}
  {\bibfnamefont {Y.}~\bibnamefont {Wang}}, \bibinfo {author} {\bibfnamefont
  {M.~H.}\ \bibnamefont {Appel}}, \bibinfo {author} {\bibfnamefont
  {S.}~\bibnamefont {Scholz}}, \bibinfo {author} {\bibfnamefont {A.~D.}\
  \bibnamefont {Wieck}}, \bibinfo {author} {\bibfnamefont {A.}~\bibnamefont
  {Ludwig}}, \bibinfo {author} {\bibfnamefont {N.}~\bibnamefont {Rotenberg}},
  \bibinfo {author} {\bibfnamefont {L.}~\bibnamefont {Midolo}}, \bibinfo
  {author} {\bibfnamefont {J.~J.}\ \bibnamefont {García-Ripoll}}, \bibinfo
  {author} {\bibfnamefont {A.~S.}\ \bibnamefont {Sørensen}},\ and\ \bibinfo
  {author} {\bibfnamefont {P.}~\bibnamefont {Lodahl}},\ }\bibfield  {title}
  {\bibinfo {title} {Dynamical photon–photon interaction mediated by a
  quantum emitter},\ }\href {https://doi.org/10.1038/s41567-022-01720-x}
  {\bibfield  {journal} {\bibinfo  {journal} {Nature Physics}\ }\textbf
  {\bibinfo {volume} {18}},\ \bibinfo {pages} {1191–1195} (\bibinfo {year}
  {2022})}\BibitemShut {NoStop}%
\bibitem [{\citenamefont {Tomm}\ \emph {et~al.}(2023)\citenamefont {Tomm},
  \citenamefont {Mahmoodian}, \citenamefont {Antoniadis}, \citenamefont
  {Schott}, \citenamefont {Valentin}, \citenamefont {Wieck}, \citenamefont
  {Ludwig}, \citenamefont {Javadi},\ and\ \citenamefont
  {Warburton}}]{tomm2023photon}%
  \BibitemOpen
  \bibfield  {author} {\bibinfo {author} {\bibfnamefont {N.}~\bibnamefont
  {Tomm}}, \bibinfo {author} {\bibfnamefont {S.}~\bibnamefont {Mahmoodian}},
  \bibinfo {author} {\bibfnamefont {N.~O.}\ \bibnamefont {Antoniadis}},
  \bibinfo {author} {\bibfnamefont {R.}~\bibnamefont {Schott}}, \bibinfo
  {author} {\bibfnamefont {S.~R.}\ \bibnamefont {Valentin}}, \bibinfo {author}
  {\bibfnamefont {A.~D.}\ \bibnamefont {Wieck}}, \bibinfo {author}
  {\bibfnamefont {A.}~\bibnamefont {Ludwig}}, \bibinfo {author} {\bibfnamefont
  {A.}~\bibnamefont {Javadi}},\ and\ \bibinfo {author} {\bibfnamefont {R.~J.}\
  \bibnamefont {Warburton}},\ }\bibfield  {title} {\bibinfo {title} {Photon
  bound state dynamics from a single artificial atom},\ }\href
  {https://doi.org/10.1038/s41567-023-01997-6} {\bibfield  {journal} {\bibinfo
  {journal} {Nature Physics}\ }\textbf {\bibinfo {volume} {19}},\ \bibinfo
  {pages} {857–862} (\bibinfo {year} {2023})}\BibitemShut {NoStop}%
\bibitem [{\citenamefont {Zhang}\ and\ \citenamefont
  {M\o{}lmer}(2020)}]{zhang2020subradiant}%
  \BibitemOpen
  \bibfield  {author} {\bibinfo {author} {\bibfnamefont {Y.-X.}\ \bibnamefont
  {Zhang}}\ and\ \bibinfo {author} {\bibfnamefont {K.}~\bibnamefont
  {M\o{}lmer}},\ }\bibfield  {title} {\bibinfo {title} {Subradiant emission
  from regular atomic arrays: Universal scaling of decay rates from the
  generalized bloch theorem},\ }\href
  {https://doi.org/10.1103/PhysRevLett.125.253601} {\bibfield  {journal}
  {\bibinfo  {journal} {Phys. Rev. Lett.}\ }\textbf {\bibinfo {volume} {125}},\
  \bibinfo {pages} {253601} (\bibinfo {year} {2020})}\BibitemShut {NoStop}%
\bibitem [{\citenamefont {Poddubny}(2020)}]{poddubny2020quasiflat}%
  \BibitemOpen
  \bibfield  {author} {\bibinfo {author} {\bibfnamefont {A.~N.}\ \bibnamefont
  {Poddubny}},\ }\bibfield  {title} {\bibinfo {title} {Quasiflat band enabling
  subradiant two-photon bound states},\ }\href
  {https://doi.org/10.1103/PhysRevA.101.043845} {\bibfield  {journal} {\bibinfo
   {journal} {Phys. Rev. A}\ }\textbf {\bibinfo {volume} {101}},\ \bibinfo
  {pages} {043845} (\bibinfo {year} {2020})}\BibitemShut {NoStop}%
\bibitem [{\citenamefont {Bakkensen}\ \emph {et~al.}(2021)\citenamefont
  {Bakkensen}, \citenamefont {Zhang}, \citenamefont {Bjerlin},\ and\
  \citenamefont {S{\o}rensen}}]{bakkensen2021photonic}%
  \BibitemOpen
  \bibfield  {author} {\bibinfo {author} {\bibfnamefont {B.}~\bibnamefont
  {Bakkensen}}, \bibinfo {author} {\bibfnamefont {Y.-X.}\ \bibnamefont
  {Zhang}}, \bibinfo {author} {\bibfnamefont {J.}~\bibnamefont {Bjerlin}},\
  and\ \bibinfo {author} {\bibfnamefont {A.~S.}\ \bibnamefont {S{\o}rensen}},\
  }\bibfield  {title} {\bibinfo {title} {Photonic bound states and scattering
  resonances in waveguide qed},\ }\href
  {https://doi.org/10.48550/arXiv.2110.06093} {\  (\bibinfo {year} {2021})},\
  \Eprint {https://arxiv.org/abs/2110.06093} {arXiv:2110.06093} \BibitemShut
  {NoStop}%
\bibitem [{\citenamefont {Calaj\'o}\ and\ \citenamefont
  {Chang}(2022)}]{calajo2022emergence}%
  \BibitemOpen
  \bibfield  {author} {\bibinfo {author} {\bibfnamefont {G.}~\bibnamefont
  {Calaj\'o}}\ and\ \bibinfo {author} {\bibfnamefont {D.~E.}\ \bibnamefont
  {Chang}},\ }\bibfield  {title} {\bibinfo {title} {Emergence of solitons from
  many-body photon bound states in quantum nonlinear media},\ }\href
  {https://doi.org/10.1103/PhysRevResearch.4.023026} {\bibfield  {journal}
  {\bibinfo  {journal} {Phys. Rev. Res.}\ }\textbf {\bibinfo {volume} {4}},\
  \bibinfo {pages} {023026} (\bibinfo {year} {2022})}\BibitemShut {NoStop}%
\bibitem [{\citenamefont {Poshakinskiy}\ and\ \citenamefont
  {Poddubny}(2023)}]{PhysRevA.108.023707}%
  \BibitemOpen
  \bibfield  {author} {\bibinfo {author} {\bibfnamefont {A.~V.}\ \bibnamefont
  {Poshakinskiy}}\ and\ \bibinfo {author} {\bibfnamefont {A.~N.}\ \bibnamefont
  {Poddubny}},\ }\bibfield  {title} {\bibinfo {title} {Bound state of distant
  photons in waveguide quantum electrodynamics},\ }\href
  {https://doi.org/10.1103/PhysRevA.108.023707} {\bibfield  {journal} {\bibinfo
   {journal} {Phys. Rev. A}\ }\textbf {\bibinfo {volume} {108}},\ \bibinfo
  {pages} {023707} (\bibinfo {year} {2023})}\BibitemShut {NoStop}%
\bibitem [{\citenamefont {Cardenas-Lopez}\ \emph {et~al.}(2023)\citenamefont
  {Cardenas-Lopez}, \citenamefont {Masson}, \citenamefont {Zager},\ and\
  \citenamefont {Asenjo-Garcia}}]{PhysRevLett.131.033605}%
  \BibitemOpen
  \bibfield  {author} {\bibinfo {author} {\bibfnamefont {S.}~\bibnamefont
  {Cardenas-Lopez}}, \bibinfo {author} {\bibfnamefont {S.~J.}\ \bibnamefont
  {Masson}}, \bibinfo {author} {\bibfnamefont {Z.}~\bibnamefont {Zager}},\ and\
  \bibinfo {author} {\bibfnamefont {A.}~\bibnamefont {Asenjo-Garcia}},\
  }\bibfield  {title} {\bibinfo {title} {Many-body superradiance and dynamical
  mirror symmetry breaking in waveguide qed},\ }\href
  {https://doi.org/10.1103/PhysRevLett.131.033605} {\bibfield  {journal}
  {\bibinfo  {journal} {Phys. Rev. Lett.}\ }\textbf {\bibinfo {volume} {131}},\
  \bibinfo {pages} {033605} (\bibinfo {year} {2023})}\BibitemShut {NoStop}%
\bibitem [{\citenamefont {Liedl}\ \emph {et~al.}(2024)\citenamefont {Liedl},
  \citenamefont {Tebbenjohanns}, \citenamefont {Bach}, \citenamefont {Pucher},
  \citenamefont {Rauschenbeutel},\ and\ \citenamefont
  {Schneeweiss}}]{PhysRevX.14.011020}%
  \BibitemOpen
  \bibfield  {author} {\bibinfo {author} {\bibfnamefont {C.}~\bibnamefont
  {Liedl}}, \bibinfo {author} {\bibfnamefont {F.}~\bibnamefont
  {Tebbenjohanns}}, \bibinfo {author} {\bibfnamefont {C.}~\bibnamefont {Bach}},
  \bibinfo {author} {\bibfnamefont {S.}~\bibnamefont {Pucher}}, \bibinfo
  {author} {\bibfnamefont {A.}~\bibnamefont {Rauschenbeutel}},\ and\ \bibinfo
  {author} {\bibfnamefont {P.}~\bibnamefont {Schneeweiss}},\ }\bibfield
  {title} {\bibinfo {title} {Observation of superradiant bursts in a cascaded
  quantum system},\ }\href {https://doi.org/10.1103/PhysRevX.14.011020}
  {\bibfield  {journal} {\bibinfo  {journal} {Phys. Rev. X}\ }\textbf {\bibinfo
  {volume} {14}},\ \bibinfo {pages} {011020} (\bibinfo {year}
  {2024})}\BibitemShut {NoStop}%
\bibitem [{\citenamefont {Goncalves}\ \emph {et~al.}(2025)\citenamefont
  {Goncalves}, \citenamefont {Bombieri}, \citenamefont {Ferioli}, \citenamefont
  {Pancaldi}, \citenamefont {Ferrier-Barbut}, \citenamefont {Browaeys},
  \citenamefont {Shahmoon},\ and\ \citenamefont {Chang}}]{goncalves2403driven}%
  \BibitemOpen
  \bibfield  {author} {\bibinfo {author} {\bibfnamefont {D.}~\bibnamefont
  {Goncalves}}, \bibinfo {author} {\bibfnamefont {L.}~\bibnamefont {Bombieri}},
  \bibinfo {author} {\bibfnamefont {G.}~\bibnamefont {Ferioli}}, \bibinfo
  {author} {\bibfnamefont {S.}~\bibnamefont {Pancaldi}}, \bibinfo {author}
  {\bibfnamefont {I.}~\bibnamefont {Ferrier-Barbut}}, \bibinfo {author}
  {\bibfnamefont {A.}~\bibnamefont {Browaeys}}, \bibinfo {author}
  {\bibfnamefont {E.}~\bibnamefont {Shahmoon}},\ and\ \bibinfo {author}
  {\bibfnamefont {D.}~\bibnamefont {Chang}},\ }\bibfield  {title} {\bibinfo
  {title} {Driven-dissipative phase separation in free-space atomic
  ensembles},\ }\href {https://doi.org/10.1103/PRXQuantum.6.020303} {\bibfield
  {journal} {\bibinfo  {journal} {PRX Quantum}\ }\textbf {\bibinfo {volume}
  {6}},\ \bibinfo {pages} {020303} (\bibinfo {year} {2025})}\BibitemShut
  {NoStop}%
\bibitem [{\citenamefont {Poshakinskiy}\ and\ \citenamefont
  {Poddubny}(2021)}]{PhysRevLett.127.173601}%
  \BibitemOpen
  \bibfield  {author} {\bibinfo {author} {\bibfnamefont {A.~V.}\ \bibnamefont
  {Poshakinskiy}}\ and\ \bibinfo {author} {\bibfnamefont {A.~N.}\ \bibnamefont
  {Poddubny}},\ }\bibfield  {title} {\bibinfo {title} {Dimerization of
  many-body subradiant states in waveguide quantum electrodynamics},\ }\href
  {https://doi.org/10.1103/PhysRevLett.127.173601} {\bibfield  {journal}
  {\bibinfo  {journal} {Phys. Rev. Lett.}\ }\textbf {\bibinfo {volume} {127}},\
  \bibinfo {pages} {173601} (\bibinfo {year} {2021})}\BibitemShut {NoStop}%
\bibitem [{\citenamefont {Shi}\ and\ \citenamefont
  {Poddubny}(2024)}]{PhysRevA.110.053707}%
  \BibitemOpen
  \bibfield  {author} {\bibinfo {author} {\bibfnamefont {J.}~\bibnamefont
  {Shi}}\ and\ \bibinfo {author} {\bibfnamefont {A.~N.}\ \bibnamefont
  {Poddubny}},\ }\bibfield  {title} {\bibinfo {title} {Multimer states in
  multilevel waveguide qed},\ }\href
  {https://doi.org/10.1103/PhysRevA.110.053707} {\bibfield  {journal} {\bibinfo
   {journal} {Phys. Rev. A}\ }\textbf {\bibinfo {volume} {110}},\ \bibinfo
  {pages} {053707} (\bibinfo {year} {2024})}\BibitemShut {NoStop}%
\bibitem [{\citenamefont {Lonigro}\ \emph {et~al.}(2021)\citenamefont
  {Lonigro}, \citenamefont {Facchi}, \citenamefont {Pascazio}, \citenamefont
  {Pepe},\ and\ \citenamefont {Pomarico}}]{lonigro2021stationary}%
  \BibitemOpen
  \bibfield  {author} {\bibinfo {author} {\bibfnamefont {D.}~\bibnamefont
  {Lonigro}}, \bibinfo {author} {\bibfnamefont {P.}~\bibnamefont {Facchi}},
  \bibinfo {author} {\bibfnamefont {S.}~\bibnamefont {Pascazio}}, \bibinfo
  {author} {\bibfnamefont {F.~V.}\ \bibnamefont {Pepe}},\ and\ \bibinfo
  {author} {\bibfnamefont {D.}~\bibnamefont {Pomarico}},\ }\bibfield  {title}
  {\bibinfo {title} {Stationary excitation waves and multimerization in arrays
  of quantum emitters},\ }\href
  {https://iopscience.iop.org/article/10.1088/1367-2630/ac2ce0} {\bibfield
  {journal} {\bibinfo  {journal} {New Journal of Physics}\ }\textbf {\bibinfo
  {volume} {23}},\ \bibinfo {pages} {103033} (\bibinfo {year}
  {2021})}\BibitemShut {NoStop}%
\bibitem [{\citenamefont {Stannigel}\ \emph {et~al.}(2012)\citenamefont
  {Stannigel}, \citenamefont {Rabl},\ and\ \citenamefont
  {Zoller}}]{stannigel2012driven}%
  \BibitemOpen
  \bibfield  {author} {\bibinfo {author} {\bibfnamefont {K.}~\bibnamefont
  {Stannigel}}, \bibinfo {author} {\bibfnamefont {P.}~\bibnamefont {Rabl}},\
  and\ \bibinfo {author} {\bibfnamefont {P.}~\bibnamefont {Zoller}},\
  }\bibfield  {title} {\bibinfo {title} {Driven-dissipative preparation of
  entangled states in cascaded quantum-optical networks},\ }\href
  {https://iopscience.iop.org/article/10.1088/1367-2630/14/6/063014/meta}
  {\bibfield  {journal} {\bibinfo  {journal} {New Journal of Physics}\ }\textbf
  {\bibinfo {volume} {14}},\ \bibinfo {pages} {063014} (\bibinfo {year}
  {2012})}\BibitemShut {NoStop}%
\bibitem [{\citenamefont {Pichler}\ \emph {et~al.}(2015)\citenamefont
  {Pichler}, \citenamefont {Ramos}, \citenamefont {Daley},\ and\ \citenamefont
  {Zoller}}]{PhysRevA.91.042116}%
  \BibitemOpen
  \bibfield  {author} {\bibinfo {author} {\bibfnamefont {H.}~\bibnamefont
  {Pichler}}, \bibinfo {author} {\bibfnamefont {T.}~\bibnamefont {Ramos}},
  \bibinfo {author} {\bibfnamefont {A.~J.}\ \bibnamefont {Daley}},\ and\
  \bibinfo {author} {\bibfnamefont {P.}~\bibnamefont {Zoller}},\ }\bibfield
  {title} {\bibinfo {title} {Quantum optics of chiral spin networks},\ }\href
  {https://doi.org/10.1103/PhysRevA.91.042116} {\bibfield  {journal} {\bibinfo
  {journal} {Phys. Rev. A}\ }\textbf {\bibinfo {volume} {91}},\ \bibinfo
  {pages} {042116} (\bibinfo {year} {2015})}\BibitemShut {NoStop}%
\bibitem [{\citenamefont {Shah}\ \emph {et~al.}(2024)\citenamefont {Shah},
  \citenamefont {Yang}, \citenamefont {Joshi},\ and\ \citenamefont
  {Mirhosseini}}]{PRXQuantum.5.030346}%
  \BibitemOpen
  \bibfield  {author} {\bibinfo {author} {\bibfnamefont {P.~S.}\ \bibnamefont
  {Shah}}, \bibinfo {author} {\bibfnamefont {F.}~\bibnamefont {Yang}}, \bibinfo
  {author} {\bibfnamefont {C.}~\bibnamefont {Joshi}},\ and\ \bibinfo {author}
  {\bibfnamefont {M.}~\bibnamefont {Mirhosseini}},\ }\bibfield  {title}
  {\bibinfo {title} {Stabilizing remote entanglement via waveguide
  dissipation},\ }\href {https://doi.org/10.1103/PRXQuantum.5.030346}
  {\bibfield  {journal} {\bibinfo  {journal} {PRX Quantum}\ }\textbf {\bibinfo
  {volume} {5}},\ \bibinfo {pages} {030346} (\bibinfo {year}
  {2024})}\BibitemShut {NoStop}%
\bibitem [{\citenamefont {Zhang}\ \emph {et~al.}(2023)\citenamefont {Zhang},
  \citenamefont {Kim}, \citenamefont {Mark}, \citenamefont {Choi},\ and\
  \citenamefont {Painter}}]{zhang2023superconducting}%
  \BibitemOpen
  \bibfield  {author} {\bibinfo {author} {\bibfnamefont {X.}~\bibnamefont
  {Zhang}}, \bibinfo {author} {\bibfnamefont {E.}~\bibnamefont {Kim}}, \bibinfo
  {author} {\bibfnamefont {D.~K.}\ \bibnamefont {Mark}}, \bibinfo {author}
  {\bibfnamefont {S.}~\bibnamefont {Choi}},\ and\ \bibinfo {author}
  {\bibfnamefont {O.}~\bibnamefont {Painter}},\ }\bibfield  {title} {\bibinfo
  {title} {A superconducting quantum simulator based on a photonic-bandgap
  metamaterial},\ }\href {https://www.science.org/doi/10.1126/science.ade7651}
  {\bibfield  {journal} {\bibinfo  {journal} {Science}\ }\textbf {\bibinfo
  {volume} {379}},\ \bibinfo {pages} {278} (\bibinfo {year}
  {2023})}\BibitemShut {NoStop}%
\bibitem [{\citenamefont {Scigliuzzo}\ \emph {et~al.}(2022)\citenamefont
  {Scigliuzzo}, \citenamefont {Calaj\`o}, \citenamefont {Ciccarello},
  \citenamefont {Perez~Lozano}, \citenamefont {Bengtsson}, \citenamefont
  {Scarlino}, \citenamefont {Wallraff}, \citenamefont {Chang}, \citenamefont
  {Delsing},\ and\ \citenamefont {Gasparinetti}}]{scigliuzzo2022controlling}%
  \BibitemOpen
  \bibfield  {author} {\bibinfo {author} {\bibfnamefont {M.}~\bibnamefont
  {Scigliuzzo}}, \bibinfo {author} {\bibfnamefont {G.}~\bibnamefont
  {Calaj\`o}}, \bibinfo {author} {\bibfnamefont {F.}~\bibnamefont
  {Ciccarello}}, \bibinfo {author} {\bibfnamefont {D.}~\bibnamefont
  {Perez~Lozano}}, \bibinfo {author} {\bibfnamefont {A.}~\bibnamefont
  {Bengtsson}}, \bibinfo {author} {\bibfnamefont {P.}~\bibnamefont {Scarlino}},
  \bibinfo {author} {\bibfnamefont {A.}~\bibnamefont {Wallraff}}, \bibinfo
  {author} {\bibfnamefont {D.}~\bibnamefont {Chang}}, \bibinfo {author}
  {\bibfnamefont {P.}~\bibnamefont {Delsing}},\ and\ \bibinfo {author}
  {\bibfnamefont {S.}~\bibnamefont {Gasparinetti}},\ }\bibfield  {title}
  {\bibinfo {title} {Controlling atom-photon bound states in an array of
  josephson-junction resonators},\ }\href
  {https://doi.org/10.1103/PhysRevX.12.031036} {\bibfield  {journal} {\bibinfo
  {journal} {Phys. Rev. X}\ }\textbf {\bibinfo {volume} {12}},\ \bibinfo
  {pages} {031036} (\bibinfo {year} {2022})}\BibitemShut {NoStop}%
\bibitem [{\citenamefont {Gong}\ and\ \citenamefont
  {Wang}(2021)}]{gong2021quantum}%
  \BibitemOpen
  \bibfield  {author} {\bibinfo {author} {\bibfnamefont {M.}~\bibnamefont
  {Gong}}\ and\ \bibinfo {author} {\bibfnamefont {e.~a.}\ \bibnamefont
  {Wang}},\ }\bibfield  {title} {\bibinfo {title} {Quantum walks on a
  programmable two-dimensional 62-qubit superconducting processor},\ }\href
  {https://doi.org/10.1126/science.abg7812} {\bibfield  {journal} {\bibinfo
  {journal} {Science}\ }\textbf {\bibinfo {volume} {372}},\ \bibinfo {pages}
  {948–952} (\bibinfo {year} {2021})}\BibitemShut {NoStop}%
\bibitem [{\citenamefont {Koll{\'a}r}\ \emph {et~al.}(2019)\citenamefont
  {Koll{\'a}r}, \citenamefont {Fitzpatrick},\ and\ \citenamefont
  {Houck}}]{kollar2019hyperbolic}%
  \BibitemOpen
  \bibfield  {author} {\bibinfo {author} {\bibfnamefont {A.~J.}\ \bibnamefont
  {Koll{\'a}r}}, \bibinfo {author} {\bibfnamefont {M.}~\bibnamefont
  {Fitzpatrick}},\ and\ \bibinfo {author} {\bibfnamefont {A.~A.}\ \bibnamefont
  {Houck}},\ }\bibfield  {title} {\bibinfo {title} {Hyperbolic lattices in
  circuit quantum electrodynamics},\ }\href@noop {} {\bibfield  {journal}
  {\bibinfo  {journal} {Nature}\ }\textbf {\bibinfo {volume} {571}},\ \bibinfo
  {pages} {45} (\bibinfo {year} {2019})}\BibitemShut {NoStop}%
\bibitem [{\citenamefont {Yu}\ \emph {et~al.}(2019)\citenamefont {Yu},
  \citenamefont {Muniz}, \citenamefont {Hung},\ and\ \citenamefont
  {Kimble}}]{yu2019two}%
  \BibitemOpen
  \bibfield  {author} {\bibinfo {author} {\bibfnamefont {S.-P.}\ \bibnamefont
  {Yu}}, \bibinfo {author} {\bibfnamefont {J.~A.}\ \bibnamefont {Muniz}},
  \bibinfo {author} {\bibfnamefont {C.-L.}\ \bibnamefont {Hung}},\ and\
  \bibinfo {author} {\bibfnamefont {H.}~\bibnamefont {Kimble}},\ }\bibfield
  {title} {\bibinfo {title} {Two-dimensional photonic crystals for engineering
  atom--light interactions},\ }\href
  {https://www.pnas.org/doi/full/10.1073/pnas.1822110116} {\bibfield  {journal}
  {\bibinfo  {journal} {Proceedings of the National Academy of Sciences}\
  }\textbf {\bibinfo {volume} {116}},\ \bibinfo {pages} {12743} (\bibinfo
  {year} {2019})}\BibitemShut {NoStop}%
\bibitem [{\citenamefont {Rui}\ \emph {et~al.}(2020)\citenamefont {Rui},
  \citenamefont {Wei}, \citenamefont {Rubio-Abadal}, \citenamefont {Hollerith},
  \citenamefont {Zeiher}, \citenamefont {Stamper-Kurn}, \citenamefont {Gross},\
  and\ \citenamefont {Bloch}}]{rui2020subradiant}%
  \BibitemOpen
  \bibfield  {author} {\bibinfo {author} {\bibfnamefont {J.}~\bibnamefont
  {Rui}}, \bibinfo {author} {\bibfnamefont {D.}~\bibnamefont {Wei}}, \bibinfo
  {author} {\bibfnamefont {A.}~\bibnamefont {Rubio-Abadal}}, \bibinfo {author}
  {\bibfnamefont {S.}~\bibnamefont {Hollerith}}, \bibinfo {author}
  {\bibfnamefont {J.}~\bibnamefont {Zeiher}}, \bibinfo {author} {\bibfnamefont
  {D.~M.}\ \bibnamefont {Stamper-Kurn}}, \bibinfo {author} {\bibfnamefont
  {C.}~\bibnamefont {Gross}},\ and\ \bibinfo {author} {\bibfnamefont
  {I.}~\bibnamefont {Bloch}},\ }\bibfield  {title} {\bibinfo {title} {A
  subradiant optical mirror formed by a single structured atomic layer},\
  }\href {https://doi.org/10.1038/s41586-020-2463-x} {\bibfield  {journal}
  {\bibinfo  {journal} {Nature}\ }\textbf {\bibinfo {volume} {583}},\ \bibinfo
  {pages} {369–374} (\bibinfo {year} {2020})}\BibitemShut {NoStop}%
\bibitem [{\citenamefont {Srakaew}\ \emph {et~al.}(2023)\citenamefont
  {Srakaew}, \citenamefont {Weckesser}, \citenamefont {Hollerith},
  \citenamefont {Wei}, \citenamefont {Adler}, \citenamefont {Bloch},\ and\
  \citenamefont {Zeiher}}]{srakaew2023subwavelength}%
  \BibitemOpen
  \bibfield  {author} {\bibinfo {author} {\bibfnamefont {K.}~\bibnamefont
  {Srakaew}}, \bibinfo {author} {\bibfnamefont {P.}~\bibnamefont {Weckesser}},
  \bibinfo {author} {\bibfnamefont {S.}~\bibnamefont {Hollerith}}, \bibinfo
  {author} {\bibfnamefont {D.}~\bibnamefont {Wei}}, \bibinfo {author}
  {\bibfnamefont {D.}~\bibnamefont {Adler}}, \bibinfo {author} {\bibfnamefont
  {I.}~\bibnamefont {Bloch}},\ and\ \bibinfo {author} {\bibfnamefont
  {J.}~\bibnamefont {Zeiher}},\ }\bibfield  {title} {\bibinfo {title} {A
  subwavelength atomic array switched by a single rydberg atom},\ }\href
  {https://doi.org/10.1038/s41567-023-01959-y} {\bibfield  {journal} {\bibinfo
  {journal} {Nature Physics}\ }\textbf {\bibinfo {volume} {19}},\ \bibinfo
  {pages} {714} (\bibinfo {year} {2023})}\BibitemShut {NoStop}%
\bibitem [{\citenamefont {Huang}\ \emph {et~al.}(2023)\citenamefont {Huang},
  \citenamefont {Yuan}, \citenamefont {Holman}, \citenamefont {Kwon},
  \citenamefont {Masson}, \citenamefont {Gutierrez-Jauregui}, \citenamefont
  {Asenjo-Garcia}, \citenamefont {Will},\ and\ \citenamefont
  {Yu}}]{HUANG2023100470}%
  \BibitemOpen
  \bibfield  {author} {\bibinfo {author} {\bibfnamefont {X.}~\bibnamefont
  {Huang}}, \bibinfo {author} {\bibfnamefont {W.}~\bibnamefont {Yuan}},
  \bibinfo {author} {\bibfnamefont {A.}~\bibnamefont {Holman}}, \bibinfo
  {author} {\bibfnamefont {M.}~\bibnamefont {Kwon}}, \bibinfo {author}
  {\bibfnamefont {S.~J.}\ \bibnamefont {Masson}}, \bibinfo {author}
  {\bibfnamefont {R.}~\bibnamefont {Gutierrez-Jauregui}}, \bibinfo {author}
  {\bibfnamefont {A.}~\bibnamefont {Asenjo-Garcia}}, \bibinfo {author}
  {\bibfnamefont {S.}~\bibnamefont {Will}},\ and\ \bibinfo {author}
  {\bibfnamefont {N.}~\bibnamefont {Yu}},\ }\bibfield  {title} {\bibinfo
  {title} {Metasurface holographic optical traps for ultracold atoms},\ }\href
  {https://doi.org/https://doi.org/10.1016/j.pquantelec.2023.100470} {\bibfield
   {journal} {\bibinfo  {journal} {Progress in Quantum Electronics}\ }\textbf
  {\bibinfo {volume} {89}},\ \bibinfo {pages} {100470} (\bibinfo {year}
  {2023})}\BibitemShut {NoStop}%
\bibitem [{\citenamefont {Marques}\ \emph {et~al.}(2021)\citenamefont
  {Marques}, \citenamefont {Shelykh},\ and\ \citenamefont
  {Iorsh}}]{marques2021bound}%
  \BibitemOpen
  \bibfield  {author} {\bibinfo {author} {\bibfnamefont {Y.}~\bibnamefont
  {Marques}}, \bibinfo {author} {\bibfnamefont {I.~A.}\ \bibnamefont
  {Shelykh}},\ and\ \bibinfo {author} {\bibfnamefont {I.~V.}\ \bibnamefont
  {Iorsh}},\ }\bibfield  {title} {\bibinfo {title} {Bound photonic pairs in 2d
  waveguide quantum electrodynamics},\ }\href
  {https://doi.org/10.1103/PhysRevLett.127.273602} {\bibfield  {journal}
  {\bibinfo  {journal} {Phys. Rev. Lett.}\ }\textbf {\bibinfo {volume} {127}},\
  \bibinfo {pages} {273602} (\bibinfo {year} {2021})}\BibitemShut {NoStop}%
\bibitem [{\citenamefont {Te\ifmmode~\check{c}\else \v{c}\fi{}er}\ \emph
  {et~al.}(2024)\citenamefont {Te\ifmmode~\check{c}\else \v{c}\fi{}er},
  \citenamefont {Di~Liberto}, \citenamefont {Silvi}, \citenamefont
  {Montangero}, \citenamefont {Romanato},\ and\ \citenamefont
  {Calaj\'o}}]{tevcer2024strongly}%
  \BibitemOpen
  \bibfield  {author} {\bibinfo {author} {\bibfnamefont {M.}~\bibnamefont
  {Te\ifmmode~\check{c}\else \v{c}\fi{}er}}, \bibinfo {author} {\bibfnamefont
  {M.}~\bibnamefont {Di~Liberto}}, \bibinfo {author} {\bibfnamefont
  {P.}~\bibnamefont {Silvi}}, \bibinfo {author} {\bibfnamefont
  {S.}~\bibnamefont {Montangero}}, \bibinfo {author} {\bibfnamefont
  {F.}~\bibnamefont {Romanato}},\ and\ \bibinfo {author} {\bibfnamefont
  {G.}~\bibnamefont {Calaj\'o}},\ }\bibfield  {title} {\bibinfo {title}
  {Strongly interacting photons in 2d waveguide qed},\ }\href
  {https://doi.org/10.1103/PhysRevLett.132.163602} {\bibfield  {journal}
  {\bibinfo  {journal} {Phys. Rev. Lett.}\ }\textbf {\bibinfo {volume} {132}},\
  \bibinfo {pages} {163602} (\bibinfo {year} {2024})}\BibitemShut {NoStop}%
\bibitem [{\citenamefont {Pedersen}\ \emph {et~al.}(2024)\citenamefont
  {Pedersen}, \citenamefont {Bruun},\ and\ \citenamefont
  {Pohl}}]{PhysRevResearch.6.043264}%
  \BibitemOpen
  \bibfield  {author} {\bibinfo {author} {\bibfnamefont {S.~P.}\ \bibnamefont
  {Pedersen}}, \bibinfo {author} {\bibfnamefont {G.~M.}\ \bibnamefont
  {Bruun}},\ and\ \bibinfo {author} {\bibfnamefont {T.}~\bibnamefont {Pohl}},\
  }\bibfield  {title} {\bibinfo {title} {Green's function approach to
  interacting lattice polaritons and optical nonlinearities in subwavelength
  arrays of quantum emitters},\ }\href
  {https://doi.org/10.1103/PhysRevResearch.6.043264} {\bibfield  {journal}
  {\bibinfo  {journal} {Phys. Rev. Res.}\ }\textbf {\bibinfo {volume} {6}},\
  \bibinfo {pages} {043264} (\bibinfo {year} {2024})}\BibitemShut {NoStop}%
\bibitem [{\citenamefont {Liang}\ \emph {et~al.}(1988)\citenamefont {Liang},
  \citenamefont {Doucot},\ and\ \citenamefont {Anderson}}]{liang1988some}%
  \BibitemOpen
  \bibfield  {author} {\bibinfo {author} {\bibfnamefont {S.}~\bibnamefont
  {Liang}}, \bibinfo {author} {\bibfnamefont {B.}~\bibnamefont {Doucot}},\ and\
  \bibinfo {author} {\bibfnamefont {P.~W.}\ \bibnamefont {Anderson}},\
  }\bibfield  {title} {\bibinfo {title} {Some new variational
  resonating-valence-bond-type wave functions for the spin-\textonehalf{}
  antiferromagnetic heisenberg model on a square lattice},\ }\href
  {https://doi.org/10.1103/PhysRevLett.61.365} {\bibfield  {journal} {\bibinfo
  {journal} {Phys. Rev. Lett.}\ }\textbf {\bibinfo {volume} {61}},\ \bibinfo
  {pages} {365} (\bibinfo {year} {1988})}\BibitemShut {NoStop}%
\bibitem [{\citenamefont {Anderson}(1973)}]{anderson1973resonating}%
  \BibitemOpen
  \bibfield  {author} {\bibinfo {author} {\bibfnamefont {P.}~\bibnamefont
  {Anderson}},\ }\bibfield  {title} {\bibinfo {title} {Resonating valence
  bonds: A new kind of insulator?},\ }\href
  {https://doi.org/https://doi.org/10.1016/0025-5408(73)90167-0} {\bibfield
  {journal} {\bibinfo  {journal} {Materials Research Bulletin}\ }\textbf
  {\bibinfo {volume} {8}},\ \bibinfo {pages} {153} (\bibinfo {year}
  {1973})}\BibitemShut {NoStop}%
\bibitem [{\citenamefont {Moessner}\ and\ \citenamefont
  {Raman}(2010)}]{moessner2010quantum}%
  \BibitemOpen
  \bibfield  {author} {\bibinfo {author} {\bibfnamefont {R.}~\bibnamefont
  {Moessner}}\ and\ \bibinfo {author} {\bibfnamefont {K.~S.}\ \bibnamefont
  {Raman}},\ }\bibfield  {title} {\bibinfo {title} {Quantum dimer models},\
  }in\ \href
  {https://link.springer.com/book/10.1007/978-3-642-10589-0#about-this-book}
  {\emph {\bibinfo {booktitle} {Introduction to frustrated magnetism:
  materials, experiments, theory}}}\ (\bibinfo  {publisher} {Springer},\
  \bibinfo {year} {2010})\ pp.\ \bibinfo {pages} {437--479}\BibitemShut
  {NoStop}%
\bibitem [{\citenamefont {Rokhsar}\ and\ \citenamefont
  {Kivelson}(1988)}]{PhysRevLett.61.2376}%
  \BibitemOpen
  \bibfield  {author} {\bibinfo {author} {\bibfnamefont {D.~S.}\ \bibnamefont
  {Rokhsar}}\ and\ \bibinfo {author} {\bibfnamefont {S.~A.}\ \bibnamefont
  {Kivelson}},\ }\bibfield  {title} {\bibinfo {title} {Superconductivity and
  the quantum hard-core dimer gas},\ }\href
  {https://doi.org/10.1103/PhysRevLett.61.2376} {\bibfield  {journal} {\bibinfo
   {journal} {Phys. Rev. Lett.}\ }\textbf {\bibinfo {volume} {61}},\ \bibinfo
  {pages} {2376} (\bibinfo {year} {1988})}\BibitemShut {NoStop}%
\bibitem [{\citenamefont {Kivelson}\ \emph {et~al.}(1987)\citenamefont
  {Kivelson}, \citenamefont {Rokhsar},\ and\ \citenamefont
  {Sethna}}]{PhysRevB.35.8865}%
  \BibitemOpen
  \bibfield  {author} {\bibinfo {author} {\bibfnamefont {S.~A.}\ \bibnamefont
  {Kivelson}}, \bibinfo {author} {\bibfnamefont {D.~S.}\ \bibnamefont
  {Rokhsar}},\ and\ \bibinfo {author} {\bibfnamefont {J.~P.}\ \bibnamefont
  {Sethna}},\ }\bibfield  {title} {\bibinfo {title} {Topology of the resonating
  valence-bond state: Solitons and high-${T}_{c}$ superconductivity},\ }\href
  {https://doi.org/10.1103/PhysRevB.35.8865} {\bibfield  {journal} {\bibinfo
  {journal} {Phys. Rev. B}\ }\textbf {\bibinfo {volume} {35}},\ \bibinfo
  {pages} {8865} (\bibinfo {year} {1987})}\BibitemShut {NoStop}%
\bibitem [{\citenamefont {Wen}\ and\ \citenamefont
  {Niu}(1990)}]{PhysRevB.41.9377}%
  \BibitemOpen
  \bibfield  {author} {\bibinfo {author} {\bibfnamefont {X.~G.}\ \bibnamefont
  {Wen}}\ and\ \bibinfo {author} {\bibfnamefont {Q.}~\bibnamefont {Niu}},\
  }\bibfield  {title} {\bibinfo {title} {Ground-state degeneracy of the
  fractional quantum hall states in the presence of a random potential and on
  high-genus riemann surfaces},\ }\href
  {https://doi.org/10.1103/PhysRevB.41.9377} {\bibfield  {journal} {\bibinfo
  {journal} {Phys. Rev. B}\ }\textbf {\bibinfo {volume} {41}},\ \bibinfo
  {pages} {9377} (\bibinfo {year} {1990})}\BibitemShut {NoStop}%
\bibitem [{\citenamefont {Savary}\ and\ \citenamefont
  {Balents}(2016)}]{savary2016quantum}%
  \BibitemOpen
  \bibfield  {author} {\bibinfo {author} {\bibfnamefont {L.}~\bibnamefont
  {Savary}}\ and\ \bibinfo {author} {\bibfnamefont {L.}~\bibnamefont
  {Balents}},\ }\bibfield  {title} {\bibinfo {title} {Quantum spin liquids: a
  review},\ }\href {https://doi.org/10.1088/0034-4885/80/1/016502} {\bibfield
  {journal} {\bibinfo  {journal} {Reports on Progress in Physics}\ }\textbf
  {\bibinfo {volume} {80}},\ \bibinfo {pages} {016502} (\bibinfo {year}
  {2016})}\BibitemShut {NoStop}%
\bibitem [{\citenamefont {Semeghini}\ \emph {et~al.}(2021)\citenamefont
  {Semeghini}, \citenamefont {Levine}, \citenamefont {Keesling}, \citenamefont
  {Ebadi}, \citenamefont {Wang}, \citenamefont {Bluvstein}, \citenamefont
  {Verresen}, \citenamefont {Pichler}, \citenamefont {Kalinowski},
  \citenamefont {Samajdar}, \citenamefont {Omran}, \citenamefont {Sachdev},
  \citenamefont {Vishwanath}, \citenamefont {Greiner}, \citenamefont
  {Vuletić},\ and\ \citenamefont {Lukin}}]{Semeghini2021}%
  \BibitemOpen
  \bibfield  {author} {\bibinfo {author} {\bibfnamefont {G.}~\bibnamefont
  {Semeghini}}, \bibinfo {author} {\bibfnamefont {H.}~\bibnamefont {Levine}},
  \bibinfo {author} {\bibfnamefont {A.}~\bibnamefont {Keesling}}, \bibinfo
  {author} {\bibfnamefont {S.}~\bibnamefont {Ebadi}}, \bibinfo {author}
  {\bibfnamefont {T.~T.}\ \bibnamefont {Wang}}, \bibinfo {author}
  {\bibfnamefont {D.}~\bibnamefont {Bluvstein}}, \bibinfo {author}
  {\bibfnamefont {R.}~\bibnamefont {Verresen}}, \bibinfo {author}
  {\bibfnamefont {H.}~\bibnamefont {Pichler}}, \bibinfo {author} {\bibfnamefont
  {M.}~\bibnamefont {Kalinowski}}, \bibinfo {author} {\bibfnamefont
  {R.}~\bibnamefont {Samajdar}}, \bibinfo {author} {\bibfnamefont
  {A.}~\bibnamefont {Omran}}, \bibinfo {author} {\bibfnamefont
  {S.}~\bibnamefont {Sachdev}}, \bibinfo {author} {\bibfnamefont
  {A.}~\bibnamefont {Vishwanath}}, \bibinfo {author} {\bibfnamefont
  {M.}~\bibnamefont {Greiner}}, \bibinfo {author} {\bibfnamefont
  {V.}~\bibnamefont {Vuletić}},\ and\ \bibinfo {author} {\bibfnamefont
  {M.~D.}\ \bibnamefont {Lukin}},\ }\bibfield  {title} {\bibinfo {title}
  {Probing topological spin liquids on a programmable quantum simulator},\
  }\href {https://doi.org/10.1126/science.abi8794} {\bibfield  {journal}
  {\bibinfo  {journal} {Science}\ }\textbf {\bibinfo {volume} {374}},\ \bibinfo
  {pages} {1242} (\bibinfo {year} {2021})}\BibitemShut {NoStop}%
\bibitem [{\citenamefont {Samajdar}\ \emph {et~al.}(2021)\citenamefont
  {Samajdar}, \citenamefont {Ho}, \citenamefont {Pichler}, \citenamefont
  {Lukin},\ and\ \citenamefont {Sachdev}}]{pnas.2015785118}%
  \BibitemOpen
  \bibfield  {author} {\bibinfo {author} {\bibfnamefont {R.}~\bibnamefont
  {Samajdar}}, \bibinfo {author} {\bibfnamefont {W.~W.}\ \bibnamefont {Ho}},
  \bibinfo {author} {\bibfnamefont {H.}~\bibnamefont {Pichler}}, \bibinfo
  {author} {\bibfnamefont {M.~D.}\ \bibnamefont {Lukin}},\ and\ \bibinfo
  {author} {\bibfnamefont {S.}~\bibnamefont {Sachdev}},\ }\bibfield  {title}
  {\bibinfo {title} {Quantum phases of rydberg atoms on a kagome lattice},\
  }\href {https://doi.org/10.1073/pnas.2015785118} {\bibfield  {journal}
  {\bibinfo  {journal} {Proceedings of the National Academy of Sciences}\
  }\textbf {\bibinfo {volume} {118}},\ \bibinfo {pages} {e2015785118} (\bibinfo
  {year} {2021})}\BibitemShut {NoStop}%
\bibitem [{\citenamefont {Zeng}\ \emph {et~al.}(2025)\citenamefont {Zeng},
  \citenamefont {Giudici},\ and\ \citenamefont
  {Pichler}}]{PhysRevResearch.7.L012006}%
  \BibitemOpen
  \bibfield  {author} {\bibinfo {author} {\bibfnamefont {Z.}~\bibnamefont
  {Zeng}}, \bibinfo {author} {\bibfnamefont {G.}~\bibnamefont {Giudici}},\ and\
  \bibinfo {author} {\bibfnamefont {H.}~\bibnamefont {Pichler}},\ }\bibfield
  {title} {\bibinfo {title} {Quantum dimer models with rydberg gadgets},\
  }\href {https://doi.org/10.1103/PhysRevResearch.7.L012006} {\bibfield
  {journal} {\bibinfo  {journal} {Phys. Rev. Res.}\ }\textbf {\bibinfo {volume}
  {7}},\ \bibinfo {pages} {L012006} (\bibinfo {year} {2025})}\BibitemShut
  {NoStop}%
\bibitem [{\citenamefont {Moessner}\ and\ \citenamefont
  {Sondhi}(2001)}]{PhysRevLett.86.1881}%
  \BibitemOpen
  \bibfield  {author} {\bibinfo {author} {\bibfnamefont {R.}~\bibnamefont
  {Moessner}}\ and\ \bibinfo {author} {\bibfnamefont {S.~L.}\ \bibnamefont
  {Sondhi}},\ }\bibfield  {title} {\bibinfo {title} {Resonating valence bond
  phase in the triangular lattice quantum dimer model},\ }\href
  {https://doi.org/10.1103/PhysRevLett.86.1881} {\bibfield  {journal} {\bibinfo
   {journal} {Phys. Rev. Lett.}\ }\textbf {\bibinfo {volume} {86}},\ \bibinfo
  {pages} {1881} (\bibinfo {year} {2001})}\BibitemShut {NoStop}%
\bibitem [{\citenamefont {Tabares}\ \emph {et~al.}(2023)\citenamefont
  {Tabares}, \citenamefont {Mu\~noz de~las Heras}, \citenamefont {Tagliacozzo},
  \citenamefont {Porras},\ and\ \citenamefont
  {Gonz\'alez-Tudela}}]{PhysRevLett.131.073602}%
  \BibitemOpen
  \bibfield  {author} {\bibinfo {author} {\bibfnamefont {C.}~\bibnamefont
  {Tabares}}, \bibinfo {author} {\bibfnamefont {A.}~\bibnamefont {Mu\~noz
  de~las Heras}}, \bibinfo {author} {\bibfnamefont {L.}~\bibnamefont
  {Tagliacozzo}}, \bibinfo {author} {\bibfnamefont {D.}~\bibnamefont
  {Porras}},\ and\ \bibinfo {author} {\bibfnamefont {A.}~\bibnamefont
  {Gonz\'alez-Tudela}},\ }\bibfield  {title} {\bibinfo {title} {Variational
  quantum simulators based on waveguide qed},\ }\href
  {https://doi.org/10.1103/PhysRevLett.131.073602} {\bibfield  {journal}
  {\bibinfo  {journal} {Phys. Rev. Lett.}\ }\textbf {\bibinfo {volume} {131}},\
  \bibinfo {pages} {073602} (\bibinfo {year} {2023})}\BibitemShut {NoStop}%
\bibitem [{\citenamefont {Murta}\ and\ \citenamefont
  {Fern\'andez-Rossier}(2024)}]{PhysRevB.109.035128}%
  \BibitemOpen
  \bibfield  {author} {\bibinfo {author} {\bibfnamefont {B.}~\bibnamefont
  {Murta}}\ and\ \bibinfo {author} {\bibfnamefont {J.}~\bibnamefont
  {Fern\'andez-Rossier}},\ }\bibfield  {title} {\bibinfo {title} {From
  heisenberg to hubbard: An initial state for the shallow quantum simulation of
  correlated electrons},\ }\href {https://doi.org/10.1103/PhysRevB.109.035128}
  {\bibfield  {journal} {\bibinfo  {journal} {Phys. Rev. B}\ }\textbf {\bibinfo
  {volume} {109}},\ \bibinfo {pages} {035128} (\bibinfo {year}
  {2024})}\BibitemShut {NoStop}%
\bibitem [{\citenamefont {Carmichael}(1999)}]{Carmichael1999}%
  \BibitemOpen
  \bibfield  {author} {\bibinfo {author} {\bibfnamefont {H.~J.}\ \bibnamefont
  {Carmichael}},\ }\href {https://doi.org/10.1007/978-3-662-03875-8} {\emph
  {\bibinfo {title} {Statistical Methods in Quantum Optics 1}}}\ (\bibinfo
  {publisher} {Springer Berlin Heidelberg},\ \bibinfo {year}
  {1999})\BibitemShut {NoStop}%
\bibitem [{\citenamefont {Breuer}\ and\ \citenamefont
  {Petruccione}(2007)}]{Breuer2007}%
  \BibitemOpen
  \bibfield  {author} {\bibinfo {author} {\bibfnamefont {H.-P.}\ \bibnamefont
  {Breuer}}\ and\ \bibinfo {author} {\bibfnamefont {F.}~\bibnamefont
  {Petruccione}},\ }\href
  {https://doi.org/10.1093/acprof:oso/9780199213900.001.0001} {\emph {\bibinfo
  {title} {The Theory of Open Quantum Systems}}}\ (\bibinfo  {publisher}
  {Oxford University {PressOxford}},\ \bibinfo {year} {2007})\BibitemShut
  {NoStop}%
\bibitem [{\citenamefont {Asenjo-Garcia}\ \emph
  {et~al.}(2017{\natexlab{a}})\citenamefont {Asenjo-Garcia}, \citenamefont
  {Moreno-Cardoner}, \citenamefont {Albrecht}, \citenamefont {Kimble},\ and\
  \citenamefont {Chang}}]{asenjo2017exponential}%
  \BibitemOpen
  \bibfield  {author} {\bibinfo {author} {\bibfnamefont {A.}~\bibnamefont
  {Asenjo-Garcia}}, \bibinfo {author} {\bibfnamefont {M.}~\bibnamefont
  {Moreno-Cardoner}}, \bibinfo {author} {\bibfnamefont {A.}~\bibnamefont
  {Albrecht}}, \bibinfo {author} {\bibfnamefont {H.~J.}\ \bibnamefont
  {Kimble}},\ and\ \bibinfo {author} {\bibfnamefont {D.~E.}\ \bibnamefont
  {Chang}},\ }\bibfield  {title} {\bibinfo {title} {Exponential improvement in
  photon storage fidelities using subradiance and ``selective radiance'' in
  atomic arrays},\ }\href {https://doi.org/10.1103/PhysRevX.7.031024}
  {\bibfield  {journal} {\bibinfo  {journal} {Phys. Rev. X}\ }\textbf {\bibinfo
  {volume} {7}},\ \bibinfo {pages} {031024} (\bibinfo {year}
  {2017}{\natexlab{a}})}\BibitemShut {NoStop}%
\bibitem [{\citenamefont {Asenjo-Garcia}\ \emph
  {et~al.}(2017{\natexlab{b}})\citenamefont {Asenjo-Garcia}, \citenamefont
  {Hood}, \citenamefont {Chang},\ and\ \citenamefont
  {Kimble}}]{asenjo2017atom}%
  \BibitemOpen
  \bibfield  {author} {\bibinfo {author} {\bibfnamefont {A.}~\bibnamefont
  {Asenjo-Garcia}}, \bibinfo {author} {\bibfnamefont {J.~D.}\ \bibnamefont
  {Hood}}, \bibinfo {author} {\bibfnamefont {D.~E.}\ \bibnamefont {Chang}},\
  and\ \bibinfo {author} {\bibfnamefont {H.~J.}\ \bibnamefont {Kimble}},\
  }\bibfield  {title} {\bibinfo {title} {Atom-light interactions in
  quasi-one-dimensional nanostructures: A green's-function perspective},\
  }\href {https://doi.org/10.1103/PhysRevA.95.033818} {\bibfield  {journal}
  {\bibinfo  {journal} {Phys. Rev. A}\ }\textbf {\bibinfo {volume} {95}},\
  \bibinfo {pages} {033818} (\bibinfo {year} {2017}{\natexlab{b}})}\BibitemShut
  {NoStop}%
\bibitem [{\citenamefont {Jenkins}\ and\ \citenamefont
  {Ruostekoski}(2012)}]{PhysRevA.86.031602}%
  \BibitemOpen
  \bibfield  {author} {\bibinfo {author} {\bibfnamefont {S.~D.}\ \bibnamefont
  {Jenkins}}\ and\ \bibinfo {author} {\bibfnamefont {J.}~\bibnamefont
  {Ruostekoski}},\ }\bibfield  {title} {\bibinfo {title} {Controlled
  manipulation of light by cooperative response of atoms in an optical
  lattice},\ }\href {https://doi.org/10.1103/PhysRevA.86.031602} {\bibfield
  {journal} {\bibinfo  {journal} {Phys. Rev. A}\ }\textbf {\bibinfo {volume}
  {86}},\ \bibinfo {pages} {031602} (\bibinfo {year} {2012})}\BibitemShut
  {NoStop}%
\bibitem [{\citenamefont {Bettles}\ \emph {et~al.}(2016)\citenamefont
  {Bettles}, \citenamefont {Gardiner},\ and\ \citenamefont
  {Adams}}]{bettles2016enhanced}%
  \BibitemOpen
  \bibfield  {author} {\bibinfo {author} {\bibfnamefont {R.~J.}\ \bibnamefont
  {Bettles}}, \bibinfo {author} {\bibfnamefont {S.~A.}\ \bibnamefont
  {Gardiner}},\ and\ \bibinfo {author} {\bibfnamefont {C.~S.}\ \bibnamefont
  {Adams}},\ }\bibfield  {title} {\bibinfo {title} {Enhanced optical cross
  section via collective coupling of atomic dipoles in a 2d array},\ }\href
  {https://doi.org/10.1103/PhysRevLett.116.103602} {\bibfield  {journal}
  {\bibinfo  {journal} {Phys. Rev. Lett.}\ }\textbf {\bibinfo {volume} {116}},\
  \bibinfo {pages} {103602} (\bibinfo {year} {2016})}\BibitemShut {NoStop}%
\bibitem [{\citenamefont {Shahmoon}\ \emph {et~al.}(2017)\citenamefont
  {Shahmoon}, \citenamefont {Wild}, \citenamefont {Lukin},\ and\ \citenamefont
  {Yelin}}]{shahmoon2017cooperative}%
  \BibitemOpen
  \bibfield  {author} {\bibinfo {author} {\bibfnamefont {E.}~\bibnamefont
  {Shahmoon}}, \bibinfo {author} {\bibfnamefont {D.~S.}\ \bibnamefont {Wild}},
  \bibinfo {author} {\bibfnamefont {M.~D.}\ \bibnamefont {Lukin}},\ and\
  \bibinfo {author} {\bibfnamefont {S.~F.}\ \bibnamefont {Yelin}},\ }\bibfield
  {title} {\bibinfo {title} {Cooperative resonances in light scattering from
  two-dimensional atomic arrays},\ }\href
  {https://doi.org/10.1103/PhysRevLett.118.113601} {\bibfield  {journal}
  {\bibinfo  {journal} {Phys. Rev. Lett.}\ }\textbf {\bibinfo {volume} {118}},\
  \bibinfo {pages} {113601} (\bibinfo {year} {2017})}\BibitemShut {NoStop}%
\bibitem [{\citenamefont {Manzoni}\ \emph {et~al.}(2018)\citenamefont
  {Manzoni}, \citenamefont {Moreno-Cardoner}, \citenamefont {Asenjo-Garcia},
  \citenamefont {Porto}, \citenamefont {Gorshkov},\ and\ \citenamefont
  {Chang}}]{manzoni2018optimization}%
  \BibitemOpen
  \bibfield  {author} {\bibinfo {author} {\bibfnamefont {M.~T.}\ \bibnamefont
  {Manzoni}}, \bibinfo {author} {\bibfnamefont {M.}~\bibnamefont
  {Moreno-Cardoner}}, \bibinfo {author} {\bibfnamefont {A.}~\bibnamefont
  {Asenjo-Garcia}}, \bibinfo {author} {\bibfnamefont {J.~V.}\ \bibnamefont
  {Porto}}, \bibinfo {author} {\bibfnamefont {A.~V.}\ \bibnamefont
  {Gorshkov}},\ and\ \bibinfo {author} {\bibfnamefont {D.~E.}\ \bibnamefont
  {Chang}},\ }\bibfield  {title} {\bibinfo {title} {Optimization of photon
  storage fidelity in ordered atomic arrays},\ }\href
  {https://doi.org/10.1088/1367-2630/aadb74} {\bibfield  {journal} {\bibinfo
  {journal} {New Journal of Physics}\ }\textbf {\bibinfo {volume} {20}},\
  \bibinfo {pages} {083048} (\bibinfo {year} {2018})}\BibitemShut {NoStop}%
\bibitem [{\citenamefont {Facchinetti}\ \emph {et~al.}(2016)\citenamefont
  {Facchinetti}, \citenamefont {Jenkins},\ and\ \citenamefont
  {Ruostekoski}}]{PhysRevLett.117.243601}%
  \BibitemOpen
  \bibfield  {author} {\bibinfo {author} {\bibfnamefont {G.}~\bibnamefont
  {Facchinetti}}, \bibinfo {author} {\bibfnamefont {S.~D.}\ \bibnamefont
  {Jenkins}},\ and\ \bibinfo {author} {\bibfnamefont {J.}~\bibnamefont
  {Ruostekoski}},\ }\bibfield  {title} {\bibinfo {title} {Storing light with
  subradiant correlations in arrays of atoms},\ }\href
  {https://doi.org/10.1103/PhysRevLett.117.243601} {\bibfield  {journal}
  {\bibinfo  {journal} {Phys. Rev. Lett.}\ }\textbf {\bibinfo {volume} {117}},\
  \bibinfo {pages} {243601} (\bibinfo {year} {2016})}\BibitemShut {NoStop}%
\bibitem [{\citenamefont {Ruostekoski}(2023)}]{PhysRevA.108.030101}%
  \BibitemOpen
  \bibfield  {author} {\bibinfo {author} {\bibfnamefont {J.}~\bibnamefont
  {Ruostekoski}},\ }\bibfield  {title} {\bibinfo {title} {Cooperative
  quantum-optical planar arrays of atoms},\ }\href
  {https://doi.org/10.1103/PhysRevA.108.030101} {\bibfield  {journal} {\bibinfo
   {journal} {Phys. Rev. A}\ }\textbf {\bibinfo {volume} {108}},\ \bibinfo
  {pages} {030101} (\bibinfo {year} {2023})}\BibitemShut {NoStop}%
\bibitem [{\citenamefont {Wootters}(1998)}]{PhysRevLett.80.2245}%
  \BibitemOpen
  \bibfield  {author} {\bibinfo {author} {\bibfnamefont {W.~K.}\ \bibnamefont
  {Wootters}},\ }\bibfield  {title} {\bibinfo {title} {Entanglement of
  formation of an arbitrary state of two qubits},\ }\href
  {https://doi.org/10.1103/PhysRevLett.80.2245} {\bibfield  {journal} {\bibinfo
   {journal} {Phys. Rev. Lett.}\ }\textbf {\bibinfo {volume} {80}},\ \bibinfo
  {pages} {2245} (\bibinfo {year} {1998})}\BibitemShut {NoStop}%
\bibitem [{\citenamefont {Plankensteiner}\ \emph {et~al.}(2015)\citenamefont
  {Plankensteiner}, \citenamefont {Ostermann}, \citenamefont {Ritsch},\ and\
  \citenamefont {Genes}}]{plankensteiner2015selective}%
  \BibitemOpen
  \bibfield  {author} {\bibinfo {author} {\bibfnamefont {D.}~\bibnamefont
  {Plankensteiner}}, \bibinfo {author} {\bibfnamefont {L.}~\bibnamefont
  {Ostermann}}, \bibinfo {author} {\bibfnamefont {H.}~\bibnamefont {Ritsch}},\
  and\ \bibinfo {author} {\bibfnamefont {C.}~\bibnamefont {Genes}},\ }\bibfield
   {title} {\bibinfo {title} {Selective protected state preparation of coupled
  dissipative quantum emitters},\ }\href
  {https://www.nature.com/articles/srep16231#citeas} {\bibfield  {journal}
  {\bibinfo  {journal} {Scientific reports}\ }\textbf {\bibinfo {volume} {5}},\
  \bibinfo {pages} {16231} (\bibinfo {year} {2015})}\BibitemShut {NoStop}%
\bibitem [{\citenamefont {Gonzalez-Tudela}\ \emph {et~al.}(2011)\citenamefont
  {Gonzalez-Tudela}, \citenamefont {Martin-Cano}, \citenamefont {Moreno},
  \citenamefont {Martin-Moreno}, \citenamefont {Tejedor},\ and\ \citenamefont
  {Garcia-Vidal}}]{PhysRevLett.106.020501}%
  \BibitemOpen
  \bibfield  {author} {\bibinfo {author} {\bibfnamefont {A.}~\bibnamefont
  {Gonzalez-Tudela}}, \bibinfo {author} {\bibfnamefont {D.}~\bibnamefont
  {Martin-Cano}}, \bibinfo {author} {\bibfnamefont {E.}~\bibnamefont {Moreno}},
  \bibinfo {author} {\bibfnamefont {L.}~\bibnamefont {Martin-Moreno}}, \bibinfo
  {author} {\bibfnamefont {C.}~\bibnamefont {Tejedor}},\ and\ \bibinfo {author}
  {\bibfnamefont {F.~J.}\ \bibnamefont {Garcia-Vidal}},\ }\bibfield  {title}
  {\bibinfo {title} {Entanglement of two qubits mediated by one-dimensional
  plasmonic waveguides},\ }\href
  {https://doi.org/10.1103/PhysRevLett.106.020501} {\bibfield  {journal}
  {\bibinfo  {journal} {Phys. Rev. Lett.}\ }\textbf {\bibinfo {volume} {106}},\
  \bibinfo {pages} {020501} (\bibinfo {year} {2011})}\BibitemShut {NoStop}%
\bibitem [{\citenamefont {Rusconi}\ \emph {et~al.}(2021)\citenamefont
  {Rusconi}, \citenamefont {Shi},\ and\ \citenamefont
  {Cirac}}]{rusconi2021exploiting}%
  \BibitemOpen
  \bibfield  {author} {\bibinfo {author} {\bibfnamefont {C.~C.}\ \bibnamefont
  {Rusconi}}, \bibinfo {author} {\bibfnamefont {T.}~\bibnamefont {Shi}},\ and\
  \bibinfo {author} {\bibfnamefont {J.~I.}\ \bibnamefont {Cirac}},\ }\bibfield
  {title} {\bibinfo {title} {Exploiting the photonic nonlinearity of free-space
  subwavelength arrays of atoms},\ }\href
  {https://doi.org/10.1103/PhysRevA.104.033718} {\bibfield  {journal} {\bibinfo
   {journal} {Phys. Rev. A}\ }\textbf {\bibinfo {volume} {104}},\ \bibinfo
  {pages} {033718} (\bibinfo {year} {2021})}\BibitemShut {NoStop}%
\bibitem [{\citenamefont {van Diepen}\ \emph {et~al.}(2025)\citenamefont {van
  Diepen}, \citenamefont {Angelopoulou}, \citenamefont {Sandberg},
  \citenamefont {Tiranov}, \citenamefont {Wang}, \citenamefont {Scholz},
  \citenamefont {Ludwig}, \citenamefont {S{\o}rensen},\ and\ \citenamefont
  {Lodahl}}]{van2025resonant}%
  \BibitemOpen
  \bibfield  {author} {\bibinfo {author} {\bibfnamefont {C.~J.}\ \bibnamefont
  {van Diepen}}, \bibinfo {author} {\bibfnamefont {V.}~\bibnamefont
  {Angelopoulou}}, \bibinfo {author} {\bibfnamefont {O.~A.~D.}\ \bibnamefont
  {Sandberg}}, \bibinfo {author} {\bibfnamefont {A.}~\bibnamefont {Tiranov}},
  \bibinfo {author} {\bibfnamefont {Y.}~\bibnamefont {Wang}}, \bibinfo {author}
  {\bibfnamefont {S.}~\bibnamefont {Scholz}}, \bibinfo {author} {\bibfnamefont
  {A.}~\bibnamefont {Ludwig}}, \bibinfo {author} {\bibfnamefont {A.~S.}\
  \bibnamefont {S{\o}rensen}},\ and\ \bibinfo {author} {\bibfnamefont
  {P.}~\bibnamefont {Lodahl}},\ }\bibfield  {title} {\bibinfo {title} {Resonant
  energy transfer and collectively driven emitters in waveguide qed},\ }\href
  {https://arxiv.org/abs/2502.17662} {\bibfield  {journal} {\bibinfo  {journal}
  {arXiv preprint arXiv:2502.17662}\ } (\bibinfo {year} {2025})}\BibitemShut
  {NoStop}%
\bibitem [{\citenamefont {Caneva}\ \emph {et~al.}(2011)\citenamefont {Caneva},
  \citenamefont {Calarco}, \citenamefont {Fazio}, \citenamefont {Santoro},\
  and\ \citenamefont {Montangero}}]{PhysRevA.84.012312}%
  \BibitemOpen
  \bibfield  {author} {\bibinfo {author} {\bibfnamefont {T.}~\bibnamefont
  {Caneva}}, \bibinfo {author} {\bibfnamefont {T.}~\bibnamefont {Calarco}},
  \bibinfo {author} {\bibfnamefont {R.}~\bibnamefont {Fazio}}, \bibinfo
  {author} {\bibfnamefont {G.~E.}\ \bibnamefont {Santoro}},\ and\ \bibinfo
  {author} {\bibfnamefont {S.}~\bibnamefont {Montangero}},\ }\bibfield  {title}
  {\bibinfo {title} {Speeding up critical system dynamics through optimized
  evolution},\ }\href {https://doi.org/10.1103/PhysRevA.84.012312} {\bibfield
  {journal} {\bibinfo  {journal} {Phys. Rev. A}\ }\textbf {\bibinfo {volume}
  {84}},\ \bibinfo {pages} {012312} (\bibinfo {year} {2011})}\BibitemShut
  {NoStop}%
\bibitem [{\citenamefont {Omran}\ \emph {et~al.}(2019)\citenamefont {Omran},
  \citenamefont {Levine}, \citenamefont {Keesling}, \citenamefont {Semeghini},
  \citenamefont {Wang}, \citenamefont {Ebadi}, \citenamefont {Bernien},
  \citenamefont {Zibrov}, \citenamefont {Pichler}, \citenamefont {Choi},
  \citenamefont {Cui}, \citenamefont {Rossignolo}, \citenamefont {Rembold},
  \citenamefont {Montangero}, \citenamefont {Calarco}, \citenamefont {Endres},
  \citenamefont {Greiner}, \citenamefont {Vuletić},\ and\ \citenamefont
  {Lukin}}]{science.aax9743}%
  \BibitemOpen
  \bibfield  {author} {\bibinfo {author} {\bibfnamefont {A.}~\bibnamefont
  {Omran}}, \bibinfo {author} {\bibfnamefont {H.}~\bibnamefont {Levine}},
  \bibinfo {author} {\bibfnamefont {A.}~\bibnamefont {Keesling}}, \bibinfo
  {author} {\bibfnamefont {G.}~\bibnamefont {Semeghini}}, \bibinfo {author}
  {\bibfnamefont {T.~T.}\ \bibnamefont {Wang}}, \bibinfo {author}
  {\bibfnamefont {S.}~\bibnamefont {Ebadi}}, \bibinfo {author} {\bibfnamefont
  {H.}~\bibnamefont {Bernien}}, \bibinfo {author} {\bibfnamefont {A.~S.}\
  \bibnamefont {Zibrov}}, \bibinfo {author} {\bibfnamefont {H.}~\bibnamefont
  {Pichler}}, \bibinfo {author} {\bibfnamefont {S.}~\bibnamefont {Choi}},
  \bibinfo {author} {\bibfnamefont {J.}~\bibnamefont {Cui}}, \bibinfo {author}
  {\bibfnamefont {M.}~\bibnamefont {Rossignolo}}, \bibinfo {author}
  {\bibfnamefont {P.}~\bibnamefont {Rembold}}, \bibinfo {author} {\bibfnamefont
  {S.}~\bibnamefont {Montangero}}, \bibinfo {author} {\bibfnamefont
  {T.}~\bibnamefont {Calarco}}, \bibinfo {author} {\bibfnamefont
  {M.}~\bibnamefont {Endres}}, \bibinfo {author} {\bibfnamefont
  {M.}~\bibnamefont {Greiner}}, \bibinfo {author} {\bibfnamefont
  {V.}~\bibnamefont {Vuletić}},\ and\ \bibinfo {author} {\bibfnamefont
  {M.~D.}\ \bibnamefont {Lukin}},\ }\bibfield  {title} {\bibinfo {title}
  {Generation and manipulation of schrödinger cat states in rydberg atom
  arrays},\ }\href {https://doi.org/10.1126/science.aax9743} {\bibfield
  {journal} {\bibinfo  {journal} {Science}\ }\textbf {\bibinfo {volume}
  {365}},\ \bibinfo {pages} {570} (\bibinfo {year} {2019})}\BibitemShut
  {NoStop}%
\bibitem [{\citenamefont {Bernien}\ \emph {et~al.}(2017)\citenamefont
  {Bernien}, \citenamefont {Schwartz}, \citenamefont {Keesling}, \citenamefont
  {Levine}, \citenamefont {Omran}, \citenamefont {Pichler}, \citenamefont
  {Choi}, \citenamefont {Zibrov}, \citenamefont {Endres}, \citenamefont
  {Greiner} \emph {et~al.}}]{levine2018probing}%
  \BibitemOpen
  \bibfield  {author} {\bibinfo {author} {\bibfnamefont {H.}~\bibnamefont
  {Bernien}}, \bibinfo {author} {\bibfnamefont {S.}~\bibnamefont {Schwartz}},
  \bibinfo {author} {\bibfnamefont {A.}~\bibnamefont {Keesling}}, \bibinfo
  {author} {\bibfnamefont {H.}~\bibnamefont {Levine}}, \bibinfo {author}
  {\bibfnamefont {A.}~\bibnamefont {Omran}}, \bibinfo {author} {\bibfnamefont
  {H.}~\bibnamefont {Pichler}}, \bibinfo {author} {\bibfnamefont
  {S.}~\bibnamefont {Choi}}, \bibinfo {author} {\bibfnamefont {A.~S.}\
  \bibnamefont {Zibrov}}, \bibinfo {author} {\bibfnamefont {M.}~\bibnamefont
  {Endres}}, \bibinfo {author} {\bibfnamefont {M.}~\bibnamefont {Greiner}},
  \emph {et~al.},\ }\bibfield  {title} {\bibinfo {title} {Probing many-body
  dynamics on a 51-atom quantum simulator},\ }\href
  {https://doi.org/10.1038/nature24622} {\bibfield  {journal} {\bibinfo
  {journal} {Nature}\ }\textbf {\bibinfo {volume} {551}},\ \bibinfo {pages}
  {579} (\bibinfo {year} {2017})}\BibitemShut {NoStop}%
\bibitem [{\citenamefont {Bintz}\ \emph {et~al.}(2024)\citenamefont {Bintz},
  \citenamefont {Liu}, \citenamefont {Hauschild}, \citenamefont {Khalifa},
  \citenamefont {Chatterjee}, \citenamefont {Zaletel},\ and\ \citenamefont
  {Yao}}]{bintz2024dirac}%
  \BibitemOpen
  \bibfield  {author} {\bibinfo {author} {\bibfnamefont {M.}~\bibnamefont
  {Bintz}}, \bibinfo {author} {\bibfnamefont {V.~S.}\ \bibnamefont {Liu}},
  \bibinfo {author} {\bibfnamefont {J.}~\bibnamefont {Hauschild}}, \bibinfo
  {author} {\bibfnamefont {A.}~\bibnamefont {Khalifa}}, \bibinfo {author}
  {\bibfnamefont {S.}~\bibnamefont {Chatterjee}}, \bibinfo {author}
  {\bibfnamefont {M.~P.}\ \bibnamefont {Zaletel}},\ and\ \bibinfo {author}
  {\bibfnamefont {N.~Y.}\ \bibnamefont {Yao}},\ }\bibfield  {title} {\bibinfo
  {title} {Dirac spin liquid in quantum dipole arrays},\ }\href
  {https://arxiv.org/abs/2406.00098} {\bibfield  {journal} {\bibinfo  {journal}
  {arXiv preprint arXiv:2406.00098}\ } (\bibinfo {year} {2024})}\BibitemShut
  {NoStop}%
\bibitem [{\citenamefont {Kastoryano}\ \emph {et~al.}(2011)\citenamefont
  {Kastoryano}, \citenamefont {Reiter},\ and\ \citenamefont
  {S\o{}rensen}}]{PhysRevLett.106.090502}%
  \BibitemOpen
  \bibfield  {author} {\bibinfo {author} {\bibfnamefont {M.~J.}\ \bibnamefont
  {Kastoryano}}, \bibinfo {author} {\bibfnamefont {F.}~\bibnamefont {Reiter}},\
  and\ \bibinfo {author} {\bibfnamefont {A.~S.}\ \bibnamefont {S\o{}rensen}},\
  }\bibfield  {title} {\bibinfo {title} {Dissipative preparation of
  entanglement in optical cavities},\ }\href
  {https://doi.org/10.1103/PhysRevLett.106.090502} {\bibfield  {journal}
  {\bibinfo  {journal} {Phys. Rev. Lett.}\ }\textbf {\bibinfo {volume} {106}},\
  \bibinfo {pages} {090502} (\bibinfo {year} {2011})}\BibitemShut {NoStop}%
\bibitem [{\citenamefont {Reiter}\ \emph {et~al.}(2012)\citenamefont {Reiter},
  \citenamefont {Kastoryano},\ and\ \citenamefont {Sørensen}}]{Reiter_2012}%
  \BibitemOpen
  \bibfield  {author} {\bibinfo {author} {\bibfnamefont {F.}~\bibnamefont
  {Reiter}}, \bibinfo {author} {\bibfnamefont {M.~J.}\ \bibnamefont
  {Kastoryano}},\ and\ \bibinfo {author} {\bibfnamefont {A.~S.}\ \bibnamefont
  {Sørensen}},\ }\bibfield  {title} {\bibinfo {title} {Driving two atoms in an
  optical cavity into an entangled steady state using engineered decay},\
  }\href {https://doi.org/10.1088/1367-2630/14/5/053022} {\bibfield  {journal}
  {\bibinfo  {journal} {New Journal of Physics}\ }\textbf {\bibinfo {volume}
  {14}},\ \bibinfo {pages} {053022} (\bibinfo {year} {2012})}\BibitemShut
  {NoStop}%
\bibitem [{\citenamefont {Lebreuilly}\ \emph {et~al.}(2017)\citenamefont
  {Lebreuilly}, \citenamefont {Biella}, \citenamefont {Storme}, \citenamefont
  {Rossini}, \citenamefont {Fazio}, \citenamefont {Ciuti},\ and\ \citenamefont
  {Carusotto}}]{Lebreuilly2017}%
  \BibitemOpen
  \bibfield  {author} {\bibinfo {author} {\bibfnamefont {J.}~\bibnamefont
  {Lebreuilly}}, \bibinfo {author} {\bibfnamefont {A.}~\bibnamefont {Biella}},
  \bibinfo {author} {\bibfnamefont {F.}~\bibnamefont {Storme}}, \bibinfo
  {author} {\bibfnamefont {D.}~\bibnamefont {Rossini}}, \bibinfo {author}
  {\bibfnamefont {R.}~\bibnamefont {Fazio}}, \bibinfo {author} {\bibfnamefont
  {C.}~\bibnamefont {Ciuti}},\ and\ \bibinfo {author} {\bibfnamefont
  {I.}~\bibnamefont {Carusotto}},\ }\bibfield  {title} {\bibinfo {title}
  {Stabilizing strongly correlated photon fluids with non-markovian
  reservoirs},\ }\href {https://doi.org/10.1103/PhysRevA.96.033828} {\bibfield
  {journal} {\bibinfo  {journal} {Phys. Rev. A}\ }\textbf {\bibinfo {volume}
  {96}},\ \bibinfo {pages} {033828} (\bibinfo {year} {2017})}\BibitemShut
  {NoStop}%
\bibitem [{\citenamefont {Ma}\ \emph {et~al.}(2019)\citenamefont {Ma},
  \citenamefont {Saxberg}, \citenamefont {Owens}, \citenamefont {Leung},
  \citenamefont {Lu}, \citenamefont {Simon},\ and\ \citenamefont
  {Schuster}}]{ma2019dissipatively}%
  \BibitemOpen
  \bibfield  {author} {\bibinfo {author} {\bibfnamefont {R.}~\bibnamefont
  {Ma}}, \bibinfo {author} {\bibfnamefont {B.}~\bibnamefont {Saxberg}},
  \bibinfo {author} {\bibfnamefont {C.}~\bibnamefont {Owens}}, \bibinfo
  {author} {\bibfnamefont {N.}~\bibnamefont {Leung}}, \bibinfo {author}
  {\bibfnamefont {Y.}~\bibnamefont {Lu}}, \bibinfo {author} {\bibfnamefont
  {J.}~\bibnamefont {Simon}},\ and\ \bibinfo {author} {\bibfnamefont {D.~I.}\
  \bibnamefont {Schuster}},\ }\bibfield  {title} {\bibinfo {title} {A
  dissipatively stabilized mott insulator of photons},\ }\href
  {https://www.nature.com/articles/s41586-019-0897-9} {\bibfield  {journal}
  {\bibinfo  {journal} {Nature}\ }\textbf {\bibinfo {volume} {566}},\ \bibinfo
  {pages} {51} (\bibinfo {year} {2019})}\BibitemShut {NoStop}%
\bibitem [{\citenamefont {Periwal}\ \emph {et~al.}(2021)\citenamefont
  {Periwal}, \citenamefont {Cooper}, \citenamefont {Kunkel}, \citenamefont
  {Wienand}, \citenamefont {Davis},\ and\ \citenamefont
  {Schleier-Smith}}]{periwal2021programmable}%
  \BibitemOpen
  \bibfield  {author} {\bibinfo {author} {\bibfnamefont {A.}~\bibnamefont
  {Periwal}}, \bibinfo {author} {\bibfnamefont {E.~S.}\ \bibnamefont {Cooper}},
  \bibinfo {author} {\bibfnamefont {P.}~\bibnamefont {Kunkel}}, \bibinfo
  {author} {\bibfnamefont {J.~F.}\ \bibnamefont {Wienand}}, \bibinfo {author}
  {\bibfnamefont {E.~J.}\ \bibnamefont {Davis}},\ and\ \bibinfo {author}
  {\bibfnamefont {M.}~\bibnamefont {Schleier-Smith}},\ }\bibfield  {title}
  {\bibinfo {title} {Programmable interactions and emergent geometry in an
  array of atom clouds},\ }\href
  {https://www.nature.com/articles/s41586-021-04156-0#citeas} {\bibfield
  {journal} {\bibinfo  {journal} {Nature}\ }\textbf {\bibinfo {volume} {600}},\
  \bibinfo {pages} {630} (\bibinfo {year} {2021})}\BibitemShut {NoStop}%
\bibitem [{\citenamefont {Weinberg}\ and\ \citenamefont
  {Bukov}(2017)}]{10.21468/SciPostPhys.2.1.003}%
  \BibitemOpen
  \bibfield  {author} {\bibinfo {author} {\bibfnamefont {P.}~\bibnamefont
  {Weinberg}}\ and\ \bibinfo {author} {\bibfnamefont {M.}~\bibnamefont
  {Bukov}},\ }\bibfield  {title} {\bibinfo {title} {{QuSpin: a Python package
  for dynamics and exact diagonalisation of quantum many body systems part I:
  spin chains}},\ }\href {https://doi.org/10.21468/SciPostPhys.2.1.003}
  {\bibfield  {journal} {\bibinfo  {journal} {SciPost Phys.}\ }\textbf
  {\bibinfo {volume} {2}},\ \bibinfo {pages} {003} (\bibinfo {year}
  {2017})}\BibitemShut {NoStop}%
\bibitem [{\citenamefont {Weinberg}\ and\ \citenamefont
  {Bukov}(2019)}]{10.21468/SciPostPhys.7.2.020}%
  \BibitemOpen
  \bibfield  {author} {\bibinfo {author} {\bibfnamefont {P.}~\bibnamefont
  {Weinberg}}\ and\ \bibinfo {author} {\bibfnamefont {M.}~\bibnamefont
  {Bukov}},\ }\bibfield  {title} {\bibinfo {title} {{QuSpin: a Python package
  for dynamics and exact diagonalisation of quantum many body systems. Part II:
  bosons, fermions and higher spins}},\ }\href
  {https://doi.org/10.21468/SciPostPhys.7.2.020} {\bibfield  {journal}
  {\bibinfo  {journal} {SciPost Phys.}\ }\textbf {\bibinfo {volume} {7}},\
  \bibinfo {pages} {020} (\bibinfo {year} {2019})}\BibitemShut {NoStop}%
\end{thebibliography}%

\end{document}